\documentclass[12pt]{article}
\usepackage{graphicx}

 \newcommand{\Sbf}{\mbox{\boldmath $S$}}
 \newcommand{\Abf}{\mbox{\boldmath $A$}}
 \newcommand{\sinc}{{\rm sinc}}
 
 \newcommand{\wmin}{\om_{{\rm min}}}
 \newcommand{\wmax}{\om_{{\rm max}}}
 \newcommand{\kr}{k_{\rho}}

 \newcommand{\Irm}{{\rm I}}
 \newcommand{\Rrm}{{\rm R}}

 \newcommand{\ra}{\rightarrow}

 \newcommand{\equ}{\; \equiv \;}
 \newcommand{\xbf}{\mbox{\boldmath $x$}}

 \newcommand{\imp}{\mbox{\boldmath $p$}}
 \newcommand{\Ebf}{\mbox{\boldmath $E$}}
 \newcommand{\kbf}{\mbox{\boldmath $k$}}
 
 \newcommand{\Hbf}{\mbox{\boldmath $H$}}

 \newcommand{\nabf}{\mbox{\boldmath $\nabla$}}

 \newcommand{\minrm}{{\rm min}}
 
 \newcommand{\maxrm}{{\rm max}}
 \newcommand{\frm}{{\rm f}}

 \newcommand{\orm}{{\rm o}}
 \newcommand{\pa}{\partial}

 \newcommand{\X}{{\rm X}}

 \newcommand{\text}{\rm}
 
 \newcommand{\drm}{{\rm d}}
 \newcommand{\grm}{{\rm g}}

 \newcommand{\Ccal}{{\cal C}}
 \newcommand{\ug}{ \; = \; }
 
 \newcommand{\ugg}{ \ = \ }
 \newcommand{\ga}{\gamma}

 \newcommand{\infi}{\infty}

 \newcommand{\la}{\lambda}

 \newcommand{\al}{\alpha}
 \newcommand{\ze}{\zeta}

 \newcommand{\bb}{\begin{equation}}
 \newcommand{\ee}{\end{equation}}
 \newcommand{\bc}{\begin{center}}
 \newcommand{\ec}{\end{center}}
 \newcommand{\bega}{\begin{eqnarray}}
 \newcommand{\ega}{\end{eqnarray}}
 \newcommand{\begae}{\begin{eqnarray*}}
 \newcommand{\egae}{\end{eqnarray*}}

 \newcommand{\h}{\hspace*{4ex}}
 \newcommand{\dis}{\displaystyle}
 
 \newcommand{\be}{\beta}

 \newcommand{\rr}{\rho}

 \newcommand{\Om}{\Omega}
 \newcommand{\om}{\omega}

 \newcommand{\cent}{\centerline}
 \newcommand{\vs}{\vspace*}

\begin{document}
\baselineskip 0.5cm

\begin{center}

{\large {\bf LOCALIZED WAVES: A NOT-SO-SHORT REVIEW}$^{\:
(\dag)}$} \footnotetext{$^{\: (\dag)}$  Work partially supported
by FAPESP (Brazil), and by MIUR and INFN (Italy). \ E-mail
addresses: \ recami@mi.infn.it ; \ mzamboni@ufabc.edu.br }

\end{center}

\vs{5mm}

\cent{ Erasmo Recami }

\vs{0.2 cm}

\cent{{\em Facolt\`a di Ingegneria, Universit\`a statale di
Bergamo, Bergamo, Italy;}} \cent{{\rm and} {\em
INFN---Sezione di Milano, Milan, Italy.}}

\vs{0.5 cm}

\centerline{\rm and}

\vs{0.3 cm}

\cent{ Michel Zamboni-Rached, }

\vs{0.2 cm}

\centerline{{\em Centro de Ci\^encias Naturais e Humanas,
Universidade Federal do ABC,}} \cent{{\em Santo Andr\'e, SP,
Brasil}}

\vs{0.5 cm}

{\bf Abstract  \ --} \ In the {\em First Part} of this paper
(which is mainly a review) we present simple, general and formal,
introductions to the ordinary gaussian waves and to the Bessel
waves, by explicitly separating the case of beams from the case of
pulses; and, afterwards, an analogous introduction is presented
for the Localized Waves (LW), pulses or beams.  Always we stress
the very different characteristics of the gaussian with respect to
the Bessel waves and to the LWs, showing the numerous important
properties of the latter:   Properties that may find application
in all fields in which an essential role is played by a
wave-equation (like electromagnetism, optics, acoustics,
seismology, geophysics, gravitation, elementary particle physics,
etc.). \ The First Part of this review ends with an {\em
Appendix,} wherein: \ (i) we recall how, in the seventies and
eighties, the geometrical methods of Special Relativity (SR)
predicted
---in the sense below specified--- the existence of the most
interesting LWs, i.e., of the X-shaped pulses; \ and \ (ii) in
connection with the circumstance that the X-shaped waves are
endowed with Superluminal group-velocities (as discussed
in the first part of this paper), we briefly mention the various
experimental sectors of physics in which Superluminal motions seem
to appear; in particular, a bird's-eye view is presented
of the experiments till now performed with evanescent waves (and/or
tunnelling photons), and with the ``localized Superluminal
solutions" to the wave equations.

\h In the {\em Second Part} of this work, we address in more detail various
theoretical approaches leading to
nondiffracting solutions of the linear wave equation in unbounded
homogeneous media, as well as some interesting applications of
these waves. After some more introductory remarks (Sec.VI), we
analyse in Section VII the general structure of the Localized
Waves, develop the so called Generalized Bidirectional
Decomposition, and use it to obtain several luminal and
Superluminal nondiffracting solutions of the wave equations. \
In Section VIII we present a method for getting a space-time focusing
by a continuous superposition of X-Shaped pulses of different
velocities. \ Section IX addresses the properties of chirped
optical X-Shaped pulses propagating in material media without boundaries.

\h Finally, in the {\em Third Part} of this paper we ``complete" our review
by investigating also the (not less interesting) case of the {\em subluminal}
Localized Solutions to the wave equations, which, among the others, will
allow us to emphasize the remarkable role of SR, in its extended,
or rather non-restricted, formulation. [For instance, the various Superluminal and
subluminal LWs are expected to be transformed one into the other by suitable Lorentz
transformations]. \ We start by studying ---by means of various
approaches--- the very peculiar topic of zero-speed waves: Namely,
of the localized fields with a {\em static}
envelope; consisting, for instance, in ``light at rest". Actually,
in Section X we show how a suitable superposition
of Bessel beams can be used to construct stationary localized wave
fields with high transverse localization, and with a longitudinal
intensity pattern that assumes any desired shape within a chosen
interval $0 \leq z \leq L$ of the propagation axis.  We have called
{\em Frozen Waves} such solutions: As we shall see,
they can have a lot of noticeable applications. \  In between,
we do not forget to briefly treat the case of not axially-symmetric
solutions, in terms of higher order Bessel beams.

\h In this review we have fixed
our attention especially on electromagnetism and optics: but
results of the present kind are valid, let us repeat, whenever an
essential role is played by a wave-equation.\\


{\em Keywords:} \ Localized waves; Nondiffracting waves;
Superluminal waves; Electromagnetic X-shaped superluminal waves;
Acoustic X-shaped supersonic waves; Focussing of pulses; Chirped
optical pulses; Electromagnetic subluminal bullets; Acoustic
subsonic bullets; Frozen Waves; Zero-speed waves; Special
Relativity; Extended Relativity; Lorentz transformations; Wave
equations; Wave propagation; Bessel beams; Optics; Microwaves;
Acoustics; Seismology; Elementary particles; Gravitational waves;
Mechanical waves; Evanescent waves; Hartman Effect; Optical or
acoustic tweezers; Optical or acoustic scalpels; Corpuscle
guiding; Tumour cells destruction.

\newpage

\cent{\Huge{\bf FIRST  PART}}

\

\

\cent{\Large{\bf LOCALIZED WAVES:}}

\

\cent{\Large{\bf A SCIENTIFIC AND HISTORICAL INTRODUCTION}}

\

\

\section{A General Introduction} 

\vs{3mm}

\subsection{Preliminary remarks}

Diffraction and dispersion are known since long to be phenomena
limiting the applications of (optical, for instance) beams or
pulses.

\h {\em Diffraction} is always present, affecting any waves that
propagate in two or three-dimensional unbounded media, even when
homogeneous.  Pulses and beams are constituted by waves travelling
along different directions, which produces a gradual {\em spatial}
broadening\cite{Born}.  This effect is really a limiting factor
whenever a pulse is needed which maintains its transverse
localization, like, e.g., in free space
communications\cite{Willebrand}, {\em image forming\/}\cite{Goodman},
optical lithography\cite{Okazaki,Ito}, electromagnetic
{\em tweezers\/}\cite{Ashkin,Curtis}, etcetera.

\h {\em Dispersion} acts on pulses propagating in material media,
causing mainly a temporal broadening: An effect known to be due to
the variation of the refraction index with the frequency, so that
each spectral component of the pulse possesses a different
phase-velocity. This entails a gradual temporal widening, which
constitutes a limiting factor when a pulse is needed which
maintains its {\em time} width, like, e.g., in communication
systems\cite{Agrawal}.

\h It is important, therefore, to develop techniques able to
reduce those phenomena. \ The so-called {\em localized waves\/}
(LW), known also as non-diffracting waves, are indeed able to
resist diffraction for a long distance in free space. \ Such
solutions to the wave equations (and, in particular, to the
Maxwell equations, under weak hypotheses) were theoretically
predicted long time ago[9-12]
(cf. also\cite{SupCh}, as well as the Appendix located at the end of this
First Part), mathematically constructed in more recent
times\cite{Lu1,PhysicaA}, and soon after experimentally
produced[16-18]. \ Today, localized waves are
well-established both theoretically and experimentally, and are
having innovative applications not only in vacuum, but also in
material (linear or non-linear) media, showing to be able to
resist also dispersion. \ As we were mentioning, their potential
applications are being intensively explored, always with
surprising results, in fields like Acoustics, Microwaves, Optics,
and are promising also in Mechanics, Geophysics, and even
Gravitational Waves and Elementary particle physics. \ Worth
noticing appear also the applications of the so-called ``Frozen
Waves", that will be presented in the Third Part of this work; while
rather interesting are the applications {\em already} obtained,
for instance, in high-resolution ultra-sound scanning of moving
organs in human body\cite{LuBiomedical,LuImaging}.

\h To confine ourselves to electromagnetism, let us recall the
present-day studies on electromagnetic
{\em tweezers\/}[21-24], optical (or
acoustic) scalpels, {\em optical guiding of atoms or (charged or
neutral) corpuscles\/}[25-27], optical
litography\cite{Erdelyi,Garces}, {\em optical (or acoustic)
images\/}\cite{Herman}, communications in free
space[30-32,14], remote optical
alignment\cite{Vasara}, {\em optical acceleration of charged
particles,} and so on.

\h In the following two Subsections we are going to set forth a
brief {\em introduction} to the theory and applications of localized
beams and localized pulses, respectively.\cite{Introd}

\h Before going on, let us explicitly remark that ---as in any review
article, for obvious reasons of space--- we had to select a few main
topics: and such a choice can only be a personal one.

\

{\em Localized (non-diffracting) {\bf beams}} --- The word {\em
beam} refers to a monochromatic solution to the considered wave
equation, with a transverse localization of its field.  To fix our
ideas, we shall explicitly refer to the optical case: But our
considerations, of course, hold for any wave equation (vectorial,
spinorial, scalar...: in particular, for the acoustic case too).

\h The most common type of optical beam is the gaussian one, whose
transverse behaviour is described by a gaussian function. But all
the common beams suffer a diffraction, which spoils the transverse
shape of their field, widening it gradually during propagation. As
an example, the transverse width of a gaussian beam doubles when
it travels a distance $z_{\rm
dif}=\sqrt{3}\pi\Delta\rho_0^2/\lambda_0$, where $\Delta\rho_0$ is
the beam initial width and $\lambda_0$ is its wavelength. One can
verify that a gaussian beam with an initial transverse aperture of
the order of its wavelength will already double its width after
having travelled a few wavelengths.

\h It was generally believed that the only wave devoid of
diffraction was the plane wave, which does not suffer any
transverse changes. Some authors had shown, actually, that it
isn't the only one. For instance, in 1941 Stratton\cite{Stratton}
obtained a monochromatic solution to the wave equation whose
transverse shape was concentrated in the vicinity of its
propagation axis and represented by a Bessel function. Such a
solution, now called a Bessel beam, was not subject to
diffraction, since no change in its transverse shape took place
with time. In ref.\cite{Courant} it was later on demonstrated how
a large class of equations (including the wave equations) admit
``non-distorted progressing waves" as solutions; while already in
1915, in ref.\cite{Bateman}, and subsequently in articles like
ref.\cite{Barut2}, it was shown the existence of soliton-like,
wavelet-type solutions to the Maxwell equations. But all such
literature did not raise the attention it deserved. In the case of
ref.\cite{Stratton}, this can be partially justified since that
(Bessel) beam was associated with an infinite power flux [as much as the
plane waves, incidentally], it being not
square-integrable in the {\em transverse} direction. An interesting
problem, therefore, was that
of investigating what it would happen to the ideal Bessel beam
solution when truncated by a finite transverse aperture.

\h Only in 1987 a heuristical answer came from the known
experiment by Durnin et al.\cite{Durnin},
when it was shown that a
realistic Bessel beam, endowed with wavelength
$\lambda_0=0.6328\;\mu$m and central spot\footnote{Let us define
the size of the central ``spot'' of a Bessel beam as the distance, along the
transverse direction $\rho$, at which the first zero occurs of the
Bessel function characterizing its transverse shape.}
$\Delta\rho_0=59\;\mu$m, passing through an aperture with radius
$R=3.5\;$mm is able to travel about $85\;$cm keeping its
transverse intensity shape approximately unchanged (in the region
$\rho << R$ surrounding its central peak). In other words, it was
experimentally shown that the transverse intensity peak, as well
as the field in the surroundings of it, do not meet any appreciable
change in shape all along a {\em large} ``depth of field". As a
comparison, let us recall once more that a gaussian beam with the
same wavelength, and with the central ``spot''\footnote{In the
case of a gaussian beam, let us define the size of its central ``spot'' as the
distance, along the transverse direction $\rho$, at which its intensity has
decayed of the factor $1/e$.} $\Delta\rho_0=59\;\mu$m, when
passing through an aperture with the same radius $R=3.5\;$mm
doubles its transverse width after $3\;$cm, and after $6\;$cm its
intensity is already diminished by a factor 10. \ Therefore, in
the considered case, a Bessel beams can travel, approximately
without deformation, a distance 28 times larger than a gaussian
beam's.

\h Such a remarkable property is due to the fact that the
transverse intensity fields (whose value decreases with increasing
$\rho$), associated with the rings which constitute the
(transverse) structure of the Bessel beam, when diffracting end
up {\em reconstructing} the beam itself, all along a large
field-depth. This depends on the Bessel beam spectrum
(wavenumber and frequency)\cite{Herman,Durnin2,Vasara}, as
explained in detail in our ref.\cite{MRH}. \ Let us stress that,
given a Bessel and a gaussian beam ---both with the same
energy $E$, the same spot
$\Delta\rho_0$ and passing through apertures with the same radius
$R$ in the plane $z=0$--- the
percentage of the total energy $E$ contained inside the central
peak region ($0 \leq \rho \leq \Delta\rho_0$) is smaller for a
Bessel than for a gaussian beam: This different energy-distribution
on the transverse plane is responsible for the reconstruction of
the Bessel-beam central peak even at large distances from the
source (and even after an obstacle, provided that its size is smaller
{\em than the aperture\/}[39-41]: a nice property possessed
also by the localized pulses we are going to examine
below\cite{Michel}).

\begin{figure}[!h]
\begin{center}
 \scalebox{1.6}{\includegraphics{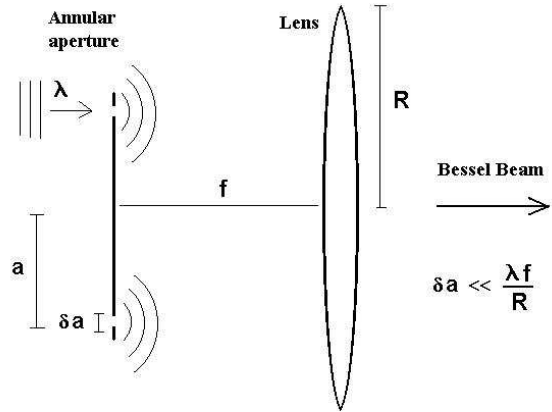}}  
\end{center}
\caption{The simple experimental set-up used by Durnin et al. for
generating a Bessel beam.} \label{fig1}
\end{figure}

\h It may be worth mentioning that most experiments carried on in
this area have been performed rapidly and with use, often, of
rather simple apparatus: The Durnin et al.'s experiment, e.g., had
recourse, for the generation of a Bessel beam, to a laser source,
an annular slit and a lens, as depicted in Fig.(\ref{fig1}). In a
sense, such an apparatus produces what can be regarded as the
cylindrically symmetric generalization of a couple of plane waves
emitted at angles $\theta$ and $-\theta$, with respect to (w.r.t.) the $z$-direction,
respectively (in which case the plane wave {\em intersection}
moves along $z$ with the speed $c/\cos \theta$). \ Of course,
these non-diffracting beams can be generated also by a a conic
lens ({\em axicon\/}) [cf., e.g., ref.\cite{Herman}], or by other
means like holographic elements [cf., e.g.,
refs.\cite{Vasara,MacDonald2}].

\h Let us stress, as already mentioned at the end of the
previous Subsection, that nowadays a lot of interesting
applications of non-diffracting beams are being investigated;
besides the Lu et al.'s ones in Acoustics. In the optical sector,
let us recall again those of using Bessel beams as optical
tweezers able to confine or move around small particles. \ In such
theoretical and application areas, a noticeable contribution is
the one presented in refs.[43-45],
wherein, by suitable superpositions of Bessel beams endowed with
the same frequency but different longitudinal wavenumbers, {\em
stationary} envelopes have been mathematically constructed in closed
form, which possess a high transverse localization and, more
important, a longitudinal intensity-shape that can be freely
chosen inside a predetermined space-interval $0 \leq z \leq L$.
For instance, a high intensity field, with a static envelope, can
be created within a tiny region, with negligible intensity
elsewhere: As already mentioned, the Third Part will deal, among the others,
with such ``Frozen Waves".

\

{\em Localized (non-diffracting) {\bf pulses}} --- As we have seen
in the previous Subsection, the existence of non-diffractive (or
localized) {\em pulses} was predicted since long: cf., once more,
refs.\cite{Bateman,Courant}, and, not less,
refs.\cite{BarutMR,SupCh}, as well as more recent articles like
refs.\cite{Barut3,Barut4}. \ The modern studies about
non-diffractive pulses (to confine ourselves, at least, to the
ones that attracted more attention) followed a development rather
independent of those on non-diffracting {\em beams}, even if both
phenomena are part of the same sector of physics: that of {\em
Localized Waves}.

\h In 1983, Brittingham\cite{Brittingham} set forth a luminal
($V=c$) solution to the wave equation (more particularly, to the
Maxwell equations) which travels rigidly, i.e., without
diffraction. The solution proposed in ref.\cite{Brittingham}
possessed however infinite energy, and once more the problem arose
of overcoming such a problem.

\h A way out was first obtained, as far as we know, by
Sezginer\cite{Sezginer}, who showed how to construct finite-energy
luminal pulses, which ---however--- do not propagate without
distortion for an infinite distance, but, as it is expected,
travel with constant speed, and approximately without deforming,
for a certain ({\em long\/}) depth of field: much longer, in this
case too, than that of the ordinary pulses like the gaussian ones.
In a series of subsequent papers[30,31,50-53], a simple
theoretical method was developed, called by those authors
``bidirectional decomposition", for constructing a new series of
non-diffracting {\em luminal} pulses.

\begin{figure}[!h]
\begin{center}
 \scalebox{1.1}{\includegraphics{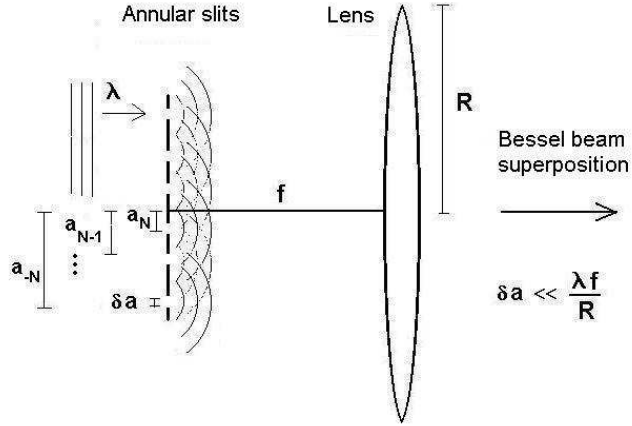}}  
\end{center}
\caption{One of the simplest experimental set-ups for generating
various kinds of Bessel beam superpositions.} \label{fig2}
\end{figure}

\h Eventually, at the beginning of the nineties, Lu et
al.\cite{Lu1,Lu2} constructed, both mathematically and
experimentally, new solutions to the wave equation in free space:
namely, an {\em X-shaped} localized pulse, with the form predicted by the
so-called extended Special Relativity\cite{BarutMR,Review}; for
the connection between what Lu et al. called ``X-waves" and
``extended" relativity see, e.g., ref.\cite{SupCh}, while brief
excerpts of that theory can be found, for instance, in
refs.[55-58,15]. \ Lu et al.'s solutions (which can be called
the ``classic" ones) were continuous superpositions of Bessel beams with
the same phase-velocity (i.e., with the same axicon
angle\cite{Friberg,PhysicaA,Lu1,Review}, $alpha$); so that they
could keep their shape for long
distances. \ Such X-shaped waves resulted to be interesting and
flexible localized solutions, and have been afterwards studied in
a number of papers, even if their velocity $V$ is supersonic or
Superluminal ($V>c$): Actually, when the phase-velocity does not
depend on the frequency, it is known that such a phase-velocity
becomes the ``group-velocity"... \ Remembering how a superposition of
Bessel beams is generated (for example, by a discrete or
continuous set of annular slits or transducers; or even by a single slit
plus a lens), it results clear
that the energy forming the Localized Waves, coming from those rings,
is transported at the ordinary speed $c$ of the plane waves in the
considered medium[60-62,15] \ (here $c$,
representing the velocity of the plane waves in the medium, is the
sound-speed in the acoustic case, and the speed of light in the
electromagnetic case; and so on). \ {\em Nevertheless}, the peak
of the LWs is {\em faster} than $c$. \ [Let us explicitly notice that, when
using {\em a lens} after an aperture located at its back focus, as
in Fig.\ref{fig2}, then a classic X-shaped pulse can be
generated even by a {\em single} annular slit,
or transducer, illuminated however by a light, or sound, pulse:
but the previous considerations about the actual transportation-speed
of the ``energy" forming the X-shaped wave remain unaffected. \
The experimental set-up depicted in Fig.2, with various annular
slits, is actually needed only for generating (X-shaped, e.g.)
pulses more complex than the classic one, namely, depending on the
co-ordinates $z$ and $t$ not only through the quantity $\zeta \equiv
z-Vt$: see the following].

\begin{figure}[!h]
\begin{center}
 \scalebox{3}{\includegraphics{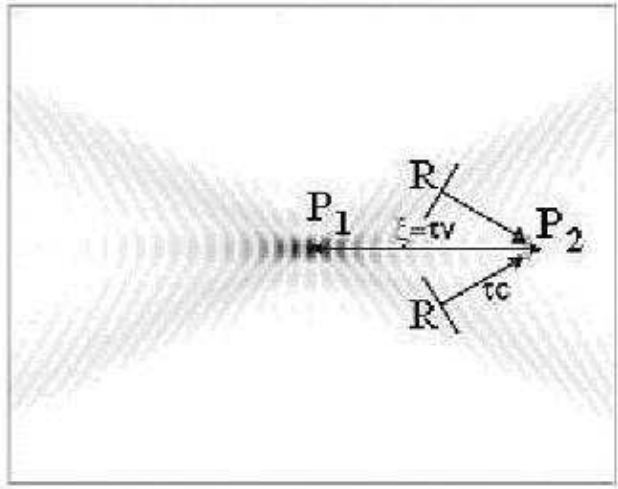}}        
\end{center}
\caption{This figure shows an X-shaped wave, that is, a localized
Superluminal pulse.  It refers to an X-wave, possessing the
velocity $V>c$, and illustrates the fact that, if its vertex or
central spot is located at $P_1$ at time $t_0$, it will reach the
position $P_2$ at a time $t+\tau$ where $\tau=|P_2-P_1|/V \; < \;
|P_2-P_1|/c$: This is something different from the illusory
``scissor effect", {\em even if the feeding energy, coming from
the regions $R$, has travelled with the ordinary speed $c$} (which
is the speed of light in the electromagnetic case, or the sound
speed in Acoustics, and so on).} \label{fig3}
\end{figure}

\h It is indeed possible to generate (besides the ``classic" X-wave
produced by Lu et al. in 1992) infinite sets of new X-shaped
waves, with their energy more and more concentrated in a spot
corresponding to the vertex region\cite{MRH}. It may therefore
appear rather intriguing that such a spot [even if no violations
of Special Relativity (SR) are obviously implied: all the results
come from Maxwell equations, or from the wave
equations\cite{Barbero,Brodowsky}]--- travels Superluminally when
the waves are electromagnetic. \ For simplicity, we shall call
``Superluminal" all
the X-shaped waves, even when the waves are acoustic. \ By
Fig.(\ref{fig3}), which refers to an X-wave possessing the
velocity $V>c$, we illustrate the fact that, if its vertex or
central spot is located at $P_1$ at time $t_1$, it will reach the
position $P_2$ at a time $t+\tau$ where $\tau=|P_2-P_1|/V \; < \;
|P_2-P_1|/c$: \ We shall discuss all these points below.

\h Soon after having mathematically and experimentally constructed
their ``classic" {\em acoustic} X-wave, Lu et al. started applying
them to ultrasonic scanning, obtaining ---as we have already said---
very high quality images. \ Subsequently, in a 1996 e-print and
report, Recami et al. (see, e.g., ref.\cite{PhysicaA} and refs.
therein) published the analogous X-shaped solutions to the Maxwell
equations: By constructing scalar Superluminal localized solutions
for each component of the Hertz potential. That showed, by the
way, that the localized solutions to the scalar equation can be
used, under very weak conditions, for obtaining localized
solutions to Maxwell's equations too (actually, Ziolkowski et
al.\cite{Ziolkowski0} had found something similar, called by them
{\em slingshot} pulses, for the simple scalar case; but their
solution had gone almost unnoticed). \ In 1997 Saari et
al.\cite{Saari97} announced, in an important paper, the production
in the lab of an X-shaped wave in the optical realm, thus proving
experimentally the existence of Superluminal electromagnetic
pulses. \ Three years later, in 2000, Ranfagni et al.\cite{Ranfagni}
produced, in an experiment of theirs, Superluminal X-shaped waves
in the microwave region [their paper aroused various criticisms,
to which those author however responded].

\

\

\section{A More Detailed Introduction} 

\h Let us refer\cite{TesiM} to the differential equation known as
homogeneous wave equation: simple, but so important in Acoustics,
Electromagnetism (Microwaves, Optics,...), Geophysics, and even,
as we said, gravitational waves and elementary particle physics:

\

\bb
 \left(\frac{\pa^2}{\pa x^2} + \frac{\pa^2}{\pa y^2} +
\frac{\pa^2}{\pa z^2} - \frac{1}{c^2}\frac{\pa^2}{\pa
t^2}\right)\, \psi(x,y,z;t) \ug 0 \; . \label{eo1}\ee

\

\h Let us write it in the cylindrical co-ordinates $(\rho,\phi,z)$
and, for simplicity's sake, confine ourselves to axially symmetric
solutions $\psi(\rho,z;t)$. \ Then, eq.(\ref{eo1}) becomes

\

\bb \left(\frac{\pa^2}{\pa\rho^2} +
\frac{1}{\rho}\frac{\pa}{\pa\rho} + \frac{\pa^2}{\pa z^2} -
\frac{1}{c^2}\frac{\pa^2}{\pa t^2} \right)\, \psi(\rho,z;t) \ug 0
\ . \label{eo2}\ee

\

\h In free space, solution $\psi(\rho,z;t)$ can be written in
terms of a Bessel-Fourier transform w.r.t. the variable
$\rho$, and two Fourier transforms w.r.t. variables $z$ and $t$, as
follows:

\

\bb \psi(\rho,z,t) \ug
\int_{0}^{\infi}\int_{-\infi}^{\infi}\int_{-\infi}^{\infi}\,k_\rr
\, J_0(k_\rr\,\rho)\,e^{ik_zz}\,e^{-i\om
t}\,\bar{\psi}(k_\rr,k_z,\om)\,\drm k_\rr\, \drm k_z\,\drm\om
\label{sg1}\ee

\

where $J_0(.)$ is an ordinary zero-order Bessel function and
$\bar{\psi}(k_\rr,k_z,\om)$ is the transform of $\psi(\rho,z,t)$.

\h Substituting eq.(\ref{sg1}) into eq.(\ref{eo2}), one obtains
that the relation, among $\om$, $k_\rr$ and $k_z \,$,

\

\bb \frac{\om^2}{c^2} \ug k_\rr^2 + k_z^2  \label{c1} \ee

\

has to be satisfied. \ As a consequence, by using condition (\ref{c1})
in eq.(\ref{sg1}), any solution to the wave equation (\ref{eo2})
can be written

\

\bb \psi(\rho,z,t) \ug
\dis{\int_{0}^{\om/c}\int_{-\infi}^{\infi}\, k_\rr \,
J_0(k_\rr\,\rho)\,e^{i\sqrt{\om^2/c^2
\,-\,k_\rr^2}\,\,z}\,e^{-i\om t}\,S(k_\rr,\om)\,\drm
k_\rr\,\drm\om} \label{sg} \ee

\

where $S(k_\rr,\om)$ is the chosen spectral function, when $k_z>0$ (and
we disregard evenescent waves).

\h The general integral solution (\ref{sg}) yields for instance
the ({\bf non-localized}) gaussian beams and pulses, to which we
shall refer for illustrating the differences of the localized
waves w.r.t. them.

\

{\em The Gaussian Beam} --- A very common (non-localized) beam is
the gaussian beam\cite{Molone}, corresponding to the spectrum

\

\bb S(k_\rr,\om) \ug 2a^2\,e^{-a^2 k_\rr^2}\,\delta(\om - \om_0) \; .
\label{eg} \ee

\

In eq.(\ref{eg}), $a$ is a positive constant, which will be shown
to depend on the transverse aperture of the initial pulse.

\begin{figure}[!h]
\begin{center}
 \scalebox{2.3}{\includegraphics{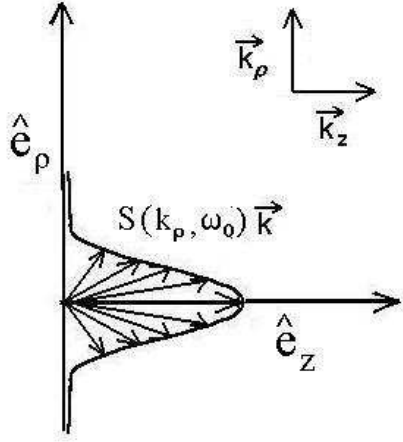}} 
\end{center}
\caption{Visual interpretation of the integral solution
(\ref{sg}), with spectral function (\ref{eg}), in terms of a
superposition of plane waves.} \label{fig4}
\end{figure}

\h Figure \ref{fig4} illustrates the interpretation of the
integral solution (\ref{sg}), with spectral function (\ref{eg}),
as a superposition of plane waves. \ Namely, from Fig.\ref{fig4}
one can easily realize that this case corresponds to plane waves
propagating in all directions (always with $k_z \geq 0$), the most
intense ones being those directed along (positive) $z$. Notice
that, in the plane-wave case, $\vec{k_z}$ is the longitudinal
component of the wave-vector, $\vec{k} = \vec{k_\rr} + \vec{k_z}$,
where $\vec{k_\rr} = \vec{k_x}+\vec{k_y}$.

\h On substituting eq.(\ref{eg}) into eq.(\ref{sg}) and adopting
the paraxial approximation, one meets the gaussian beam

\

\bb \psi_{\rm gauss}(\rho,z,t) \ug \frac{\dis{2a^2\,{\rm
exp}\left(\dis{\frac{-\rho^2}{4(a^2 +
i\,z/2k_0)}}\right)}}{\dis{2(a^2 +
i\,z/2k_0)}}\,\,\dis{e^{ik_0(z-ct)}}\;\; , \label{fg}\ee

\

where $k_0=\om_0/c$.  We can verify that such a beam, which
suffers transverse diffraction, doubles its initial width
$\Delta\rho_0 = 2a$ after having travelled the distance $z_{{\rm
dif}} \ug \sqrt{3}\,k_0 \Delta\rho_0^2 /2$, called diffraction
length. The more concentrated a gaussian beam happens to be, the
more rapidly it gets spoiled.

\

{\em The Gaussian Pulse} --- The most common (non-localized) {\em
pulse} is the gaussian pulse, which is got from eq.(\ref{sg}) by
using the spectrum\cite{chirped}

\

\bb S(k_\rr,\om) \ug
\frac{2ba^2}{\sqrt{\pi}}e^{-a^2k_\rr^2}e^{-b^2(\om-\om_0)^2}
\label{epg} \ee

\

where $a$ and $b$ are positive constants.  Indeed, such a pulse is
a superposition of gaussian beams of different frequency.

\h Now, on substituting eq.(\ref{epg}) into eq.(\ref{sg}), and
adopting once more the paraxial approximation, one gets the
gaussian pulse:

\

\bb \psi(\rho,z,t) \ug \frac{a^2\,{\rm
exp}\left(\dis{\frac{-\rho^2}{4(a^2+iz/2k_0)} }\right){\rm
exp}\left(\dis{\frac{-(z-ct)^2}{4c^2b^2}}
\right)}{a^2+iz/2k_0}\;\;, \label{pg} \ee

\

endowed with speed $c$ and temporal width $\Delta t = 2b$, and
suffering a progressing enlargement of its transverse width, so
that its initial value gets doubled already at position $z_{{\rm
dif}} \ug \sqrt{3}\,k_0 \Delta\rho_0^2/2 \;$, \ with $\Delta\rho_0 =
2a$.

\

\subsection{The localized solutions}

Let us finally go on to the construction of the two most renowned
localized waves\cite{TesiM}: the Bessel beam, and the ordinary X-shaped
pulse.

\h First of all, it is interesting to observe that, when
superposing (axially symmetric) solutions of the wave equation
in the vacuum, three spectral parameters, $(\om, \ k_\rr, \ k_z)$,
come into the play, which have however to satisfy the constraint
(\ref{c1}), deriving from the wave equation itself. Consequently,
only two of them are independent: and we choose\footnote{Elsewhere
we chose $\om$ and $k_z$.} here $\om$ and $k_\rr$. \ Such a
possibility of choosing $\om$ and $k_\rr$ was already apparent in the
spectral functions generating gaussian beams and pulses, which
consisted in the product of two functions, one depending only on
$\om$ and the other on $k_\rr$.

\h We are going to see that further particular relations between $\om$ and
$k_\rr$ [or, analogously, between $\om$ and $k_z$] can be moreover
enforced, in order to get interesting and unexpected results, such
as the {\em localized waves}.

\

{\em The Bessel beam} --- Let us start by imposing a {\em linear}
coupling between $\om$ and $k_\rr$ (it could be actually
shown\cite{Durnin2} that it is the unique coupling leading to
localized solutions).

\h Namely, let us consider the spectral function

\

\bb S(k_\rr,\om) \ug \frac{\delta(k_\rr -
\dis{\frac{\om}{c}}\sin\theta)}{k_\rr}\,\,\delta(\om - \om_0)
\;\;, \label{eb} \ee

\

which implies that $k_\rr = (\om\sin\theta)/c$, \ with $0 \leq
\theta \leq \pi/2$: \ A relation that can be regarded as a
space-time coupling. Let us add that this linear constraint
between $\om$ and $k_\rr$, together with relation (\ref{c1}),
yields \ $k_z = (\om\cos\theta)/c$. This is an important fact, since
it has been shown elsewhere\cite{TesiM,MRH} that an {\em ideal}
localized wave {\em must} contain a coupling of the type $\om=V
k_z + b$, where $V$ and $b$ are arbitrary constants.

\begin{figure}[!h]
\begin{center}
 \scalebox{2}{\includegraphics{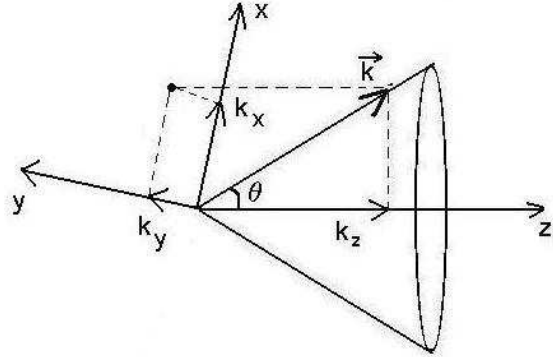}} 
\end{center}
\caption{The axially-symmetric Bessel beam is created by the
superposition of plane waves whose wave vectors lay on the surface
of a cone having the propagation axis as its symmetry axis and
angle equal to $\theta$ ("axicon angle").} \label{fig5}
\end{figure}

\h The interpretation of the integral function (\ref{sg}), this
time with spectrum (\ref{eb}), as a superposition of plane
waves is visualized in Figure \ref{fig5}: which shows that an
axially-symmetric Bessel beam is nothing but the result of the
superposition of plane waves whose wave vectors lay on the surface
of a cone having the propagation line as its symmetry axis and an
opening angle equal to $\theta$; such $\theta$ being called the
{\em axicon angle}.

\h By inserting eq.(\ref{eb}) into eq.(\ref{sg}), one gets the
mathematical expression of the so-called Bessel beam:

\

\bb \psi(\rho,z,t) \ug J_0\left(\frac{\om_0}{c}\sin\theta\,\,\rho
\right)\,{\rm
exp}\left[i\,\,\frac{\om_0}{c}\cos\theta\,\left(z-\frac{c}{\cos\theta}t\right)\right] \; .
\label{fb} \ee

\

\h This beam possesses phase-velocity $v_{\rm ph}=c/\cos\theta$,
and field transverse shape represented by a Bessel function
$J_0(.)$ so that its field in concentrated in the surroundings of
the propagation axis $z$. Moreover, eq.(\ref{fb}) tells us that
the Bessel beam keeps its transverse shape (which is therefore
invariant) while propagating, with central ``spot'' $\Delta\rho =
2.405 c /(\om\sin\theta)$.

\h The ideal Bessel beam, however, is not square-integrable in the
transverse direction, and is therefore associated with an infinite
power flux: \ i.e., it
cannot be experimentally produced.

\h But we can have recourse to truncated Bessel beams, generated
by finite apertures. In this case the (truncated) Bessel beams are
still able to travel a long distance while maintaining their
transfer shape, as well as their speed, approximately
unchanged\cite{Durnin,Durnin3,Overfelt}: That is to say, they
still possess a large depth of field. For instance, the field-depth
of a Bessel beam generated by a circular finite aperture
with radius $R$ is given by

\

\bb Z_{\rm max} \ug \frac{R}{\tan\theta} \; , \ee

\begin{figure}[!h]
\begin{center}
 \scalebox{2.5}{\includegraphics{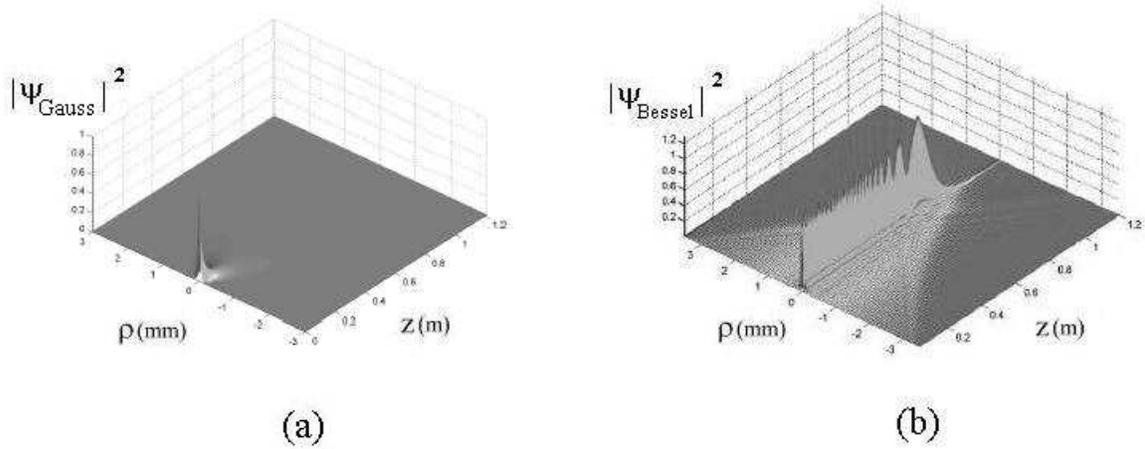}}  
\end{center}
\caption{Comparison between a gaussian (a) and a truncated Bessel
beam (b). One can see that the gaussian beam doubles its initial
transverse width already after $3\;$cm, while after $6\;$cm its
intensity decays of a factor 10.  By contrast, the Bessel beam
does approximately keep its transverse shape till the distance
$85\;$cm.} \label{fig6}
\end{figure}

\

where $\theta$ is the beam axicon angle. In the finite aperture
case, the Bessel beam cannot be represented any longer by
eq.(\ref{fb}), and one has to calculate it by the scalar
diffraction theory: By using, for example, Kirchhoff's or
Rayleigh-Sommerfeld's diffraction integrals. But till the
distance $Z_{\rm max}$ one may still use eq.(\ref{fb}) for
approximately describing the beam, at least in the vicinity of the
axis $\rho=0$, that is, for $\rho << R$. To realize how much a
truncated Bessel beam succeeds in resisting diffraction, let us
consider also a gaussian beam, with the same frequency and central
``spot", and compare their field-depths. In particular, let us
assume for both beams $\lambda = 0.63\;\mu$m and initial central
``spot'' size $\Delta\rho_0 = 60\;\mu$m. The Bessel beam will
possess axicon angle $\theta=\arcsin[2.405
c/(\om\Delta\rho_0)]=0.004\;$rad. Figure \ref{fig6} depicts the
behaviour of the two beams for a circular aperture
with radius $3.5\;$mm. \ We can verify how the gaussian beam
doubles its initial transverse width already after $3\;$cm, and
after $6\;$cm its intensity has become an order of magnitude
smaller. By contrast, the truncated Bessel beam keeps its
transverse shape until the distance $Z_{\rm max}=R/\tan\theta=
85\;$cm. Afterwards, the Bessel beam rapidly decays, as a
consequence of the sharp cut performed on its aperture (such cut
being responsible also for the intensity oscillations suffered by
the beam along its propagation axis, and for the fact that
eventually the feeding waves, coming from the aperture, at a
certain point get faint).

\h The zeroth-order (axially symmetric) Bessel beam is nothing but
one example of localized beam. Further examples are the higher
order (not cylindrically symmetric) Bessel beams

\bb \psi(\rho,\phi,z;t) \ug
J_{\nu}\left(\frac{\om_0}{c}\sin\theta\,\,\rho \right)\,{\rm
exp}(i\nu\phi) \; {\rm
exp}\left(i\,\,\frac{\om_0}{c}\cos\theta\,\left(z-\frac{c}
{\cos\theta}t\right)\right)\;\; , \label{fbn} \ee

or the Mathieu beams\cite{Dartora1}, and so on.

\

{\em The Ordinary X-shaped Pulse} --- Following the same procedure
adopted in the previous subsection, let us construct pulses by
using spectral functions of the type

\

\bb S(k_\rr,\om) \ug \frac{\delta(k_\rr -
\dis{\frac{\om}{c}}\sin\theta)}{k_\rr}\,\,F(\om) \; , \label{ex} \ee

\

where this time the Dirac delta function furnishes the spectral
space-time coupling $k_\rr = (\om\sin\theta)/c$. \ Function $F(\om)$
is, of course, the frequency spectrum; it is left for the moment
undetermined.

\h On using eq.(\ref{ex}) into eq.(\ref{sg}), one obtains

\

\bb \psi(\rho,z,t) \ug \int_{-\infty}^{\infty}\,F(\om)\,
J_0\left(\frac{\om}{c}\sin\theta\,\,\rho \right)\,{\rm
exp}\left(\frac{\om}{c}\cos\theta\,\left(z-\frac{c}
{\cos\theta}t\right)\right) \, \drm\om \ . \label{px} \ee

\

It is easy to see that $\psi$ will be a pulse of the type

\

\bb \psi \ug \psi(\rho,z-Vt) \ee

\

with a speed $V=c/\cos\theta$ independent of the frequency
spectrum $F(\om)$.

\h Such solutions are known as X-shaped pulses, and are {\em
localized} (non-diffractive) waves in the sense that they do
obviously maintain their spatial shape during propagation
(see., e.g., refs.\cite{Lu1,PhysicaA,MRH} and refs. therein;
as well as the following).

\

\h {\em At this point, some remarkable observations are to be stressed:}

\

(i) When a pulse consists in a superposition of waves (in this
case, Bessel beams) all endowed with the same phase-velocity
$V_{\rm ph}$ (in this case, with the same axicon angle)
independent of their frequency, {\em then} it is known that the
phase-velocity (in this case $V_{\rm ph}=c/\cos\theta$)
becomes\cite{Majorana,MRF} the group-velocity $V$: That is, $V=
c/\cos\theta >c$. In this sense, the X-shaped waves are called
``Superluminal localized pulses" (cf., e.g., ref.\cite{PhysicaA}
and refs. therein). \ [For simplicity, the group-velocity we are
talking about\cite{MRH,RFG,Reports2,RFG} can be regarded as the peak-velocity. \
Here, let us only add the observations: \ (a) that the group-velocity for
a pulse, in general, is well defined only when the pulse has a clear bump in space; but
it can be calculated by the approximate, simple relation \
$V \simeq \drm \om / \drm k$ \
only when some extra conditions are satisfied (namely, when $\om$
as a function of $k$ is also clearly bumped); \ and \ (b) that the group
velocity can
a priori be evaluated through the mentioned, customary derivation of
$\om$ with respect to the wavenumber for the infinite total
energy solutions; whilst, for the finite total energy Superluminal solutions,
the group-velocity cannot be calculated through such an elementary relation,
since in those cases it does not even exist a one-to-one function
$\om = \om(k_z)$].

\

\begin{figure}[!h]
\begin{center}
 \scalebox{1.8}{\includegraphics{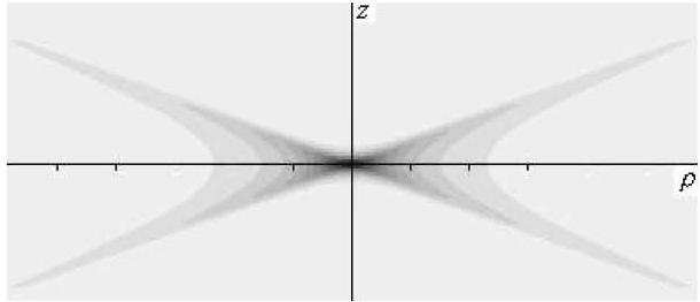}}  
\end{center}
\caption{The truncated X-waves considered in this paper, as
predicted by SR (all wave-equations have an intrinsic {\em
relativistic} structure!), must have a leading cone in addition to
the rear cone; such a leading cone having a role for the peak
stability\cite{Lu1}: For example, when producing a finite conic
wave truncated both in space and in time, the theory of SR
suggested to have recourse, in the {\em simplest} case, to a
dynamic antenna emitting a radiation cylindrically symmetrical in
space and symmetric in time, for a better approximation to what
Courant and Hilbert\cite{Courant} called an ``undistorted
progressing wave". See the following, in the text. } \label{fig7}
\end{figure}

(ii) Such pulses, even if their group-velocity is Superluminal, do
not contradict standard physics, having been found in what
precedes on the basis of the wave equations ---in particular, of
Maxwell equations\cite{Ziolk,PhysicaA}--- only.  Indeed, as we
shall better see in the historical Appendix following below at the end of the
First Part, their existence can be understood within Special Relativity
itself[9,13,15,55-57], on the
basis of its ordinary Postulates\cite{Review}. Actually, let us
repeat it once more, they are fed by waves originating at the aperture
and carrying energy with the standard speed $c$ of the medium (the
light-velocity in the electromagnetic case, and the sound-velocity
in the acoustic\cite{Lu2} case).
 \ We can become convinced about the possibility of realizing Superluminal
X-shaped pulses by imagining the simple ideal case of a negligibly
sized Superluminal source $S$ endowed with speed $V>c$ in vacuum,
and emitting electromagnetic waves $W$ (each one travelling with
the invariant speed $c$). The electromagnetic waves will result to
be internally tangent to an enveloping cone $C$ having $S$ as its
vertex, and as its axis the propagation line $z$ of the
source\cite{Review,SupCh}: {\em This is completely analogous to
what happens for an airplane that moves in air with constant
supersonic speed}. \ The waves $W$ interfere mainly negatively
inside the cone $C$, and constructively on its surface. \ We can
place a plane detector orthogonally to $z$, and record magnitude
and direction of the $W$ waves that hit on it, as (cylindrically
symmetric) functions of position and of time. \ It will be enough,
then, to replace the plane detector with a plane antenna which
{\em emits} ---instead of recording--- exactly the same (axially
symmetric) space-time pattern of waves $W$, for constructing a
cone-shaped electromagnetic wave $C$ that will propagate with the
Superluminal speed $V$ (of course, without a source any longer at
its vertex...): \ even if each wave $W$ travels with the invariant
speed $c$. \ {\em Once more, this is exactly what would happen in
the case of a supersonic airplane} (in which case $c$ is the sound
speed in air: for simplicity, assume the observer to be at rest
with respect to the air). \ For further details, see the quoted
references. \ Actually, by suitable superpositions, and {\em
interference}, of speed-$c$ waves, one can obtain pulses more and
more localized in the vertex region\cite{MRH}: That is, very
localized field-``blobs" which travels with Superluminal
group-velocity. \ This has nothing to do apparently with the illusory
``scissors effect'', since such blobs, along their field-depth,
are a priori able, e.g., to get two successive (weak) detectors,
located at a distance $L$, to click after a time {\em smaller} than
$L/c$. Incidentally, an analysis of the above-mentioned case (that
of a supersonic plane or a Superluminal charge) led, as
expected\cite{Review}, to the simplest type of ``X-shaped
pulse"\cite{SupCh}. \ It might be useful, finally, to recall that
SR (even the wave-equations have an internal {\em relativistic}
structure!) implies considering also the forward cone: cf.
Fig.\ref{fig7}. The truncated X-waves considered in this paper,
for instance, must have a leading cone in addition to the rear
cone; such a leading cone having a role for the peak
stability\cite{Lu1}: For example, in the approximate case in which
we produce a finite conic wave truncated both in space and in
time, the theory of SR suggested the bi-conic shape (symmetrical
in space with respect to the vertex $S$) to be a better
approximation to a rigidly travelling wave (so that SR suggests to
have recourse to a dynamic antenna emitting a radiation
cylindrically symmetric in space and symmetric in time, for a
better approximation to an ``undistorted progressing wave").

\

(iii)  Any solutions that depend on $z$ and on $t$ only through
the quantity $z-Vt$, like eq.(\ref{px}), will appear with a constant
shape to an observer travelling along $z$ with the speed $V$. That is,
such a solution will propagate rigidly with speed $V$.
This does further explain why our X-shaped pulses, after having been
produced, will travel almost rigidly at speed $V$ (in this case, a
faster-than-light group-velocity), all along their depth of field. \
To be even clearer, let us consider a generic function, depending
on $z-Vt$ with $V>c$, and show, {\em by explicit calculations
involving the Maxwell equations only}, that it obeys the scalar
wave equation. \ Following Franco Selleri\cite{Selleri}, let us
consider, e.g., the wave function

\

\bb \Phi(x,y,z,t) \ug {\dis{\frac{a}
{{\sqrt{[b-ic(z-Vt)]^2+(V^2-c^2)(x^2+y^2)}}}}} \label{sellerieq10}
\ee

\

with $a$ and $b$ non-zero constants, $c$ the ordinary speed of
light, and $V>c$ \ [incidentally, this wave function is nothing
but the classic X-shaped wave in cartesian co-ordinates].  Let us
naively verify that it is a solution to the wave equation

\

\bb  \nabf^2 \Phi(x,y,z,t) - {\frac{1}{c^2}} \; {\frac
{\pa^2\Phi(x,y,z,t)}{\pa^2t}} \ug 0 \ . \label{sellerieq11} \ee

\

On putting

\bb R \, \equiv \,
\dis{{{\sqrt{[b-ic(z-Vt)]^2+(V^2-c^2)(x^2+y^2)}}}} \; ,
\label{sellerieq12} \ee

one can write $\Phi = a/R$ and evaluate the second derivatives

\

$$\frac{1}{a} \, \frac{\pa^2\Phi}{\pa^2z} \ug
\frac{c^2}{R^3} - \frac{3c^2}{R^5}\,[b-ic(z-Vt)]^2 \ ;$$

$$\frac{1}{a} \, \frac{\pa^2\Phi}{\pa^2x} \ug  - \frac{V^2-c^2}{R^3} \, + \,
3\left(V^2-c^2 \right)^2 \; \frac{x^2}{R^5} \ ;$$

$$\frac{1}{a} \, \frac{\pa^2\Phi}{\pa^2y} \ug - \frac{V^2-c^2}{R^3} \, + \,
3\left(V^2-c^2 \right)^2 \; \frac{y^2}{R^5} \;$$

$$\frac{1}{a} \, \frac{\pa^2\Phi}{\pa^2t} \ug
\frac{c^2V^2}{R^3} - \frac{3c^2V^2}{R^5}\,[b-ic(z-Vt)]^2 \ ;$$

\

wherefrom

\

$$\frac{1}{a} \, \left[ \frac{\pa^2\Phi}{\pa^2z} - \frac{1}{c^2}
\frac{\pa^2\Phi}{\pa^2t} \right] \ug - \frac{V^2-c^2}{R^3} +
3\left(V^2-c^2 \right)^2 \; \frac{[b-ic(z-Vt)]^2}{R^5} \ ,$$

and

$$\frac{1}{a} \, \left[ \frac{\pa^2\Phi}{\pa^2x} + \frac{\pa^2\Phi}{\pa^2y}
\right] \ug -2 \, \frac{V^2-c^2}{R^3} + 3\left(V^2-c^2 \right)^2
\; \frac{x^2+y^2}{R^5} \ .$$

\

From the last two equations, remembering the previous definition,
one finally gets

$$\frac{1}{a} \, \left[ {\frac{\pa^2\Phi}{\pa^2z}} + {\frac{\pa^2\Phi}{\pa^2x}}
+ {\frac{\pa^2\Phi}{\pa^2y}} - {\frac{1}{c^2}} \;
{\frac{\pa^2\Phi}{\pa^2t}} \right] \ug 0$$

that is nothing but the (d'Alembert) wave equation
(\ref{sellerieq11}), {\bf q.e.d.} \  In conclusion, function
$\Phi$ is a solution of the wave equation even if it does
obviously represent a pulse (Selleri says ``a signal"!) propagating
with Superluminal speed.

\h At this point, the reader should be however warned that all the
{\em subluminal} LWs, solutions of the ordinary {\em homogeneous}
wave equation, have appeared till now to present sungularities
whenever they depend on $z$ and $t$ only via the quantity
$\zeta \equiv z-Vt$: \ This is still an open, interesting research
topic, which is related also to analogous results met in gravitation
physics.

\

\h After the previous three important comments, let us go back to
our evaluations with regard to the X-type solutions to the wave
equations. \ Let us now consider in eq.(\ref{px}), for instance,
the particular frequency spectrum $F(\om)$ given by

\

\bb F(\om) \ug H(\om)\,\frac{a}{V}\,\,\,{\rm
exp}\left(-\frac{a}{V}\,\om\right) \; , \label{fx} \ee

\

where $H(\om)$ is the Heaviside step-function and $a$ a positive
constant. Then, eq.(\ref{px}) yields

\

\bb \psi(\rho,\zeta) \, \equiv \, X \ug \frac{a}{\sqrt{(a -
i\zeta)^2 + \left(\frac{V^2}{c^2}-1\right)\rho^2}} \;, \label{ox}
\ee

\

with $\zeta \equiv z - Vt$.  This solution (\ref{ox}) is the well-known
ordinary, or ``classic", X-wave, which constitutes a simple
example of X-shaped pulse.\cite{Lu1,PhysicaA} \ Notice that
function (\ref{fx}) contains mainly low frequencies, so that the
classic X-wave is suitable for low frequencies only.

\begin{figure}[!h]
\begin{center}
 \scalebox{3.2}{\includegraphics{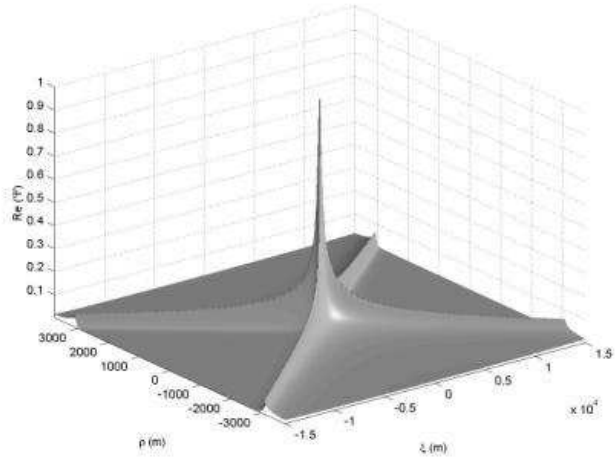}}  
\end{center}
\caption{Plot of the real part of the ordinary X-wave, evaluated
for $V=1.1\,c$ with $a=3\;$m .} \label{fig8}
\end{figure}

\h Figure \ref{fig8} depicts (the real part of) an ordinary
X-wave with $V=1.1\,c$ and $a=3\;$m.

\h Solutions (\ref{px}), and in particular the pulse (\ref{ox}),
have got an infinite field-depth, and an infinite energy as well.
Therefore, as it was done in the Bessel beam case, one should
pass to truncated pulses, originating from a finite aperture.
Afterwards, our truncated pulses will keep their spatial shape
(and their speed) all along the depth of field

\bb Z \ug \frac{R}{\tan\theta} \; , \ee

where, as before, $R$ is the aperture radius and $\theta$ the
axicon angle.

\

{\em Some Further Observations} --- Let us put forth some further
observations.

\h It is not strictly correct to call non-diffractive the
localized waves, since diffraction affects, more or less, all
waves obeying eq.(\ref{eo1}). {\em However}, all localized waves
(both beams and pulses) possess the remarkable
``self-reconstruction" property: That is to say, the localized
waves, when diffracting during propagation, do immediately
re-build their shape[39-41] (even after
obstacles with size much larger than the characteristic wave-lengths,
provided it is smaller ---as we know--- than the aperture size),
due to their particular spectral structure [as it is shown
more in detail, e.g., in the book {\em Localized
Waves} (J.Wiley; Jan.2008)]. In particular, the
``ideal localized waves'' (with infinite energy and field-depth)
are able to re-build themselves for an infinite time;
while, as we have seen, the finite-energy (truncated) ones can do
it, and thus resist the diffraction effects, only along a certain
depth of field...

\h Let us stress again that the interest of the localized waves
(especially from the point of view of applications) lies in the
circumstance that they are almost non-diffractive, rather than in
their group-velocity: From this point of view, Superluminal,
luminal, and subluminal localized solutions are equally
interesting and suited to important applications.

\h Actually, the localized waves are not restricted to the
(X-shaped, Superluminal) ones corresponding to the integral
solution (\ref{px}) to the wave equation; and, as we were already
saying, three classes of localized pulses exist: the Superluminal
(with speed $V > c$), the luminal ($V=c$), and the subluminal
($V<c$) ones; all of them with, or without, axial symmetry, and
corresponding in any case to a single, unified mathematical
background. \ This issue will be touched again in the present
review.

\begin{figure}[!h]
\begin{center}
 \scalebox{2.1}{\includegraphics{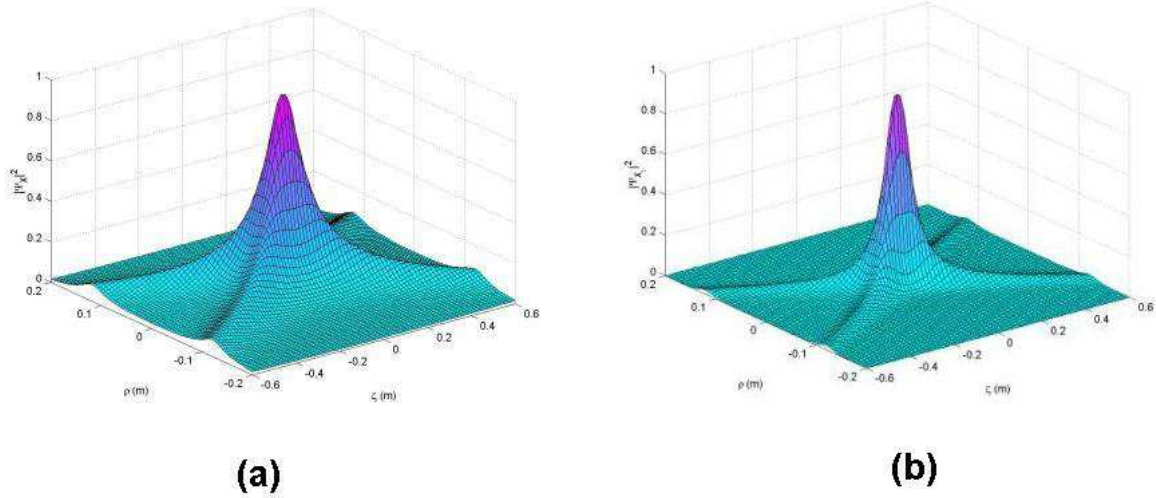}}    
 \end{center}
\caption{In Fig.(a) it is represented (in arbitrary units) the
square magnitude of the ``classic", $X$-shaped Superluminal
localized solution (SLS) to the wave equation, with $V=5c$ and
$a=0.1 \;$m. \ Families of infinite SLSs however exists, which
generalize the classic $X$-shaped solution; for instance, a family
of SLSs obtained\cite{MRH} by suitably differentiating the classic
X-wave: Fig.(b) depicts the first of them (corresponding to the
first differentiation) with the same parameters. \ As we said, the
successsive solutions in such a family are more and more localized
around their vertex.  Quantity $\rho$ is the distance in meters
from the propagation axis $z$, while quantity $\ze$ is the
``$V$-cone" variable[ref.\cite{MRH}] (still in meters) $\ze \equiv
z-Vt$, with $V \geq c$. \ Since all these solutions depend on $z$
only via the variable $\ze$, they propagate ``rigidly", i.e., as
we know, without distortion (and are called ``localized", or
non-diffracting, for such a reason).  Here we are assuming
propagation in the vacuum (or in a homogeneous medium). }
\label{fig9}
\end{figure}

\h Incidentally, we have addressed elsewhere topics \ as:  (i)
the construction of infinite families of generalizations of the
classic X-shaped wave [with energy more and more concentrated
around the vertex: cf., e.g., Figs.\ref{fig9}, taken from
ref.\cite{MRH}]; as \ (ii) the behaviour of some finite
total-energy Superluminal localized solutions (SLS); \ (iii) the
techniques for building up new series of SLS's to the Maxwell equations
suitable for arbitrary frequencies and bandwidths; \
(iv) questions related with the case of dispersive (and even lossy) media;
 \ (v) the construction of (infinite or finite energy) Superluminal LWs
propagating down waveguides or coaxial cables; \ (vi) finding out
Localized Solutions also to equations different from the wave
equation, as Schroedinger's; \ (vii) using the above tecniques
for constructing, in General Relativity, new exact solutions for
gravitational ways. \  In the
Second Part of this paper we shall come back to some
(few) of those points. \ Let us add that X-shaped waves have been
easily produced also in nonlinear media\cite{Conti}.

\h A more technical introduction to the subject of localized waves
(particularly w.r.t. the Superluminal X-shaped ones) can be found
in the Second Part of this review, and in papers like ref.\cite{JSTQE}.

\h Before going on to the Second Part of this paper,
let us end the present First Part by a historical (theoretical and experimental)
{\em Appendix.}

\

\

\

\

\

\centerline{\huge{\bf  An APPENDIX to the First Part:}}

\

\

\centerline{\large{\bf A HISTORICAL (THEORETICAL AND
EXPERIMENTAL)}}

\

\centerline{\large{\bf APPENDIX}}

\

\

\h In this mainly ``historical" Appendix, written as far as
possible in a (partially) self-consistent form, we shall first
refer ourselves, from the theoretical point of view, to the most
intriguing localized solutions to the wave equation: the
Superluminal ones (SLS), and in particular the X-shaped pulses. As
a start, we shall recall their geometrical interpretation
within SR. Afterwards, to help resolving possible doubts, we shall
seize the opportunity, given by this Appendix, for
presenting a bird's-eye view of the various {\em experimental}
sectors of physics in which Superluminal motions seem to appear:
In particular, of the experiments with evanescent waves (and/or
tunnelling photons), and with the SLS's we are more interested in
here. In some parts of this Appendix the propagation-line is called
$x$, and no longer $z$, without originating, however, any
interpretation problems.

\

\section {An Introduction to the APPENDIX}  

The question of Superluminal ($V^{2}>c^{2}$) objects or waves has
a long story. Still in pre-relativistic times, one meets various
relevant papers, from those by J.J.Thomson to the interesting ones
by A.Sommerfeld. It is well-known, however, that with SR the
conviction spread out that the speed $c$ of light in vacuum was
the upper limit of any possible speed. For instance, R.C.Tolman in
1917 believed to have shown by his ``paradox'' that the existence
of particles endowed with speeds larger than $c$ would have
allowed sending information into the past. \ Our problem started
to be tackled again only in the fifties and sixties, in particular
after the papers\cite{ECGS1} by E.C.George Sudarshan et al., and,
later on\cite{Rev1974,North-Holl}, by one of the present authors
with R.Mignani et al., as well as  ---to confine ourselves at
present to the theoretical researches--- by H.C.Corben and others.
The first experimental attempts were performed by T.Alv\"{a}ger et
al.

\h We wish to face the still unusual issue of the possible
existence of Superluminal wavelets, and objects ---within standard
physics and SR, as we said--- since at least four different
experimental sectors of physics  {\em seem} to support such a
possibility [apparently confirming some long-standing theoretical
predictions\cite{Review,BarutMR,ECGS1,North-Holl}]. The
experimental review will be necessarily short, but we shall
provide the reader with enough bibliographical
information, limited for brevity's sake to the last century only
(i.e., up-dated till the year 2000 only). \

\

\section{APPENDIX: Historical Recollections - Theory} 

\begin{figure}[!h]
\begin{center}
 \scalebox{0.8}{\includegraphics{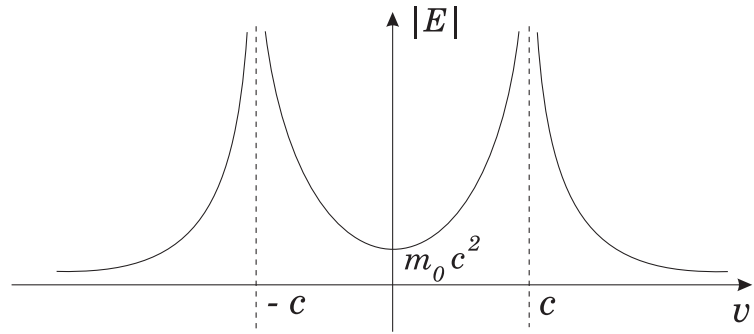}}    
\end{center}
\caption{Energy of a free object as a function of its
speed.\cite{ECGS1,
 Rev1974,Review}}
\label{fig10}
\end{figure}

\h A simple theoretical framework was long ago
proposed\cite{ECGS1,Review,Rev1974}, merely based on the
space-time geometrical methods of SR, which appears to incorporate
Superluminal waves and objects, and in a sense predicts\cite{BarutMR} among
the others the Superluminal X-shaped waves, without violating the
Relativity principles. A suitable choice of the Postulates of SR
(equivalent of course to the other, more common, choices) is the
following one: (i) the standard Principle of Relativity; and (ii)
space-time homogeneity and space isotropy. \ It follows that one
and only one {\em invariant}  speed exists; and experience shows
that invariant speed to be the light-speed, $c$, in vacuum: The
essential role of $c$ in SR being just due to its invariance, and
not to the fact that it be a maximal, or minimal, speed. No sub-
or Super-luminal objects or pulses can be endowed with an
invariant speed: so that their speed cannot play in SR the same
essential role played the light-speed $c$ in vacuum. Indeed, the
speed $c$ turns out to be also a {\em limiting} speed: but any limit
possesses two sides, and can be approached a priori both from
below and from above: See Fig.\ref{fig10}. \ As E.C.G.Sudarshan
put it, from the fact that no one could climb over the Himalayas
ranges, people of India cannot conclude that there are no people
North of the Himalayas... Indeed, speed-$c$  photons exist, which
are born, live and die just ``at the top of the mountain," without
any need for performing the impossible task of accelerating from
rest to the light-speed. \ [Actually, the ordinary formulation of
SR has been too much restricted: For instance, even leaving Superluminal
speeds aside, it can be easily so widened as to include
antimatter\cite{Review,FoP87,RFG}].

\h An immediate consequence is that the quadratic form \ $c^{2}
\drm t^{2} - \drm \xbf^{2} \equiv \drm x_\mu \drm x^\mu$, called
$\drm s^{2}$, \ with $\drm \xbf^{2} \equiv \drm x^{2}+\drm
y^{2}+\drm z^{2}$, \ results to be invariant, {\em except for its
sign}. \ Quantity $\drm s^{2}$, let us recall, is the
four-dimensional length-element square, along the space-time path
of any object. \ In correspondence with the positive (negative)
sign, one gets the subluminal (Superluminal) Lorentz
``transformations" [LT]. More specifically, the ordinary subluminal LTs are
known to leave, e.g., the quadratic forms $\drm x_\mu \drm x^\mu$, $\drm
p_\mu \drm p^\mu$ and  $\drm x_\mu \drm p^\mu$ exactly invariant,
where the $p_\mu$ are the component of the energy-impulse
four-vector; while the Superluminal LTs, by contrast, have to
change (only) the sign of such quadratic forms. This is enough for
deducing some important consequences, like the one that a
Superluminal charge has to behave as a magnetic monopole, in the
sense specified in ref.\cite{Review} and refs. therein.

\begin{figure}[!h]
\begin{center}
 \scalebox{1.76}{\includegraphics{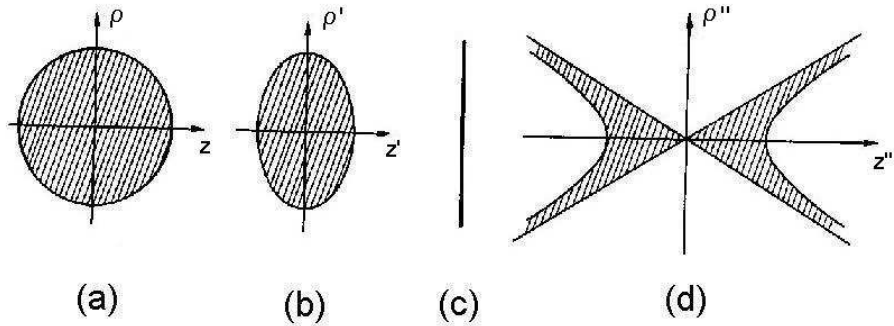}}  
\end{center}
\caption{An intrinsically spherical (or pointlike, at the limit)
object appears in the vacuum as an ellipsoid contracted along the
motion direction when endowed with a speed $v<c$. \ By contrast,
if endowed with a speed $V>c$ (even if the $c$-speed barrier
cannot be crossed, neither from the left nor from the right), it
would appear\cite{BarutMR,Review} no longer as a particle, but
rather as an ``X-shaped" wave travelling rigidly: Namely, as
occupying the region delimited by a double cone and a two-sheeted
hyperboloid ---or as a double cone, at the limit--, and moving
without distortion in the vacuum, or in a homogeneous medium, with
Superluminal speed $V$ [the square cotangent of the cone
semi-angle being $(V/c)^2-1$]. For simplicity, a space axis is
skipped. \ This figure is taken from refs.\cite{BarutMR,Review}. }
\label{fig11}
\end{figure}

\h A more important consequence, for us, is ---see
Fig.\ref{fig11}--- that the simplest subluminal object, namely a
spherical particle at rest (which appears as ellipsoidal, due to
Lorentz contraction, at subluminal speeds $v$), will
appear\cite{BarutMR,Review,PhysicaA} as occupying the
cylindrically symmetrical region bounded by a two-sheeted rotation
hyperboloid and an indefinite double cone, as in Fig.(11d), for
Superluminal speeds $V$. In the limiting case of a point-like particle, one
obtains only a double cone.

\begin{figure}[!h]
\begin{center}
 \scalebox{.96}{\includegraphics{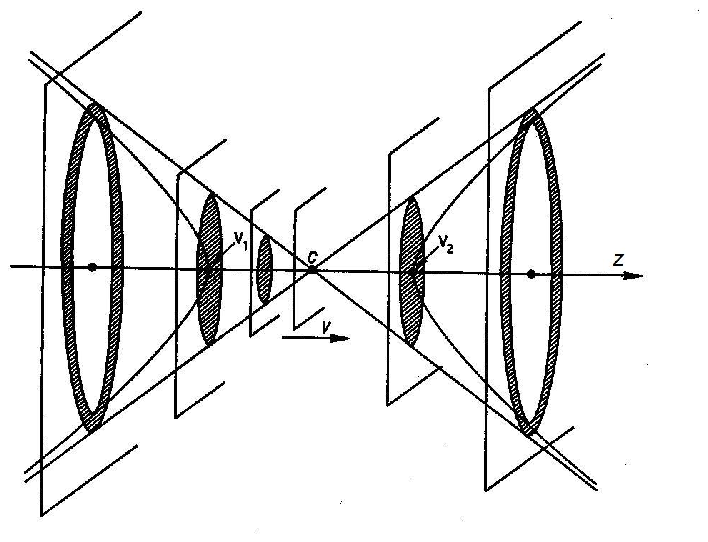}}  
\end{center}
\caption{Here we show the {\em intersections} of the Superluminal
object $T$ represented in Fig.(11d) with planes $P$ orthogonal to
its motion line (the $z$-axis). \ For simplicity, we assumed again
the object to be spherical in its rest-frame, and the cone vertex
$C$ to coincide with the origin $O$ for $t=0$. \ Such
intersections evolve in time so that the same pattern appears on a
second plane ---shifted by $\Delta x$--- after the time $\Delta t
= \Delta x / V$. \ On each plane, as time elapses, the
intersection is therefore predicted by (extended) SR to be a
circular ring which, for negative times, goes on shrinking until
it reduces to a circle and then to a point (for $t=0$); \
afterwards, such a point becomes again a circle and then a
circular ring that goes on
broadening\cite{BarutMR,Review,PhysicaA}. \ This picture is taken
from refs.\cite{BarutMR,Review}. \ [Notice that, if the object is
not spherical when at rest (but, e.g., is ellipsoidal in its own
rest-frame), then the axis of $T$  will no longer coincide with
$x$, but its direction will depend on the speed $V$ of the tachyon
itself]. \ For the case in which the space extension of the
Superluminal object $T$ is finite, see refs.\cite{BarutMR} }
\label{fig12}
\end{figure}

Such a result is got by writing down the equation of the
{\em world-tube} of a subluminal particle, and transforming it simply by
changing the sign of the quadratic forms entering that equation. Thus,
in 1980-1982, it was predicted\cite{BarutMR} that the simplest
Superluminal object appears (not as a particle, but as a field or
rather) as a wave: namely, as an ``X-shaped pulse", the cone
semi-angle $\al$ being given (with $c=1$) by \ ${\rm cotg} \, \al =
\sqrt{V^2 - 1}$. Such X-shaped pulses will move {\em rigidly} with
speed $V$ along their motion direction: In fact, any ``X-pulse"
can be regarded at each instant of time as the (Superluminal)
Lorentz transform of a spherical object, which of course moves in
vacuum ---or in a homogeneous medium--- without any deformation as
time elapses. \ The three-dimensional picture of Fig.(11d) appears
in Fig.\ref{fig12}, where its annular intersections with a
transverse plane are shown (cf. refs.\cite{BarutMR}). \ The
X-shaped waves here considered are merely the simplest ones: if
one starts not from an intrinsically spherical or point-like
object, but from a non-spherically symmetric particle, or from a
pulsating (contracting and dilating) sphere, or from a particle
oscillating back and forth along the motion direction, then their
Superluminal Lorentz transforms would result to be more and more
complicated. The above-seen ``X-waves", however, are typical for a
Superluminal object, as much as the spherical or point-like shape is
typical, let us repeat, for a subluminal object.

\h Incidentally, it has been believed for a long time that
Superluminal objects would have allowed sending information into
the past; but such problems with causality seem to be solvable
within SR. Once SR is generalized in order to include Superluminal
objects or pulses, no signal travelling backward in time is
apparently left.  For a solution of those causal paradoxes, see
refs.\cite{FoP87,RFG,ECGS1} and references therein.

\h When addressing the problem, within this elementary
Appendix, of the production of an X-shaped pulse like the one
depicted in Fig.\ref{fig12} (maybe truncated, in space and in
time, by use of a finite antenna radiating for a finite time), all
the considerations expounded under point (ii) of the subsection
{\em The Ordinary X-shaped Pulse} become in order: And, here, we
simply refer to them.  Those considerations, together with the
present ones (related, e.g., to Fig.\ref{fig12}), suggest the
simplest antenna to consist in a series of concentric annular
slits, or transducers [like in Fig.\ref{fig2}], which suitably
radiate following specific time patterns: See, e.g.,
refs.\cite{Mjosaatoappear} and refs. therein. \ Incidentally, the
above procedure can lead to a very simple type of X-shaped wave,
as investigated below.

\ From the present point of view, it is rather interesting to note
that, during the last fifteen years, X-shaped waves have been {\em
actually} found as solutions to the Maxwell and to the wave
equations [let us repeat that the form of any wave equations is
intrinsically relativistic]. {\em In order to see more deeply the
connection existing between what predicted by SR} (see, e.g.,
Figs.\ref{fig11},\ref{fig12}) {\em and the localized X-waves
mathematically, and experimentally, constructed in recent times,}
let us tackle below, in detail, the problem of the (X-shaped)
field created by a Superluminal electric charge\cite{SupCh}, by
following a paper recently appeared in Physical Review E.\footnote{At
variance with the old
times ---e.g., at the beginning of the seventies our papers on
similar subjects were always rejected by the most important journals---,
things have now changed as to superluminal
motions: For instance, the paper of ours quoted in Ref.\cite{SupCh},
submitted in 2002 to PRL, was diverted to PRE, but was eventually {\em published}
therein in 2004, even if dealing ---as we said--- with the
X-shaped field generated by a superluminal electric charge...}

\

\subsection{The particular X-shaped field associated with a Superluminal
charge}

It is well-known by now that Maxwell equations admit of
wavelet-type solutions endowed with arbitrary group-velocities ($0
< v_\grm < \infi$). \ We shall again confine ourselves, as above,
to the localized solutions, rigidly moving: and, more in
particular, to the Superluminal ones (SLS), the most interesting
of which resulted to be X-shaped, as we have already seen. \ The SLSs
have been actually produced in a number of experiments, always by
suitable interference of ordinary-speed waves. \ In this
subsection we show, by contrast, that even a Superluminal charge
creates an electromagnetic X-shaped wave, in agreement with what
predicted\cite{BarutMR,Review} within SR.  In fact, on the basis of
Maxwell equations, one is able to evaluate the field associated
with a Superluminal charge (at least, under the rough
approximation of pointlikeness): As announced in what precedes, it
results to constitute a very simple example of {\em true} X-wave.

\h Indeed, the theory of SR, when based on the {\em ordinary}
Postulates but not restricted to subluminal waves and objects,
i.e., in its extended version, predicted the simplest X-shaped
wave to be the one corresponding to the electromagnetic field
created by a Superluminal charge\cite{Folman,SupCh}. \ It seems
really important evaluating such a field, at least approximately,
by following ref.\cite{SupCh}.

\

{\em The toy-model of a pointlike Superluminal charge} ---  Let us
start by considering, formally, a pointlike Superluminal charge,
even if the hypothesis of pointlikeness (already unacceptable in
the subluminal case) is totally inadequate in the Superluminal
case\cite{Review}. \ Then, let us consider the ordinary
vector-potential $A^\mu$ and a current density $j^\mu \equiv
(0,0,j_z;j^\orm)$ flowing in the $z$-direction (notice that the
motion line is still the axis $z$). On assuming the fields to be
generated by the sources only, one has that $A^\mu \equiv
(0,0,A_z;\phi)$, which, when adopting the Lorentz gauge, obeys the
equation $A^\mu = j^\mu$. \ We can write such non-homogeneous wave
equation in the cylindrical co-ordinates $(\rho,\theta,z;t)$; for
axial symmetry [which requires a priori that $A^\mu =
A^\mu(\rho,z;t)$], when choosing the ``$V$-cone variables" $\ze
\equ z-Vt; \ \eta \equ z+Vt \;$, with \ $V^2 > c^2$, we
arrive\cite{SupCh} at the equation

\

\bb \dis{\left[-\rho {\pa \over {\pa\rho}} \left(\rho {\pa \over
{\pa\rho}}\right) + \frac{1}{\ga^2} \frac{\pa^2}{\pa \ze^2} +
\frac{1}{\ga^2} \frac{\pa^2}{\pa \eta^2} - 4 \frac{\pa^2}{\pa\ze
\pa\eta} \right] \; A^\mu(\rho,\ze,\eta) \ug j^\mu(\rho,\ze,\eta)}
\ , \label{17} \ee

\

where $\mu$ assumes the two values $\mu = 3,0$ only, so that \
$A^\mu \equiv (0,0,A_z;\phi)$, \ and \ $\ga^2 \equiv
[V^2-1]^{-1}$. \ [Notice that, whenever convenient, we set $c=1$].
\ Let us now suppose $A^\mu$ to be actually independent of $\eta$,
namely, $A^\mu = A^\mu (\rho, \ze)$. \ Due to eq.(\ref{17}), we
shall have $j^\mu = j^\mu (\rho, \ze) \,$ too; and therefore $j_z
= V j^0$ (from the continuity equation), and $A_z = V \phi / c$
(from the Lorentz gauge). \ Then, by calling $\psi \equiv A_z$, we
end up in two equations\cite{SupCh}, which allow us to analyse the
possibility and consequences of having a Superluminal pointlike
charge, $e$, travelling with constant speed $V$ along the $z$-axis
($\rho = 0$) in the positive direction, in which case $j_z = e\,
V\, \delta(\rho)/{\rho} \; \delta(\ze)$. \ Indeed, one of those
two equations becomes the hyperbolic equation

\

\bb \dis{\left[-\frac{1}{\rho}\frac{\pa}{\pa\rho}\left(\rho
\frac{\pa}{\pa\rho}\right) + \frac{1}{\ga^2}
\frac{\pa^2}{\pa\ze^2}\right] \, \psi \ug e V \, \frac{\delta(\rho
)}{\rho} \, \delta(\ze)} \label{18}\ee

\

which can be solved\cite{SupCh} in few steps. First, by applying
(with respect to the variable $\rho$) the Fourier-Bessel (FB)
transformation \ $f(x) \ug \dis{\int_0^{\infty} \Om f(\Om) J_0(\Om
x) \, \drm\Om}$, \ quantity $J_0(\Om x)$ being the ordinary
zero-order Bessel function. \ Second, by applying the ordinary
Fourier transformation with respect to the variable $\ze \,$
(going on, from $\ze$, to the variable $\om$). \ And, third, by
finally performing the corresponding {\em inverse} Fourier and FB
transformations. \ Afterwards, it is enough to have recourse to
formulae (3.723.9) and (6.671.7) of ref.\cite{ref[14]}, still with
$\ze \equiv z - Vt \,$, for being able to write down the solution
of eq.(\ref{18}) in the form

\

\hfill{$ \left\{\begin{array}{clr}

\psi(\rho,\ze) \ug 0 \ \ \ \ \ \ \ \ \ \ \ \ \ \ \ \ \ \ \ \ \ \ \
\ \ \ \ \ \ \
{\rm {for}} \ \ \ 0 < \ga\mid{\ze}\mid < \rho \\

\\

\dis{\psi(\rho,\ze) \ug \dis{e \; \frac{V}{\sqrt{\ze^2 -
\rho^2(V^2-1)}}}} \ \ \ \ \ {\rm {for}} \ \ \ 0 \le \rho <
\ga\mid{\ze}\mid \ .

\end{array} \right.
$\hfill} (25)

\

\begin{figure}[!h]
\begin{center}
 \scalebox{0.8}{\includegraphics{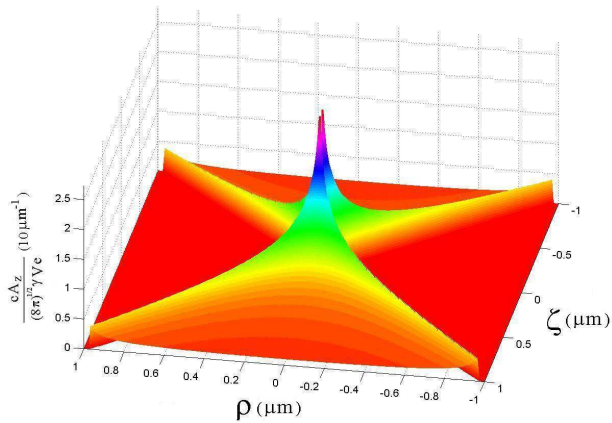}}  
\end{center}
\caption{Behaviour of the field $\psi \equiv A_z$  generated by a
charge supposed to be Superluminal, as a function of $\rho$ and
$\ze \equiv z-Vt$, evaluated for $\ga = 1$ (i.e., for $V = c
\sqrt{2}$): According to ref.\cite{SupCh} \ [Of course, we skipped
the points at which $\psi$ must diverge: namely, the vertex and
the cone surface]. }
\label{fig13}
\end{figure}

In Fig.\ref{fig13} we show our solution $A_z \equiv \psi$, as a
function of $\rho$ and $\ze$, evaluated for $\ga = 1$ (i.e., for
$V = c \sqrt{2}$). Of course, we skipped the points at which $A_z$
{\em must} diverge, namely the vertex and the cone surface.

\h For comparison, one may recall that the {\em classic} X-shaped
solution\cite{Lu1} of the {\em homogeneous} wave-equation ---which
is shown, e.g., in Figs.\ref{fig8}, \ref{fig9}, \ref{fig12}--- has
the form (with $a > 0$):

\

\hfill{$ \dis{X \ug {\frac{V}{\sqrt{(a-i\ze)^2 + \rho^2(V^2-1)}}}}
\; . $\hfill} (26)

\setcounter{equation}{26}

\

The second one of eqs.(25) includes expression (26), given by the
spectral parameter\cite{MRH,meiodisp} $a = 0$, which indeed
corresponds to the non-homogeneous case [the not negligible fact that for
$a=0$ these equations differ for an imaginary unit\cite{Review,chmonPLA}
will be discussed elsewhere].

\begin{figure}[!h]
\begin{center}
 \scalebox{1.08}{\includegraphics{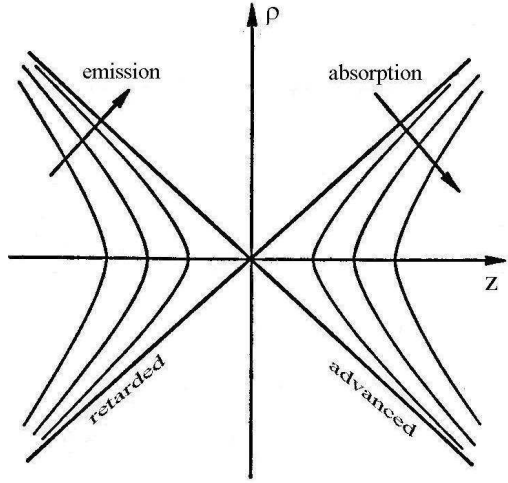}}  
\end{center}
\caption{The spherical equipotential surfaces of the electrostatic
field created by a charge at rest get transformed into two-sheeted
rotation-hyperboloids, contained inside an unlimited double-cone,
when the charge travels at Superluminal speed (cf.
refs.\cite{SupCh,Review}). This figures shows, among the others,
that a Superluminal charge travelling at constant speed, in a
homogeneous medium like the vacuum, does {\em not} lose
energy\cite{Folman}. \ Let us mention, incidentally, that this
double cone has nothing to do with the Cherenkov cone\cite{Comments2008}. \ [The
present picture is a reproduction of figure 27, appeared in 1986
at page 80 of ref.\cite{Review}]. }
\label{fig14}
\end{figure}

\h It is rather important, at this point, to notice that such a
solution, eq.(25), does represent a wave existing only inside the
(unlimited) double cone $\Ccal$ generated by the rotation around
the $z$-axis of the straight lines $\rho = \pm \ga\ze$: \ This too
is in full agreement with the predictions of the extended theory
of SR. \ For the explicit evaluation of the electromagnetic fields
generated by the Superluminal charge (and of their boundary values
and conditions) we confine ourselves here to merely quoting
ref.\cite{SupCh}. \ Incidentally, the same results found by
following the above procedure can be obtained by starting from the
four-potential associated with a subluminal charge (e.g., an
electric charge at rest), and then applying to it the suitable
Superluminal Lorentz ``transformation". \ One should also notice
that this double cone does not have much to do with the Cherenkov
cone\cite{Review,Folman,Comments2008}; and that a Superluminal charge travelling at
constant speed, in the vacuum, does {\em not} lose energy: See,
e.g., Fig.\ref{fig14} [which reproduces figure 27 at page 80 of
ref.\cite{Review}].

\h Outside the cone $\Ccal$, i.e., for $0 < \ga\mid{\ze}\mid <
\rho$, we get as expected no field, so that one meets a field
discontinuity when crossing the double-cone surface. Nevertheless,
the boundary conditions imposed by Maxwell equations are satisfied
by our solution (25), since at each point of the cone surface the
electric and the magnetic field are both tangent to the cone: also
for a discussion of this point we refer to quotation\cite{SupCh}.

\h Here, let us stress that, when $V \ra \infi$, and therefore $\ga \ra 0$, the
electric field tends to vanish, while the magnetic field tends to
the value $H_\phi = -\pi e / \rho^2$: This does agree once more
with what expected from extended SR, which predicted Superluminal
charges to behave (in a sense) as magnetic monopoles. In the
present paper we can only mention such a circumstance, and
refer to citations\cite{Rev1974,Review,North-Holl,chmonPLA}, and
papers quoted therein.

\

\

\section{APPENDIX: A Glance at the Experimental State-of-the-Art} 

\h Extended relativity can allow a better understanding of many
aspects also of {\em ordinary} physics\cite{Review}, even if
Superluminal objects (tachyons) did not exist in our cosmos as
asymptotically free objects. \ Anyway, at least three or four
different experimental sectors of physics seem to suggest the
possible existence of faster-than-light motions, or, at least, of
Superluminal group-velocities. \ We are going to put forth in the
following some information about the experimental results obtained
in {\em two} of those different physics sectors, with a mere
mention of the others.

\

\h {\em Neutrinos} -- First: A long series of experiments, started
in 1971, seems to show that the square ${m_{0}}^{2}$ of the mass
$m_{0}$ of muon-neutrinos, and more recently of electron-neutrinos
too, is negative; which, if confirmed, would mean that (when using
a na\"{i}ve language, commonly adopted) such neutrinos possess an
``imaginary mass'' and are therefore tachyonic, or mainly
tachyonic\cite{neutrinos,Review,Giannetto}.  \ [In extended SR, however,
the dispersion relation for a free Superluminal object does become \
$\om^2-\kbf^2=-\Om^2$, \ or \ $E^2-\imp^2=-m_\orm^{2}$, \ and
there is {\em no} need at all, therefore, of imaginary masses].

\

\h {\em Galactic Micro-quasars} -- Second: As to the {\em
apparent} Superluminal expansions observed in the core of
quasars\cite{quasars} and, recently, in the so-called galactic
micro-quasars\cite{microquasars}, we shall not really deal with
that problem, too far from the other topics of this paper; without
mentioning that for those astronomical observations there exist also
orthodox interpretations, based on ref.\cite{Rees}, that are still
accepted by the majority of the astrophysicists. \ For a
theoretical discussion, see ref.\cite{Rodono}.  Here, let us
only emphasize that simple geometrical considerations in Minkowski
space show that a {\em single} Superluminal source of light would
appear\cite{Rodono,Review}: \ (i) initially, in the ``optical
boom'' phase (analogous to the acoustic ``boom'' produced by an
airplane travelling with constant supersonic speed), as an intense
source which suddenly comes into view; and which, afterwards, \
(ii) seems to split into TWO objects receding one from the other
with speed \ $V>2c$ [all this being similar to what has been
actually observed, according to refs.\cite{microquasars}].

\

\begin{figure}[!h]
\begin{center}
 \scalebox{1.43}{\includegraphics{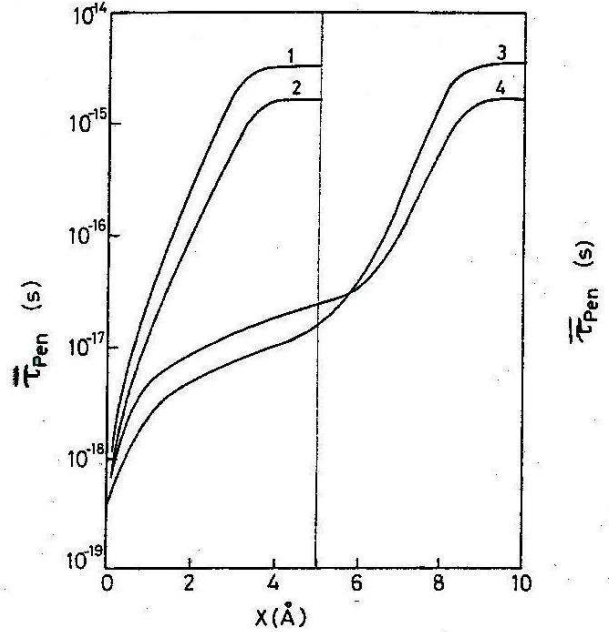}}  
\end{center}
\caption{Behaviour of the average ``penetration time" (in seconds)
spent by a tunnelling wavepacket, as a function of the penetration
depth (in {\aa}ngstroms) down a potential barrier (from Olkhovsky
et al., ref.\cite{Physique}). \ According to the predictions of
quantum mechanics, the wavepacket speed inside the barrier
increases in an unlimited way for opaque barriers; and the total
tunnelling time does {\em not} depend on the barrier
width.\cite{Reports1,Physique} } \label{fig15}
\end{figure}

\h {\em Evanescent waves and ``tunnelling photons''} -- Third:
Within quantum mechanics (and precisely in the {\em tunnelling}
processes), it had been shown that the tunnelling time ---firstly
evaluated as a simple Wigner's ``phase time'' and later on
calculated through the analysis of the wavepacket behaviour--- does
not depend\cite{Reports1,Physique} on the barrier width in the case of
opaque barriers (``Hartman effect''). This implies Superluminal
and arbitrarily large group-velocities $V$ inside long enough
barriers: see Fig.\ref{fig15}. \

\begin{figure}[!h]
\begin{center}
 \scalebox{1.6}{\includegraphics{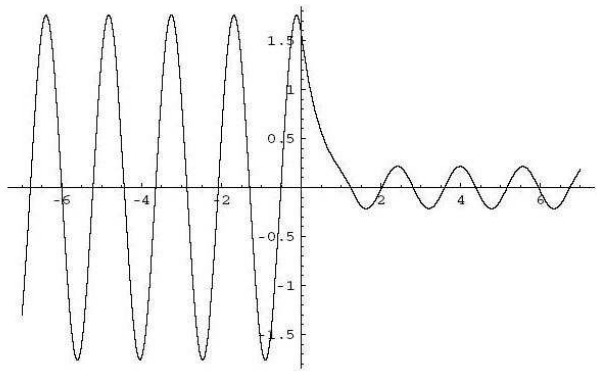}}  
\end{center}
\caption{Picture of the damping taking place inside a barrier
(from ref.\cite{RFG}): this damping does reduce the amplitude of
the tunnelling wavepacket, imposing a practical limit on the adoptable
barrier length.}
\label{fig16}
\end{figure}

Experiments that may verify this prediction by, say, electrons or
neutrons are difficult and rare\cite{Reports2,ORZ2}. Luckily
enough, however, the Schroedinger equation in the presence of a
potential barrier is mathematically identical to the Helmholtz
equation for an electromagnetic wave propagating, for instance,
down a metallic waveguide (along the $z$-axis): as shown, e.g., in
refs.\cite{ref[13]}; \ and a barrier height $U$ bigger than the
electron energy $E$ corresponds (for a given wave frequency) to a
waveguide of transverse size lower than a cut-off value. A segment
of ``undersized" guide ---to go on with our example--- does
therefore behave as a barrier for the wave ({\em photonic barrier\/}), as
well as any other photonic band-gap filters. \ The wave assumes
therein ---like a particle inside a quantum barrier--- an
imaginary momentum or wavenumber and, as a consequence, results
exponentially damped along $x$ [see, e.g. Fig.\ref{fig16}]: It
becomes an {\em evanescent} wave (going back to normal
propagation, even if with reduced amplitude, when the narrowing
ends and the guide returns to its initial transverse size). \
Thus, a tunnelling experiment can be simulated by having recourse
to evanescent waves (for which the concept of group velocity can
be properly extended: see the first one of refs.\cite{RFG}).

\begin{figure}[!h]
\begin{center}
 \scalebox{1.95}{\includegraphics{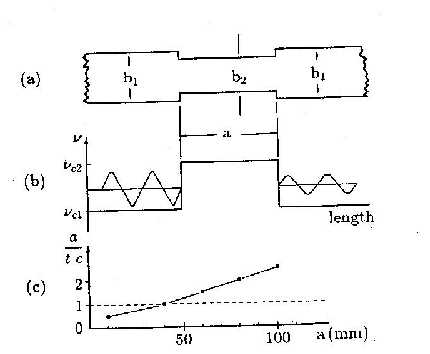}}  
\end{center}
\caption{Simulation of tunnelling by experiments with evanescent
{\em classical} waves (see the text), which were predicted to be
Superluminal also on the basis of extended SR\cite{ref19}. The
figure shows one of the measurement results by Nimtz et
al.\cite{Nimtz}; that is, the average beam speed while crossing
the evanescent region ( = segment of undersized waveguide, or
``barrier") as a function of its length. As theoretically
predicted\cite{Reports1,ref19}, such an average speed exceeds $c$
for long enough ``barriers". \ Further results appeared in
ref.\cite{Longhi}, and are reported below: see Figs.\ref{fig20}
and \ref{fig21} in the following .} \label{fig17}
\end{figure}

\begin{figure}[!h]
\begin{center}
 \scalebox{0.53}{\includegraphics{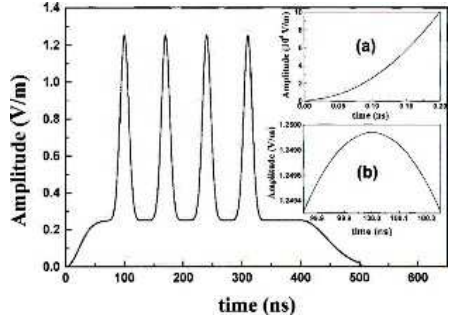}}  
\end{center}
\caption{The delay of a wavepacket crossing a barrier (cf., e.g.,
Fig.\ref{fig17} is due to the initial discontinuity. We then
performed suitable numerical simulations\cite{Barbero} by
considering an (indefinite) {\em undersized} waveguide, and
therefore eliminating any geometric discontinuity in its
cross-section.   \ This figure shows the envelope of the initial
signal. \ Inset (a) depicts in detail the initial part of this
signal as a function of time, while inset (b) depicts the gaussian
pulse peak centered at $t = 100$ ns .} \label{fig18}
\end{figure}

\begin{figure}[!h]
\begin{center}
 \scalebox{0.52}{\includegraphics{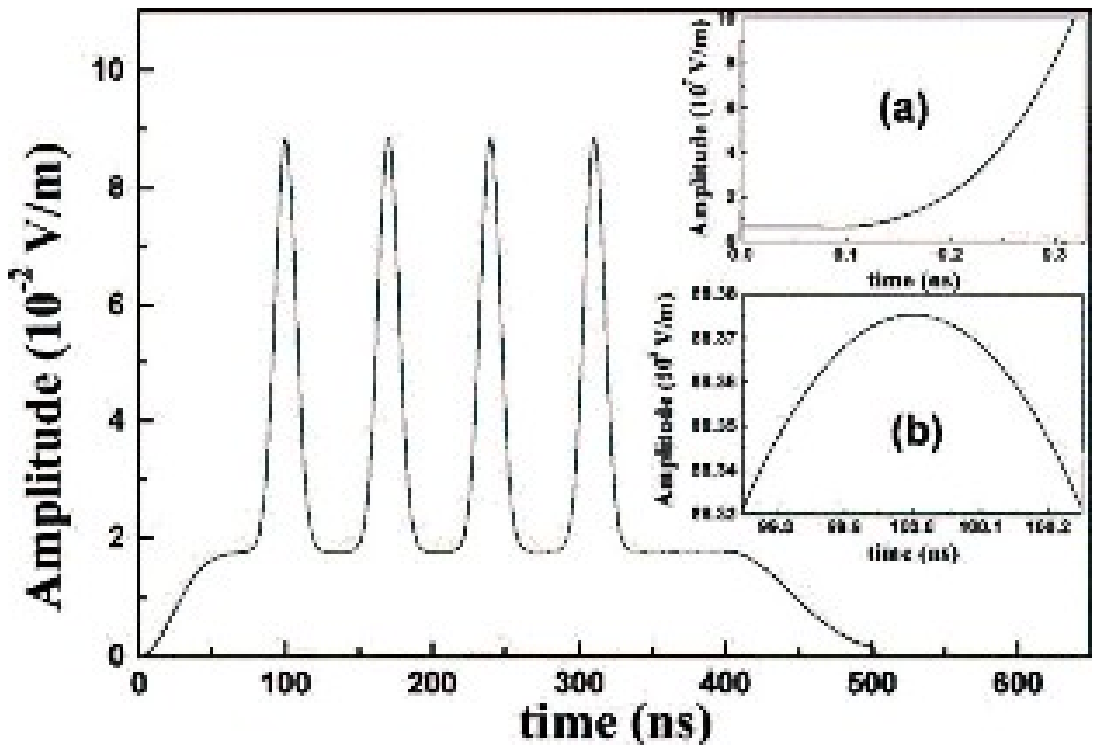}} 
\end{center}
\caption{Envelope of the signal in the previous figure
(Fig.\ref{fig18}) after having travelled a distance $L = 32.96$ mm
through the mentioned undersized waveguide. \ Inset (a) shows in
detail the initial part (in time) of such arriving signal, while
inset (b) shows the peak of the gaussian pulse that had been
initially modulated by centering it at $t = 100$ ns. \ One can see
that its propagation took {\em zero} time, so that the signal
travelled with infinite speed. \ The numerical simulation has been
based on Maxwell equations only. \ Going on from Fig.18 to Fig.19
one verifies that the signal strongly lowered its amplitute:
However, the width of each peak did not change (and this might
have some relevance when thinking of a Morse alphabet
``transmission": see the text) .} \label{fig19}
\end{figure}

\h The fact that evanescent waves travel with Superluminal speeds
(cf., e.g., Fig.\ref{fig17}) has been actually {\em verified} in a
series of famous experiments. Namely, various experiments,
performed since 1992 onwards by G.Nimtz et al. in
Cologne\cite{Nimtz}, by R.Chiao, P.G.Kwiat and A.Steinberg at
Berkeley\cite{Chiao}, by A.Ranfagni and colleagues in
Florence\cite{Ranfagni}, and by others in Vienna, Orsay, Rennes,
etcetera\cite{afterChiao}, verified that ``tunnelling photons" travel with
Superluminal group velocities [Such experiments raised a great
deal of interest\cite{Chiao2}, also within the non-specialized
press, and were reported in Scientific American, Nature, New
Scientist, etc.]. \ Let us further remark that also extended SR had
predicted\cite{ref19} evanescent waves to be endowed with
faster-than-$c$ speeds; the whole matter appears to be therefore
theoretically selfconsistent. \ The debate in the current
literature does not refer to the experimental results (which can
be correctly reproduced even by numerical
simulations\cite{Barbero,Brodowsky} based on Maxwell equations
only: Cf. Figs.\ref{fig18},\ref{fig19}), but rather to the
question whether they allow, or do {\em not} allow, sending
signals or information with Superluminal speed (see, e.g.,
refs.\cite{debates}).

\h In the above-mentioned experiments one meets a substantial
attenuation of the considered pulses ---cf. Fig.\ref{fig16}---
during tunnelling (or during propagation in an absorbing medium):
However, by employing ``gain doublets", it has been recently
reported the observation of undistorted pulses propagating with
Superluminal group-velocity with a {\em small} change in amplitude
(see, e.g., ref.\cite{Wang}).

\begin{figure}[!h]
\begin{center}
 \scalebox{0.8}{\includegraphics{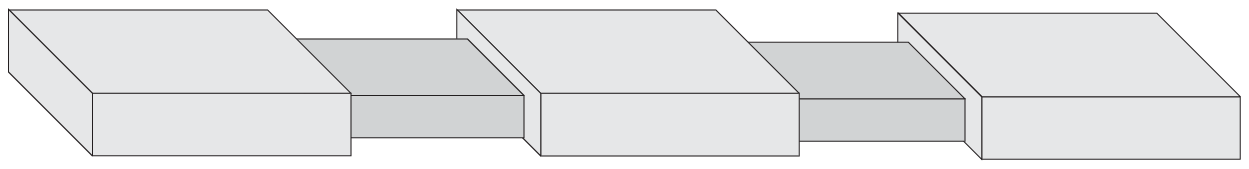}}  
\end{center}
\caption{Very interesting experiments have been performed with TWO
successive barriers, i.e., with two evanescence regions: For
example, with two gratings in an optical fiber. \ This
figure\cite{RFG} refers to the interesting experiment\cite{Nimtz3}
performed with microwaves travelling along a metallic waveguide:
the waveguide being endowed with {\em two} classical barriers
(undersized guide segments). See the text .} \label{fig20}
\end{figure}

\h  Let us emphasize that some of the most interesting experiments
of this series seem to be the ones with TWO or more ``barriers"
(e.g., with two gratings in an optical fiber\cite{Longhi}, or with
two segments of undersized waveguide separated by a piece of
normal-sized waveguide\cite{Nimtz3}: Fig.\ref{fig20}).

\

\

\begin{figure}[!h]
\begin{center}
 \scalebox{3.8}{\includegraphics{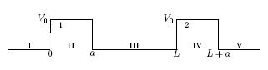}}  
\end{center}
\caption{Scheme of the (non-resonant) tunnelling process, through
{\em two} successive (opaque) quantum barriers. Far from
resonances, the (total) {\em phase time} for tunnelling through
the two potential barriers does depend neither on the barrier
widths {\em nor on the distance between the barriers}
(``generalized Hartman effect")\cite{twobar,Reports2,predic2b}. \
See the text.} \label{fig21}
\end{figure}

\h For suitable frequency bands ---namely, for ``tunnelling" far from
resonances---, it was found by us that the total crossing time
does not depend on the length of the intermediate (normal) guide:
that is, that the beam speed along it is
infinite\cite{twobar,Nimtz3,Reports2}. \ This does agree with what
predicted by Quantum Mechanics for the non-resonant tunnelling
through two successive opaque barriers\cite{twobar}: Fig.\ref{fig21}. \
Such a prediction has been verified first theoretically, by
Y.Aharonov et al.\cite{twobar}, and then, a second time,
experimentally: by taking advantage of the circumstance that
evanescence regions can consist in a variety of photonic band-gap
materials or gratings (from multilayer dielectric mirrors, or
semiconductors, to photonic crystals). Indeed, the best
experimental confirmation has come by having recourse to two
gratings in an optical fiber\cite{Longhi}: see Figs.\ref{fig22}
and 23; in particular, the rather peculiar (and quite
interesting) results represented by the latter.

\setcounter{figure}{21}
\begin{figure}[!h]
\begin{center}
 \scalebox{1.65}{\includegraphics{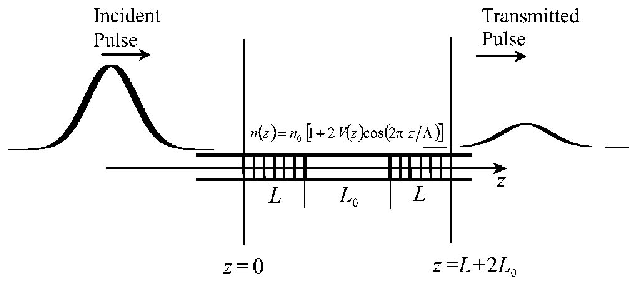}} 
\end{center}
\caption{Realization of the quantum-theoretical set-up represented
in Fig.\ref{fig21}, by using, as classical (photonic) barriers,
two gratings in an optical fiber\cite{predic2b}. \ The
corresponding experiment has been performed by Longhi et
al.\cite{Longhi} }
\label{fig22}             
\end{figure}

\

\begin{figure}[!h]
\begin{center}
 \scalebox{1.2}{\includegraphics{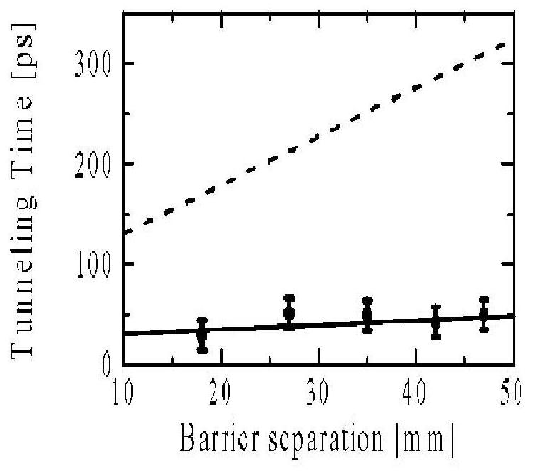}} 
\caption{Off-resonance tunnelling time versus barrier separation
for the rectangular symmetric DB FBG structure considered in
ref.\cite{Longhi} (see Fig.22). The solid line is the theoretical
prediction based on group delay calculations; {\em the dots are
the experimental points} as obtained by time delay measurements
[the dashed curve is the expected transit time from input to
output planes for a pulse tuned far away from the stopband of the
FBGs]. \ The experimental results\cite{Longhi} do confirm ---as
well as the early ones in refs.\cite{Nimtz3}]--- the theoretical
prediction  of a ``generalized Hartman Effect": in particular, the
independence of the total tunnelling time from the distance
between the two barriers. }
\end{center}
\label{fig23}
\end{figure}

\

\h  We cannot skip a further topic ---which, being delicate,
should not appear, probably, in a brief overview like this---
since it is presently arising more and more interest\cite{Wang}. \
Even if all the ordinary causal paradoxes seem to be
solvable\cite{FoP87,Review,RFG}, nevertheless one has to bear in
mind that (whenever it is met an object, ${\cal O}$, travelling
with Superluminal speed) one may have to deal with negative
contributions to the {\em tunnelling
times\/}\cite{negativeTh,Review,Reports2}: and this should not be
regarded as unphysical. In fact, whenever an ``object'' (particle,
electromagnetic pulse,,...) ${\cal O}$ {\em
overcomes\/}\cite{FoP87,Review} the infinite speed with respect to
a certain observer, it will afterwards appear to the same observer
as the ``{\em anti}-object'' $\overline{{\cal O}} $ travelling in
the opposite {\em space} direction\cite{ECGS1,Review,FoP87}. \ For
instance, when going on from the lab to a frame ${\cal F}$ moving
in the {\em same} direction as the particles or waves entering the
barrier region, the object ${\cal O}$ penetrating through the
final part of the barrier (with almost infinite
speed\cite{Physique,Reports1,Barbero,Reports2}, like in Figs.15)
will appear in the frame ${\cal F}$ as an anti-object
$\overline{{\cal O}}$ crossing that portion of the barrier {\em in
the opposite space-direction\/}\cite{FoP87,Review,ECGS1}. In the
new frame ${\cal F}$, therefore, such anti-object $\overline{{\cal
O}}$ would yield a {\em negative} contribution to the tunnelling
time: which could even result, in total, to be negative. \ For any
clarifications, see the quoted references. \ Let us stress, here,
that even the appearance of such negative times had been predicted
within SR itself\cite{negativeTh}, on the basis of its ordinary postulates;
and has been recently confirmed by quantum-theoretical evaluations
too\cite{Reports2,Petrillo}. \ (In the case of a non-polarized
beam,, the wave anti-packet coincides with the initial wave
packet; if a photon is however endowed with helicity $\lambda
=+1$, the anti-photon will bear the opposite helicity $\lambda
=-1$). \ From the theoretical point of view, besides the
above-quoted papers (in particular refs.\cite{Reports2,Reports1}),
see more specifically refs.\cite{negativeTh2}. \ On the (very
interesting!) {\em experimental} side, see the intriguing papers
\cite{negativeExp}.

\h  Let us {\em add} here that, via quantum interference effects,
it is possible to obtain dielectrics with refraction indices very
rapidly varying as a function of frequency, also in three-level
atomic systems, with almost complete absence of light absorption
(i.e., with quantum induced transparency)\cite{Alzetta}. \ The
group velocity of a light pulse propagating in such a medium can
decrease to very low values, either positive or negative, with
{\em no} pulse distortion. \ It is known that experiments have
been performed both in atomic samples at room temperature, and in
Bose-Einstein condensates, which showed the possibility of
reducing the speed of light to a few meters per second. \ Similar,
but negative group velocities, implying a propagation with
Superluminal speeds thousands of time higher than the previously
mentioned ones, have been recently predicted also in the presence
of such an ``electromagnetically induced transparency'', for light
moving in a rubidium condensate\cite{Artoni}. \ Finally, let us
recall that faster-than-$c$ propagation of light pulses can be
(and has been, in same cases) observed also by taking advantage of the
anomalous dispersion near an absorbing line, or nonlinear and
linear gain lines ---as already seen---, or nondispersive
dielectric media, or inverted two-level media, as well as of some
parametric processes in nonlinear optics (cf., e.g., G.Kurizki et
al.'s works).

\

{\bf D)} \ {\bf Superluminal Localized Solutions (SLS) to the wave
equations. The ``X-shaped waves"} -- The fourth sector (to leave
aside the others) is not less important. It came into fashion
again, when it was rediscovered in a series of remarkable works
that any wave equation ---to fix the ideas, let us think of the
electromagnetic case--- admit also solutions as much sub-luminal
as Super-luminal (besides the luminal ones, having speed $c/n$). \
Let us recall, indeed, that, starting from pioneering works as
H.Bateman's, it had slowly become known that all wave equations
admit soliton-like (or rather wavelet-type) solutions with
sub-luminal group velocities. \ Subsequently, also Superluminal
solutions started to be written down (in one case\cite{Barut2}
just by the mere application of a Superluminal Lorentz
``transformation"\cite{Review}).

\begin{figure}[!h]
\begin{center}
 \scalebox{0.8}{\includegraphics{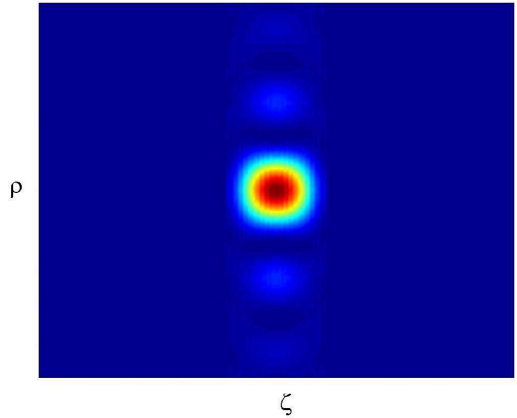}} 
\end{center}
\caption{The wave equations possess pulse-type solutions that, in
the subluminal case, are ball-like, in agreement with
Fig.\ref{fig11}.  \ For comments, see the text. }
\label{fig24}
\end{figure}

\h  As we know, a remarkable feature of some new solutions of
these (which attracted much attention for their possible
applications) is that they propagate as localized, non-diffracting
pulses, also because of their self-reconstruction property. \  It
is easy to realize the practical importance, for instance, of a
radio transmission carried out by localized beams, independently
of their speed; but non-diffracting wave packets can be of use even
in theoretical physics for a reasonable representation of
elementary particles; and so on. \ Incidentally, from the point of
view of elementary particles, it can be a source of meditation the
fact that the wave equations possess pulse-type solutions that, in
the subluminal case, are ball-like (cf. Fig.\ref{fig24}): this can
have a bearing on the corpuscle/wave duality problem met in
quantum physics (besides agreeing, e.g., with Fig.\ref{fig11}). Further
comments on this point are to be found below.

\h  At the cost of repeating ourselves, let us emphasize once more
that, within extended SR, since 1980 it had been found that
---whilst the simplest subluminal object conceivable is a small
sphere, or a point in the limiting case--- the simplest
Superluminal objects results by contrast to be (see
refs.\cite{BarutMR}, and Figs.\ref{fig11} and \ref{fig12} of this
paper) an ``X-shaped'' wave, or a double cone as its limit, which
moreover travels without deforming ---i.e., rigidly--- in a
homogeneous medium. \ It is not without meaning that the most
interesting localized solutions to the wave equations happened to
be just the Superluminal ones, and with a shape of that kind. \
Even more, since from Maxwell equations under simple hypotheses
one goes on to the usual {\em scalar} wave equation for each
electric or magnetic field component, one expected the same
solutions to exist also in the field of acoustic waves, of seismic
waves, and of gravitational waves too: and this has already been
demonstrated in the literature for all those cases, and especially
in Acoustics. \ Actually,
such pulses (as suitable superpositions of Bessel beams) were
mathematically constructed for the first time, by Lu et al. {\em
in Acoustics\/}: and were then called ``X-waves'' or rather
X-shaped waves. \ (One should not forget that, however, LWs have been
constructed in exact form even for other equations, as Schroedinger's
and as Einstein's).

\begin{figure}[!h]
\begin{center}
 \scalebox{1.2}{\includegraphics{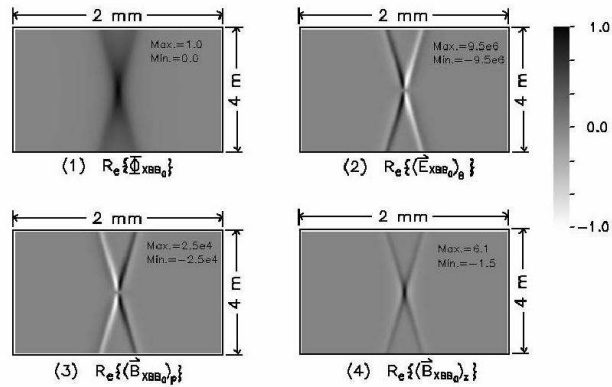}} 
\end{center}
\caption{Real part of the Hertz potential and of the field
components of the localized electromagnetic (``classic", axially
symmetric) X-shaped wave predicted, and first mathematically
constructed for the electromagnetic case, in refs.\cite{PhysicaA}.
\ For the meaning of the various panels, see the quoted
references. \ The dimension of each panel is 4 m (in the radial
direction) ${\times}$ 2 mm (in the propagation direction). \ [The
values shown on the right-top corner of each panel represent the
maxima and the minima of the images before normalization for
display (MKSA units)] .}
\label{fig25}
\end{figure}

\h  It is indeed important for us that the X-shaped waves have
been really produced in experiments, both with acoustic and with
electromagnetic waves; that is, X-pulses were produced that, in
their medium, travel undistorted with a speed larger than sound,
in the first case, and than light, in the second case. \ In
Acoustics, the first experiment was performed by Lu et al.
themselves in 1992, at the Mayo Clinic (and their papers received
the first IEEE 1992 award). \ In the electromagnetic case,
certainly more intriguing, Superluminal localized X-shaped
solutions were first mathematically constructed (cf., e.g.,
Fig.\ref{fig25}) in refs.\cite{PhysicaA}, and later on
experimentally produced by Saari et al.\cite{Saari97} in 1997

\begin{figure}[!h]
\begin{center}
 \scalebox{1.82}{\includegraphics{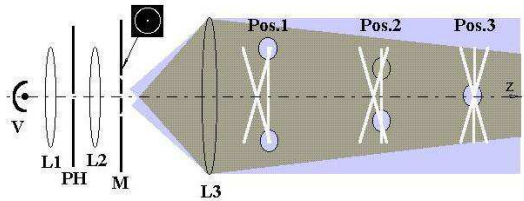}}  
\end{center}
\caption{Scheme of the experiment by Saari et al., who announced
(PRL of 24 Nov.1997) the production in optics of the beams
depicted in the previous Fig.\ref{fig25}. \  In the present figure
one can see what it was shown by the experimental results: Namely,
that the ``X-shaped" waves are Superluminal: indeed, they, running
after plane waves (the latter regularly travelling with speed $c$),
do catch up with the considered plane waves. \ An analogous
experiment has been later on performed with microwaves at Florence
by Ranfagni et al. (PRL of 22 May 2000). }
\label{fig26}
\end{figure}

\begin{figure}[!h]
\begin{center}
 \scalebox{2.6}{\includegraphics{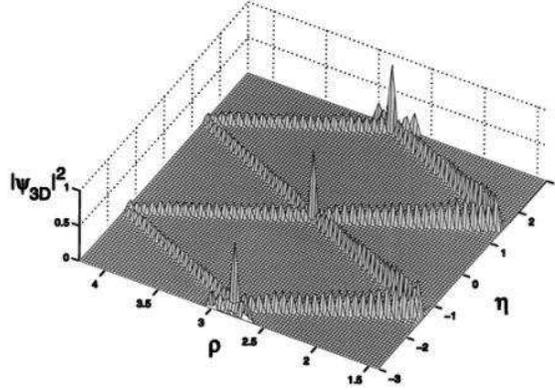}} 
\caption{In this figure a couple of elements are depicted of one
of the trains of X-shaped pulses, mathematically constructed in
ref.\cite{coaxial}, which propagate down a coaxial guide (in the
TM case): This picture is just taken from ref.\cite{coaxial}, \
but analogous X-pulses exist (with infinite or finite total
energy) for propagation along a cylindrical, normal-sized metallic
waveguide]. }
\end{center}
\label{fig27}
\end{figure}

at Tartu by visible light (Fig.\ref{fig26}), and more recently by
Ranfagni et al. at Florence by
microwaves\cite{Ranfagni}. \ In the theoretical sector the
activity has been not less intense, in order to build up ---for
example--- analogous new solutions with finite total energy or
more suitable for high frequencies, on one hand, and localized
solutions Superluminally propagating even along a normal waveguide
(cf. Fig.\ref{fig27}), on another hand, and so on.

\

\h  Let us eventually recall the problem of producing an X-shaped
Superluminal wave like the one in Fig.12, but truncated
---of course-- in space and in time (by use of a finite antenna,
radiating for a finite time): in such a situation, the wave is
known to keep its localization and Superluminality only till a
certain depth of field [i.e., as long as they are fed by the waves
arriving (with speed $c$) from the antenna], decaying abruptly
afterwards.\cite{Durnin,MRH} \ Let us add that various authors, taking
account, e.g., of the time needed for fostering such Superluminal
waves, have concluded that these localized Superluminal pulses are
unable to transmit information faster than $c$.
 \ Many of these questions have been discussed in what
precedes; for further details, see the second of refs.\cite{PhysicaA}.

\h Anyway, the existence of the X-shaped Superluminal (or
Super-sonic) pulses seem to constitute, together, e.g., with the
Superluminality of evanescent waves, a confirmation of extended
SR: a theory\cite{Review} based on the ordinary postulates of SR
and that consequently does not appear to violate any of the
fundamental principles of physics. \ It is curious moreover,  that
one of the first applications of such X-waves (that takes
advantage of their propagation without deformation) has been
accomplished in the field of medicine, and precisely ---as we
know--- of ultrasound scanners\cite{LuBiomedical,LuImaging}; while
the most important applications of the (subluminal!) Frozen Waves
will very probably affect, once more, human health problems like
the cancerous ones.

\h After the ``digression" constituted by the above Appendix, let
us go on to the Second Part of this work, with a slightly more
technical\cite{Capitulo} review about the physical and mathematical
characteristics of the Localized Waves and about some interesting
applications. [In the Third Part we shall deal with the ones
endowed with zero speed, i.e., with a static envelope, and, more
in general, with the subluminal LWs].

\newpage

\cent{\Huge{\bf SECOND PART}}

\

\

\cent{\Large{\bf STRUCTURE OF THE NONDIFFRACTING WAVES}}

\

\cent{\Large{\bf AND SOME INTERESTING APPLICATIONS}}

\

\

\section{Foreword} 

Since the early works[113-116] on the so-called nondiffracting
waves (or ``Localized Waves"), a great deal of results has been
published on this important subject, from both the theoretical and
the experimental point of view. Initially, the theory was
developed taking into account only free space; however, in recent
years, it has been extended for more complex media exhibiting
effects such as dispersion[117-119], nonlinearity\cite{[8]},
anisotropy\cite{[9]} and losses\cite{[10]}. Such extensions have
been carried out along with the development of efficient methods
for obtaining nondiffracting beams and pulses in the subluminal,
luminal and Superluminal regimes[123-130].

\h This Second Part addresses some theoretical methods related to
nondiffracting solutions of the linear wave equation in unbounded
homogeneous media, as well as to some interesting applications of
such solutions.\cite{Capitulo}

\h The usual cylindrical coordinates $(\rho,\phi,z)$ will be used
here. We already know that in these coordinates the linear wave equation
is written as

\bb
\frac{1}{\rho}\frac{\pa}{\pa\rho}\left(\rho\frac{\pa\Psi}{\pa\rho}
\right) + \frac{1}{\rho^2}\frac{\pa^2\Psi}{\pa\phi^2} +
\frac{\pa^2\Psi}{\pa z^2} - \frac{1}{c^2}\frac{\pa^2\Psi}{\pa t^2}
\ug 0 \label{S2eo}\ee

In Section VII we analyse the general structure of the Localized
Waves, develop the so called Generalized Bidirectional
Decomposition, and use it to obtain several luminal and
Superluminal nondiffracting wave solutions of eq.(\ref{S2eo}).

\h In Section VIII we develop a kind of space-time focusing method
by a continuous superposition of X-Shaped pulses of different
velocities.

\h Section IX addresses the properties of chirped optical X-Shaped
pulses propagating in material media without boundaries.

\h Subsequently, on the basis of what expounded in Section VII, we shall show
at the beginning of the Third Part (in Section X) how a suitable superposition of
Bessel beams can be even used to obtain {\em stationary} localized wave
fields with high transverse localization, and whose longitudinal
intensity pattern can assume any desired shape within a chosen
interval $0 \leq z \leq L$ of the propagation axis.

\

\h For containing the length of this review, we had, obviously, to skip many
interesting results. \ Let us just mention, for example, that rather simple
{\em analytic} expressions, capable of describing the longitudinal (on-axis)
evolution of axially-symmetric nondiffracting pulses, have been recently
worked out in ref.\cite{JosaMlast*} even for pulses {\em truncated} by
finite apertures.  By comparing what easily provided by such
expressions, for several situations (involving subluminal, luminal,
or Superluminal localized pulses), with the results obtained by numerical
evaluations of the Rayleigh-Sommerfeld diffraction integrals, an exellent
agreement has been found. Therefore, those new closed-form expressions
dispense with the need of time-consuming numerical simulations (and provide
an effective tool for finding out the most important properties of the
truncated localized pulses).

\

\section{Spectral structure of the Localized Waves and the Generalized
Bidirectional Decomposition} 

An effective way to understand the concept of the (ideal)
nondiffracting waves is furnishing a precise mathematical definition
of these solutions, so to extract the necessary spectral
structure from them.

\h Intuitively, an ideal nondiffracting wave (beam or pulse) can be
defined as a wave capable of maintaining indefinitely its spatial
form (apart from local variations) while propagating.

\h We can express such a characteristic by saying that a localized
wave has to possess the property[124,125]

\bb \Psi(\rho,\phi,z,t) \ug \Psi(\rho,\phi,z + \Delta z_0,t +
\frac{\Delta z_0}{V}) \label{S2def} \ee

where $\Delta z_0$ is a
certain length and $V$ is the pulse propagation speed that here
can assume any value: $0\leq V \leq \infty$.

\h In terms of a Fourier Bessel expansion, we can write a function
$\Psi(\rho,\phi,z,t)$ as

\bb \Psi(\rho,\phi,z,t) \ug
\sum_{n=-\infty}^{\infty}\left[\int_{0}^{\infty}\drm \kr\,\int_{-\infi}^{\infty}\drm k_z\,\int_{-\infi}^{\infty}
\drm \om\,\kr A_n^{'}(\kr,k_z,\om) J_n(\kr \rho)e^{ik_z z}e^{-i\om t}e^{i n
\phi} \right] \, . \label{S2geral1} \ee

On using the translation property of the Fourier transforms
$\emph{T}[f(x+a)] = {\rm exp}(ika)\emph{T}[f(x)]$, we have that
$A_n^{'}(\kr,k_z,\om)$ and ${\rm exp}[i(k_z\Delta z_0 -\om\Delta
z_0/V)]A_n^{'}(\kr,k_z,\om)$ are the Fourier Bessel transforms of
the l.h.s and r.h.s. functions in eq.(\ref{S2def}). And from this
same equation we can get[124,125] the fundamental constraint
linking the angular frequency $\om$ and the longitudinal
wavenumber $k_z$:

\bb \om \ug Vk_z + 2m\pi \frac{V}{\Delta z_0} \label{S2cond2} \ee

with $m$ an integer. Obviously, this constraint can be satisfied
by means of the spectral functions $A_n^{'}(\kr,k_z,\om)$.

\h Now, let us explicitly mention that constraint (\ref{S2cond2})
does not imply any breakdown of the wave equation. In
fact, when inserting expression (\ref{S2geral1}) in the wave
equation (\ref{S2eo}), one gets that

\bb \frac{\om^2}{c^2} \ug k_z^2 + \kr^2 \label{S2cond3} \; . \ee

So, to obtain a solution of the wave equation by
(\ref{S2geral1}), the spectrum $A_n^{'}(\kr,k_z,\om)$ must possess
the form

\bb A_n^{'}(\kr,k_z,\om) \ug A_n(k_z,\om)\,\delta\left[\kr^2 -
\left(\frac{\om^2}{c^2} - k_z^2 \right)\right] \label{S2sepc1} \ee

where $\delta(.)$ is the Dirac delta function. With this, we can
write a solution of the wave equation as

\bb \Psi(\rho,\phi,z,t) \ug
\sum_{n=-\infty}^{\infty}\left[\int_{0}^{\infty}\drm \om\,\int_{-\om/c}^{\om/c}\drm k_z
A_n(k_z,\om) J_n\left(\rho\sqrt{\frac{\om^2}{c^2} - k_z^2}
\right)e^{ik_z z}e^{-i\om t}e^{i n \phi} \right] \label{S2geral2}
\ee

where we have considered positive angular frequencies only.

\h Equation (\ref{S2geral2}) is a superposition of Bessel beams and
it is understood that the integrations in the $(\om,k_z)$ plane
are confined to the region $0\leq \om \leq \infty$ and $-\om/c
\leq k_z \leq \om/c \,$.

\h Now, to obtain an ideal nondiffracting wave, the spectra
$A_n(k_z,\om)$ must obey the fundamental constraint
(\ref{S2cond2}), and so we write

\bb A_n(k_z,\om) \ug
\sum_{m=-\infty}^{\infty}\,S_{nm}(\om)\delta\left[\om - (Vk_z +
b_m) \right] \label{S2speclw} \ee

where $b_m$ are constants representing the terms $2m\pi V/\Delta
z_0$ in eq.(\ref{S2cond2}), and $S_{nm}(\om)$ are arbitrary
frequency spectra.

\h By inserting eq.(\ref{S2speclw}) into eq.(\ref{S2geral2}), we get a general
integral form of the ideal nondiffracting wave (\ref{S2def}):

\bb \Psi(\rho,\phi,z,t) \ug
\sum_{n=-\infty}^{\infty}\,\sum_{m=-\infty}^{\infty}
\psi_{nm}(\rho,\phi,z,t) \label{S2suppsinm}\ee

with

\bb \begin{array}{clcr} \psi_{nm}(\rho,\phi,z,t) \ug &
\dis{e^{-ib_m
z/V}\,\int_{(\wmin)_m}^{(\wmax)_m}\, d\om \, S_{nm}(\om)} \\

\\

& \times \dis{
J_n\left(\rho\sqrt{\left(\frac{1}{c^2}-\frac{1}{V^2}\right)\om^2 +
\frac{2b}{V^2}\om - \frac{b^2}{V^2}}
\right)e^{i\frac{\om}{V}(z-Vt)}e^{i n \phi}} \end{array}
\label{S2psinm} \ee where $\wmin$ and $\wmax$ depend on the values
of V:

\begin{itemize}
    \item{for subluminal $(V<c)$ localized waves: $b_m>0$,
    $(\wmin)_m=cb_m/(c+V)$ and $(\wmax)_m=cb_m/(c-V)$};
    \item{for luminal $(V=c)$ localized waves: $b_m>0$,
    $(\wmin)_m=b_m/2$ and $(\wmax)_m=\infty$};
    \item{for Superluminal $(V>c)$ localized waves: $b_m\geq 0$,
    $(\wmin)_m=cb_m/(c+V)$ and $(\wmax)_m=\infty$. Or $b_m<0$,
    $(\wmin)_m=cb_m/(c-V)$ and $(\wmax)_m=\infty$}.
\end{itemize}

It is important to notice that each $\psi_{nm}(\rho,\phi,z,t)$ in
the superposition (\ref{S2suppsinm}) is a truly nondiffracting
wave (beam or pulse) and the superposition of them,
(\ref{S2suppsinm}), is just the most general form to represent a
nondiffracting wave defined by eq.(\ref{S2def}). Due to this fact,
the search for methods capable of providing analytic solutions for
$\psi_{nm}(\rho,\phi,z,t)$, eq.(\ref{S2psinm}), becomes an
important task.

\h Let us recall that equation (\ref{S2psinm}) is also a Bessel
beam superposition, but with constraint (\ref{S2cond2})
linking their angular frequencies and longitudinal wavenumbers.

\h In spite of the fact that expression (\ref{S2psinm})
represents ideal nondiffracting waves, it is difficult to obtain
closed analytic solutions from it. Due to this, we are going to
develop a method capable of overcoming such a difficulty, providing
several interesting localized wave solutions (luminal and
Superluminal) of arbitrary frequencies, including some solutions
endowed with finite energy.

\subsection{The Generalized Bidirectional Decomposition}

For reasons that will be clear soon, instead of dealing with the
integral expression (\ref{S2suppsinm}), our starting point is the
general expression (\ref{S2geral2}). \ Here, for simplicity, we shall
restrict ourselves to axially
symmetric solutions, adopting the spectral functions

\bb A_n(k_z,\om) \ug \delta_{n0}A(k_z,\om) \ee

where $\delta_{n0}$ is the Kronecker delta.

\h In this way, we get the following general solution (considering
positive angular frequencies only), which describes axially symmetric
waves:

\bb \Psi(\rho,\phi,z,t) \ug
\int_{0}^{\infty}\drm \om\,\int_{-\om/c}^{\om/c}\drm k_z A(k_z,\om)
J_0\left(\rho\sqrt{\frac{\om^2}{c^2} - k_z^2}\right)e^{ik_z
z}e^{-i\om t}\label{S2geral4} \ee

As we have seen, we can obtain ideal nondiffracting waves, given that
the spectrum $A(k_z,\om)$ satisfies the linear relationship
(\ref{S2cond2}). Therefore, it becomes natural to choose new spectral
parameters, in place of $(\om,k_z)$, that make easier to
implement the mentioned constraint[124,125]. \ With this in mind, let
us choose the new spectral parameters $(\al,\be)$

\bb \al \equiv \frac{1}{2V}(\om + Vk_z) \;;\;\;\; \be \equiv
\frac{1}{2V}(\om - Vk_z) \; . \label{S2ab} \ee

\emph{Let us consider here only luminal $(V=c)$ and Superluminal
$(V>c)$ nondiffracting pulses}.

\h With the change of variables (\ref{S2ab}) in the integral
solution(\ref{S2geral4}), and considering $(V \geq c)$, the
integration limits on $\al$ and $\be$ have to satisfy the three
inequalities

\bb \left\{ \begin{array}{clcr} 0 < \al + \be < \infty \\

\\

\al \geq \dis{\frac{c-V}{c+V}}\be   \\

\\

\al \geq \dis{\frac{c+V}{c-V}}\be \end{array}\right.\label{S2inq2}
\ee

Let us suppose both $\al$ and $\be$ to be positive $[\al,\,\be\geq
0]$. The first inequality in (\ref{S2inq2}) is then satisfied;
while the coefficients $(c-V)/(c+V)$ and $(c+V)/(c-V)$ entering
relations (\ref{S2inq2}) are both negatives (since $V \geq c$). As a
consequence, the other two inequalities in (\ref{S2inq2}) result
to be automatically satisfied. In other words, the integration
limits $0\leq \al \leq \infty$ and $0\leq \be \leq \infty$ are
 contained" into the limits (\ref{S2inq2}) and are therefore
acceptable. Indeed, they constitute a rather suitable choice for
facilitating all the subsequent integrations.

\h Therefore, instead of eq.(\ref{S2geral4}), we shall consider the
(more easily integrable) Bessel beam superposition in the new
variables [with $V \geq c$]

\bb\Psi(\rho,\ze,\eta) = \int_{0}^{\infi}
 \drm \al \int_{0}^{\infi}\drm \be A(\al,\be)\,
J_0\left(\rho\,\sqrt{\left(\frac{V^2}{c^2}-1\right)
 (\al^2+\be^2)+2\left(\frac{V^2}{c^2}+1\right)\al\be} \right)
e^{i\al\ze} e^{-i\be\eta}\label{S2geral5}\ee

where we have defined

\bb \zeta \equiv z-Vt\;;\;\;\;\;\;\;\; \eta \equiv z+Vt \; . \ee

The present procedure is a generalization of the so-called
``bidirectional decomposition" technique\cite{[11]}, which was
devised in the past for $V = c$.

\h From the new spectral parameters defined in transformation
(\ref{S2ab}), it is easy to see that the constraint
(\ref{S2cond2}), i.e. $\om = Vk_z + b$, is implemented just by
making

\bb A(k_z,\om) \rightarrow  A(\al,\be) \ug S(\al)\delta(\be-\be_0)
\label{S2specab} \ee

with $\beta_0=b/2V$. The delta function
$\delta(\be-\be_0)$ in the spectrum (\ref{S2specab}) means that we
are integrating Bessel beams along the continuous line $\om = Vk_z
+ 2V\be_0$ and, in this way, the function $S(\al)$ will give the
frequency dependence of the spectrum: $S(\al)\rightarrow S(\om/V -
\be_0)$.

\h This method constitutes a simple, natural way for obtaining
pulses with field concentration on $\rho=0$ and at $\zeta=0
\rightarrow z=Vt$.

\h Now, it is important to emphasize\cite{[13]} that, when
$\be_0>0$ in (\ref{S2specab}), superposition (\ref{S2geral5}) gets
contributions from both backward and forward travelling Bessel beams,
corresponding to the
frequency intervals $V\be_0 \leq \om < 2V\be_0$ (where $k_z<0$)
and $2V\be_0 \leq \om \leq \infty$ (where $k_z\geq 0$),
respectively. \ Nevertheless, we can obtain physical solutions when
rendering the
contribution of the backward-travelling components negligible, by choosing
suitable weight functions $S(\al)$.

\h It is also worth noticing that we adopted the new spectral
parameters $\al$ and $\be$ just to obtain (closed-form) analytic
localized wave solutions: The spectral characteristics
of these new solutions can be brought into evidence by using
transformations (\ref{S2ab}) and writing the corresponding
spectrum in terms of the usual $\om$ and $k_z$ spectral
parameters.

\h In the following, we consider some cases with $\be_0=0$ and
$\be_0>0$.

\

\subsubsection{Closed analytic expressions describing some ideal
nondiffracting pulses}

\

\

Let us first consider, in eq.(\ref{S2geral5}),
spectra of the type (\ref{S2specab}) with $\be_0 = 0$:

\bb A(\al,\be) \ug aV\,\delta(\be)e^{-aV\al} \label{S2spec1} \ee

\bb A(\al,\be) \ug aV\,\delta(\be)J_0(2d\sqrt{\al})e^{-aV\al}
\label{S2spec2} \ee

\bb A(\al,\be) \ug \delta(\be)\frac{\sin(d\al)}{\al}e^{-aV\al} \; ,
\label{S2spec4} \ee

$a>0$ and $d$ being constants.

\h One can obtain from the above spectra the following Superluminal
LW solutions, respectively:

--- from spectrum (\ref{S2spec1}), we can use the identity (6.611.1)
in ref.\cite{[19]}, obtaining the well known ordinary X xave
solution (also called X-shaped pulse)

 \bb \Psi(\rho,\zeta) \; \equiv \; X \ug \frac{aV}{\sqrt{(aV-i\ze)^2 +
\left(\frac{V^2}{c^2}-1\right)\rho^2}} \; ; \label{S2X}\ee

--- by using spectrum (\ref{S2spec2}) and the identity (6.6444) of
ref.\cite{[19]}, one gets

\bb \Psi(\rho,\zeta) \ug X \ J_0\left(\sqrt{\frac{V^2}{c^2}-1}
\;\; (aV)^{-2}d^2X^2\rho\right){\rm exp}\left[-(aV-i\zeta) \,
(aV)^{-2}d^2 \, X^2\right] \; ; \ee

--- the Superluminal nondiffracting pulse

\bb \begin{array}{clcr} \Psi(\rho,\ze) \ug \sin^{-1} \, \left[
2\dis{\frac{d}{aV}}\left(\dis{ \sqrt{X^{-2}+(d/aV)^2+2\rho
d(aV)^{-2} \sqrt{V^2/c^2-1} } } \right.\right. \\

\\

 + \left.\left. \dis{ \sqrt{X^{-2}+(d/aV)^2-2\rho d(aV)^{-2}
 \sqrt{V^2/c^2-1}} } \;\right)^{-1} \right]\end{array} \ee

is obtained from spectrum (\ref{S2spec4}) by using identity (6.752.1)
of ref.\cite{[19]}  for $a>0$ and $d>0$.

\h From the previous discussion, we get to know that any solutions obtained
from spectra of the type (\ref{S2specab}) with $\be_0=0$ are free
from noncausal (backward travelling) components.

\h In addition, \emph{when $\be_0=0$}, we can see that the pulsed
solutions depend on $z$ and $t$ through $\zeta=z-Vt$ only, and so
propagate rigidly, i.e. without distortion. Such pulses can be
transversally localized {\em only} if $V>c$, because if $V=c$ the
function $\Psi$ has to obey the Laplace equation on the transverse
planes[124,125].

\h Many others Superluminal localized waves can be easily
constructed\cite{[13]} from the above solutions just by taking the
derivatives (of any order) with respect to $\zeta$. It is also
possible to show\cite{[13]} that the new solutions, obtained in
this way, have their spectra shifted towards higher frequencies.

\

\h Now, let us pass to consider, in eq.(\ref{S2geral5}), a
spectrum of the type (\ref{S2specab}) with $\be_0 > 0$:

\bb A(\al,\be) \ug aV\delta(\be - \be_0) \, e^{-aV\al}
\label{S2specab2} \ee

with $a$ a positive constant.

\h As we have seen, the presence of the delta function, with the
constant $\be_0>0$, implies that we are integrating (summing)
Bessel beams along the continuous line $\om = Vk_z + 2V\be_0$.
Now, the function $S(\al)=aV{\rm exp}(-aV\om)$ entails that we are
considering a frequency spectrum of the type $S(\om) \propto {\rm
exp}(-a\om)$, and therefore with a bandwidth given by $\Delta\om =
1/a$.

\h Since $\be_0>0$, the interval $V\be_0 \leq \om < 2V\be_0$ (or,
equivalently, in this case, $0\leq \al < \be_0$), corresponds to
backward Bessel beams, i.e., negative values of $k_z$. However, we
can get physical solutions when making the contribution of this
frequency interval negligible. In our case, this can be obtained by
making $a\be_0 V<<1$, so that the exponential decay of the
spectrum $S$ with respect to $\om$ is very slow, and the
contribution of the interval $\om\geq 2V\be_0$ (where $k_z\geq 0$)
largely overruns the $V\be_0 \leq \om < 2V\be_0$ (where $k_z<0$)
contribution.

\h Incidentally, let us note that, once we ensure the causal behaviour of
the pulse by making $aV\be_0<<1$ in (\ref{S2specab2}), we have
that $\Delta\al=1/aV>>\be_0$, and one can therefore simplify the
argument of the Bessel function, in the integrand of superposition
(\ref{S2geral5}), by neglecting the term $(V^2/c^2 -1)\be_0^2$.
With this, the superposition (\ref{S2geral5}), with the spectrum
(\ref{S2specab2}), can be written as

\bb \Psi(\rho,\zeta,\eta)  \approx  a V e^{-i\be_0\eta}
\int_{0}^{\infi} \drm \al
J_0\left(\rho \sqrt{\left(\frac{V^2}{c^2}-1\right)
 \al^2+2\left(\frac{V^2}{c^2}+1\right)\al\be_0} \right)
e^{i\al\ze} e^{-aV\al} \label{S2geral6}\ee

Now, we can use identity (6.616.1) of ref.\cite{[19]} and obtain
the new localized Superluminal solution called\cite{[13]}
Superluminal Focus Wave Mode (SFWM):

\bb \Psi_{\rm SFWM}(\rho,\ze,\eta) \ug {\rm e}^{-i\be_0\eta} \; X
\; \dis{\exp\left[ \frac{\be_0(V^2+c^2)}{V^2-c^2} \, \left(
(aV-i\ze)- a\,VX^{-1} \right) \right]}\label{S2SFWM}\ee

where, as before, $X$ is the ordinary X-pulse (\ref{S2X}). The
center of the SFWM is located on $\rho=0$ and $\zeta=0\,$ (i.e. at
 $z=Vt$). The intensity, $|\Psi|^2$, of this pulse propagates rigidly,
it being a
 function of $\rho$ and $\zeta$ only. However, the complex function
 $\Psi_{\rm SFWM}$ (i.e. its real and imaginary parts) propagate
 with {\em local} variations, recovering their whole three
 dimensional form after each space and time interval $\Delta
z_0 =
 \pi/\be_0$ and $\Delta t_0 = \pi/\be_0 V$.

\h The SFWM solution written above, for $V \longrightarrow c^+$
 reduces to the well known Focus Wave Mode (FWM) solution\cite{[11]},
 travelling with speed $c$:

\bb \Psi_{\rm FWM}(\rho,\ze,\eta) \ug ac \dis{\frac{{\rm
e}^{-i\be_0\eta}}{ac-i\ze} \;
 \exp\left[-\frac{\be_0\rho^2}{ac-i\ze}\right]} \ . \ee

Let us also emphasize that, since $\be_0>0$, spectrum
(\ref{S2specab2}) results to correspond to angular frequencies
$\om \geq V\be_0$. Thus, our new solution can be used to construct
also {\em high frequency} pulses.

\

\subsubsection{Finite energy nondiffracting pulses}

\

In this subsection, we shall show how to get finite energy
localized wave pulses, that can propagate for long
distances while maintaining their spatial resolution, i.e., that
possess a large depth of field.

\h As we have seen, ideal nondiffracting waves can be constructed by
superposing Bessel beams [cf. eq.(\ref{S2geral4}) for cylindrical
symmetry] with a spectrum $A(\om,k_z)$ that satisfies a linear
relationship between $\om$ and $k_z$. In the general bidirectional
decomposition method, this can be obtained by using spectra of the
type (\ref{S2specab}) in superposition (\ref{S2geral5}).

\h Solutions of that type possess an infinity depth of field:
however, they are endowed with infinite energy[123,125]. To overcome
this problem, we can truncate an ideal nondiffracting wave by a
finite aperture, and the resulting pulse will have finite energy
and a finite field-depth.  Even so, such field-depths may be very
large when compared with those of ordinary waves.

\h The problem in the present case is that the resulting field has to be
calculated by diffraction integrals (such as the well known
Rayleigh-Sommerfeld formula) and, in general, a closed analytic
formula for the resulting pulse cannot be obtained.

\h However, there is another way to construct localized pulses
with finite energy\cite{[13]}. Namely, by using spectra
$A(\om,k_z)$ in eq.(\ref{S2geral4}) whose domains are not restricted
exactly to the straight line $\om = Vk_z + b$, but are defined
in the surroundings of that line, wherein the spectra should have their main
values concentrated (in other words, any spectrum has to be well localized in
the vicinity of that line).

\h Similarly, in terms of the generalized bidirectional decomposition
given in eq.(\ref{S2geral5}), finite energy nondiffracting wave
pulses can be constructed by adopting spectral
functions $A(\al,\be)$ well localized in the vicinity of the
line $\be=\be_0$, quantity $\be_0$ being a constant.

\h To exemplify this method, let us consider the following spectrum

\bb A(\al,\be) \ug \left\{\begin{array}{clr}
 &a\,q\,V\,e^{-aV\al}e^{-q(\be-\be_0)}\ \ \ \ \ \ \ \ \ \ & {\rm for} \
 \be \geq \be_0  \\

 \\

 &0 \ \ \ \ \ \ \ \ \ \ \ \ \ \ \ \ \ \ & {\rm for} \
 0 \leq \be < \be_0
 \end{array} \right. \label{S2specSMPS}\ee

in superposition (\ref{S2geral5}), quantities $a$ and $q$ being free
positive constants and $V$ the peak's pulse velocity (here $V\geq
c$).

\h It is easy to see that the above spectrum is zero in the region
above the $\be=\be_0$ line, while it decays in the region below
(as well as along) such a line. We can concentrate this spectrum
on $\be=\be_0$ by choosing values of $q$ in such a way that
$q\be_0>>1$. The faster the spectrum decay takes place in the
region below the $\be=\be_0$ line, the larger the field-depth of
the corresponding pulse results to be.

\h Besides this, once we choose $q\be_0>>1$ to obtain pulses with
a large field-depth, we can also minimize the contribution of the
noncausal (backward) components by choosing $aV\be_0<<1$;  analogously
with what we did and obtained for the SFWM case.

\h Again in analogy with the SFWM case, when we choose $q\be_0>>1$
(i.e., a long field-depth) and $aV\be_0<<1$ (a minimal contribution of
the backward components), one is allowed to simplify the argument of the Bessel
function, in the integrand of superposition (\ref{S2geral5}), by
neglecting the term $(V^2/c^2 -1)\be_0^2$.

\h With the help of the observations above, one can write the superposition
(\ref{S2geral5}), with spectrum (\ref{S2specSMPS}), as:

\bb \begin{array}{clcr} \Psi(\rho,\zeta,\eta) \, \approx & a\,q\,V
\dis{\int_{\be_0}^{\infi} \drm \be\,\int_{0}^{\infi} \drm \al\,
J_0\left(\rho\,\sqrt{\left(\frac{V^2}{c^2}-1\right)\al^2+
2\left(\frac{V^2}{c^2}+1\right)\al\be}\, \right)}\\

\\

& \times\,\dis{ e^{-i\be\eta}
e^{i\al\ze}e^{-q(\be-\be_0)}\,e^{-aV\al}}\end{array}
\label{S2intSMPS}\ee

and, using identity (6.616.1) given in ref.\cite{[19]}, we get

\bb \Psi(\rho,\zeta,\eta) \, \approx  q\,X \int_{\be_0}^{\infi}
\drm \be\,\dis{e^{-q(\be-\be_0)} e^{-i\be\eta}{\rm
exp}\left[\be\,\frac{V^2+c^2}{V^2-c^2}\left(aV-i\zeta -
aVX^{-1}\right) \right]}, \label{S2intSMPS2} \ee

which can be viewed as a superposition of the SFWM pulses (see
eq.(\ref{S2SFWM})).

\h The above integration can be easily performed and results\cite{[13]}
in the so called Superluminal Modified Power Spectrum (SMPS)
pulse:

\bb \Psi_{\rm SMPS}(\rho,\ze,\eta) \ug q\,X \,\frac{{\rm
exp}[(Y-i\eta)\be_0]}{q-(Y-i\eta)} \label{S2SMPS}   \ee

where $X$ is the ordinary X-pulse (\ref{S2X}) and $Y$ is defined by

 \bb Y \ \equiv \ \dis{\frac{V^2+c^2}{V^2-c^2} \; \left[(aV-i\ze)-
aVX^{-1} \right]} \; . \ee

The SMPS pulse is a {\em Superluminal} localized wave, with field
concentration around $\rho=0$ and $\zeta=0\,$ (i.e., $z=Vt$),
and with {\em finite} total energy. We will show that the depth of
field, $Z$, of this pulse is given by $Z_{\rm SMPS} = q/2$.

\h An interesting property of the SMPS pulse is related to its
transverse width (the transverse spot size at the pulse center).
It can be shown from eq.(\ref{S2SMPS}) that, for the cases where
$aV<<1/\be_0$ and $q\be_0>>1$, i.e., for the cases considered by
us, the transverse spot size, $\Delta\rho$, of the pulse center
($\zeta=0$) is determined by the exponential function in
(\ref{S2SMPS}) and is given by

\bb  \Delta\rho \ug c\,\sqrt{\frac{aV}{\be_0(V^2+c^2)} +
\frac{V^2-c^2}{4\be_0^2(V^2+c^2)^2}} \; , \ee

which clearly does not
depend on $z$, and so remains constant during its propagation. In
other words, in spite of the fact that the SMPS pulse suffers an
intensity decrease during propagation, it preserves its
transverse spot size. This interesting characteristic is not
met in ordinary pulses, like the gaussian ones, where the
amplitude of the pulse decreases and the width increases by the same
factor.

\begin{figure}[!h]
\begin{center}
 \scalebox{0.8}{\includegraphics{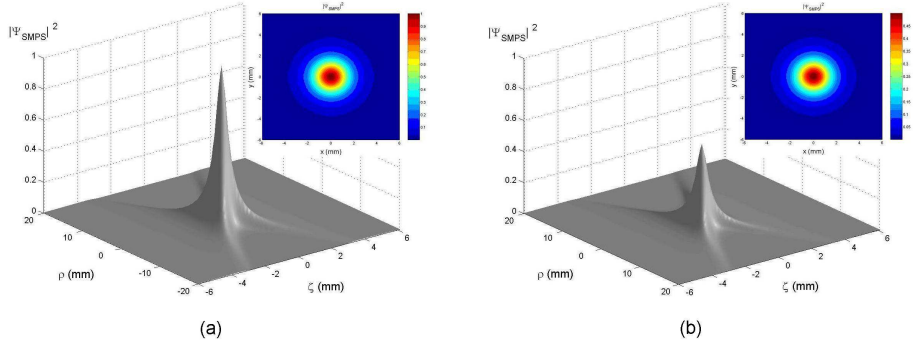}}
\end{center}
\caption{Representation of a Superluminal Modified Power Spectrum
pulse, eq.(\ref{S2SMPS}). \ Its total  energy is {\em finite}
(even without any truncation), and so it gets deformed
 while propagating, since its amplitude decreases with time. \ In
Fig.\ref{S2MHRFig1sec2}a we
 represent, for $t=0$, the pulse corresponding to $\be_0=33\,{\rm m}^{-1}$, $V=1.01 c$,
$a=10^{-12}\,$s and $q=10^5\,$m. \ In Fig.\ref{S2MHRFig1sec2}b it
is depicted the same pulse after having travelled $50 \; $km.}
\label{S2MHRFig1sec2}
\end{figure}

\h Figure \ref{S2MHRFig1sec2} shows the intensity of a SMPS pulse, with
$\be_0=33\,{\rm m}^{-1}$, $V=1.01 c$, $a=10^{-12}\,$s and
$q=10^5\,$m, at two different moments, for $t=0$ and after
$50\,$km of propagation, where, as one can see, the pulse becomes
less intense (precisely, with half of its initial peak intensity). \
In spite of the intensity decrease, the pulse maintains its
transverse width, as one can see from the 2D plots in
Fig.(\ref{S2MHRFig1sec2}), which show the field intensities in the
transverse sections at $z=0$ and $z=q/2=50$ km.

\

\h Other three important well known \emph{finite energy}
nondiffracting solutions can be obtained directly from the SMPS
pulse:

--- The first one, obtained from (\ref{S2SMPS}) by making $\be_0=0$,
is the so called\cite{[13]} Superluminal Splash Pulse (SSP)

\bb \Psi_{\rm SSP}(\rho,\ze,\eta) \ug \frac{q\,X}{q+i\eta-Y} \; .
\label{S2SSP} \ee

--- The other two pulses are luminal.  By taking the limit $V
\rightarrow c^+$ in the SMPS pulse (\ref{S2SMPS}), we get the well
known\cite{[11]} \emph{luminal} Modified Power Spectrum (MPS)
pulse

\bb \Psi_{\rm MPS}(\rho,\ze,\eta) \ug
\frac{a\,q\,c\,e^{-i\be_0\eta}}{(q+i\eta)(ac-i\zeta) +
\rho^2}\,{\rm exp}\left(\frac{-\be_0\rho^2}{ac-i\zeta}\right) \; ;
\label{S2MPS}\ee

finally, by taking the limit $V \rightarrow c^+$ and making
$\be_0=0$ in the SMPS pulse [or, equivalently, by making $\be_0=0$
in the MPS pulse (\ref{S2MPS}), or,instead, by taking the limit $V
\rightarrow c^+$ in the SSP (\ref{S2SSP})], we obtain the well
known\cite{[11]} \emph{luminal} Splash Pulse (SP) solution

\bb \Psi_{\rm SP}(\rho,\ze,\eta) \ug
\frac{a\,q\,c}{(q+i\eta)(ac-i\zeta) + \rho^2} \; . \label{S2SP}\ee

It is also interesting to notice that the X and SFWM pulses can be
obtained from the SSP and SMPS pulses (respectively) by making $q
\rightarrow \infty$ in Eqs.(\ref{S2SSP}) and (\ref{S2SMPS}). As a
matter of fact, the solutions SSP and SMPS can be viewed as the finite
energy versions of the X and SFWM pulses, respectively.

\

\

\emph{Some characteristics of the SMPS pulse:}

\h Let us examine the on-axis ($\rho=0$) behaviour of the SMPS pulse. \
On $\rho=0$ we have

\bb \Psi_{\rm SMPS}(\rho=0,\zeta,\eta) \ug aqV
e^{-i\be_0z}[(aV-i\zeta)(q+i\eta)]^{-1} \; . \label{S2SMPSr0}\ee

From this expression, we can show that the longitudinal
localization $\Delta z$, for $t=0$, of the SMPS pulse square-magnitude is

\bb \Delta z \ug 2aV \; . \ee

If we now define the field-depth $Z$ as the distance over which
the pulse's peak intensity is at least $50\%$ of its initial
value\footnote{We can expect that, while the pulse peak-intensity
is maintained, the pulse keeps its spatial form too.}, then we can obtain,
from (\ref{S2SMPSr0}), the depth of field

\bb Z_{\rm SMPS} \ug \frac{q}{2} \; , \label{S2ZSMPS} \ee

which depends
only on $q$, as we expected since $q$ regulates the concentration
of the spectrum around the line $\om = Vk_z + 2V\be_0$.

\h Now, let us examine the maximum amplitude $M$ of the real part of
(\ref{S2SMPSr0}), which for $z=Vt$ writes ($\zeta=0$ and
$\eta=2z$):

\bb M_{\rm SMPS} \equiv {\rm Re}[\Psi_{\rm SMPS}(\rho=0,z=Vt)] \ug
\frac{\cos(2\be_0z) - 2(z/q)\sin(2\be_0z)}{1 + 4(z/q)^2} \; .
\label{S2MSMPS}\ee

\

\h Initially, for $z=0$, $t=0$, one has $M=1$ and can also infer
that:

\

(i) when $z/q<<1$, namely when $z<<Z$, equation (\ref{S2MSMPS})
becomes

\bb M_{\rm SMPS} \approx \cos(2\be_0 z)\;\;\;\;\;{\rm for}\;\;
z<<Z \ee

and the pulse's peak actually oscillates harmonically with
``wavelength'' $\Delta z_0 = \pi / \be_0$ and ``period'' $\Delta
t_0 = \pi/V\be_0$, all along its field-depth;

\

(ii) when $z/q >>1$, namely $z>>Z$, equation (\ref{S2MSMPS})
becomes

\bb M_{{\rm SMPS}} \approx - \frac{\sin(2\be_0 z)}{2z/q} \;\;\;\;{\rm
for}\;\; z>>Z  \; . \ee

Therefore, beyond its depth of field, the pulse goes on oscillating
with the same $\Delta z_0$, but its maximum amplitude decays
proportionally to $z$.

\h In the next two Sections we are going to see
applications of the localized wave pulses.

\

\section{Space-Time Focusing of X-shaped Pulses} 

In this Section we are going to show how one can in general use
any known Superluminal solution to obtain from it a large number
of analytic expressions for space-time focused waves, endowed with
a very strong intensity peak at the desired location. The method
presented here is a natural extension of that developed by
A.Shaarawi et al.\cite{[20]}, where the space-time focusing was
achieved by superimposing a discrete number of ordinary X-waves,
characterized by different values $\theta$ of the axicon angle.

\h In this Section, based on ref.\cite{[21]}, we shall go on to
more efficient superpositions for varying velocities $V$, related
to $\theta$ through the known[115,116,54] relation
$V=c/\cos\theta$. This enhanced focusing scheme has the advantage
of yielding analytic (closed-form) expressions for the
spatio-temporally focused pulses.

\h Let us start by considering an axially symmetric ideal
nondiffracting Superluminal wave pulse $\psi(\rho ,z-Vt)$ in a
dispersionless medium, where $V = c/\cos\theta>c$ is the pulse
velocity, $\theta$ being the axicon angle. As we have seen in the
previous Section, pulses like these can be obtained by a suitable
frequency superposition of Bessel beams. \ Suppose that we have
now $N$ waves of the type $\psi_n(\rho,z-V_n (t-t_n))$, with
different velocities $c<V_1<V_2<..<V_N$, and emitted at
(different) times $t_n$; quantities $t_n$ being constants, while
$n=1,2,...N$.  The center of each pulse is located at  $z = V_n (t
- t_n)$. To obtain a highly focused wave, we need all the wave
components $\psi_n(\rho,z-V_n (t-t_n))$ to reach the given point, $z
= z_\frm$, at the same time $t = t_\frm$. On choosing $t_1=0$ for
the slowest pulse $\psi_1$, it is easily seen that the peak of
this pulse reaches the point $z = z_\frm$ at the time $t_\frm =
z_\frm / V_1$. So we obtain that, for each $\psi_n$, the instant
of emission $t_n$ must be

\bb t_n \ug \left(\frac{1}{V_1}- \frac{1}{V_n}\right)z_\frm \; .
\label{S3tn} \ee

With this in mind, we can construct other exact solutions to the wave
equation, given by

\bb \Psi(\rho,z,t) \ug \int_{V_\minrm }^{V_\maxrm }\, \drm V \,
A(V) \, \psi\left(\rho,z-V\left(t-\left(\frac{1}{V_\minrm }-
\frac{1}{V}\right) z_\frm \right)\right) \label{S3int} \ , \ee

where $V$ is the velocity of the wave  $\psi(\rho,z-Vt)$ which enters the
integrand of (\ref{S3int}).  While integrating, $V$ is considered
as a continuous variable in the interval $[V_\minrm ,V_\maxrm ]$.
In eq.(70), function $A(V)$ is the velocity distribution that
specifies the contribution of each wave component (with velocity
$V$) to the integration. The resulting wave $\Psi(\rho,z,t)$ can
have a more or less strong amplitude peak at $z = z_\frm$, at time
$t_\frm = z_\frm / V_\minrm$, depending on $A(V)$ and on the
difference $V_\maxrm - V_\minrm$. Let us notice that also the
resulting wavefield will propagate with a Superluminal peak
velocity, depending on $A(V)$ too. In the cases when the
velocity-distribution function is well concentrated around a
certain velocity value, one can expect the wave (\ref{S3int}) to
increase its magnitude and spatial localization while propagating.
Finally, the pulse peak acquires its maximum amplitude and
localization at the chosen point $z = z_\frm$, and at time $t=
z_\frm / V_\minrm$, as we know. Afterwards, the wave suffers a
progressing spreading, and a decreasing of its amplitude.

\subsection{Focusing Effects by Using Ordinary X-Waves}

Here, we present a specific example by integrating (\ref{S3int})
over the standard, classic\cite{[4]} X-waves, $X =
aV[(aV-i(z-Vt))^2 + (V^2/c^2 -1)\rho^2]^{-1/2}$. When using this
ordinary X-wave, the largest spectral amplitudes are obtained for
low frequencies. For this reason, one may expect that the
solutions considered below will be suitable mainly for low
frequency applications. Let us choose, then, the function $\psi$
in the integrand of eq.(\ref{S3int}) to be $\psi(\rho,z,t) \equiv
X(\rho, z - V(t-(1/V_\minrm -1/V)z_\frm))$, viz.:

\bb \psi(\rho,z,t) \equiv X \ug
\dis{\frac{aV}{\sqrt{\left[aV-i\left( z - V\left( t -
\left(\frac{1} {V_\minrm }-\frac{1}{V}   \right)z_\frm \right)
\right)\right]^2 + \left(\frac{V^2}{c^2} -1 \right)\rho^2 }}} \; .
\label{S3psiX} \ee

After some manipulations, one obtains the analytic {\em integral
solution}

\bb \Psi(\rho,z,t) \ug \int_{V_\minrm }^{V_\maxrm }\,
\dis{\frac{aV\,A(V)} {\sqrt{PV^2 + QV + R}} }\drm V \label{S3intx}
\ee with

\bb
\begin{array}{l}
P \equiv \left[ \left( a+i\left(t-\frac{z_\frm}{V_\minrm
}\right)\right)^2
+ \frac{\rho^2}{c^2}\right]\\
\\
Q \equiv 2\left(t-\frac{z_\frm}{V_\minrm } - ai\right)(z-z_\frm) \\
\\
R \equiv \left[-(z-z_\frm)^2 - \rho^2 \right] \; . \label{S3P}
\end{array}
\ee

\

\h In what follows, we illustrate the behaviour of some new
spatio-temporally focused pulses, by taking into consideration
a few different velocity distributions $A(V)$. These new pulses are
closed analytic \emph{exact} solutions of the wave equation.

\

\emph{First example:}

\

Let us consider our integral solution (\ref{S3intx}) with $A(V) =
1\;{\rm s/m}$. In this case, the contribution of the X-waves is
the same for all velocities in the allowed range $[V_\minrm ,
V_\maxrm]$. \ On using identity 2.264.2 listed in ref.\cite{[19]},
we get the particular solution

\begin{figure}[!h]
\begin{center}
 \scalebox{1.1}{\includegraphics{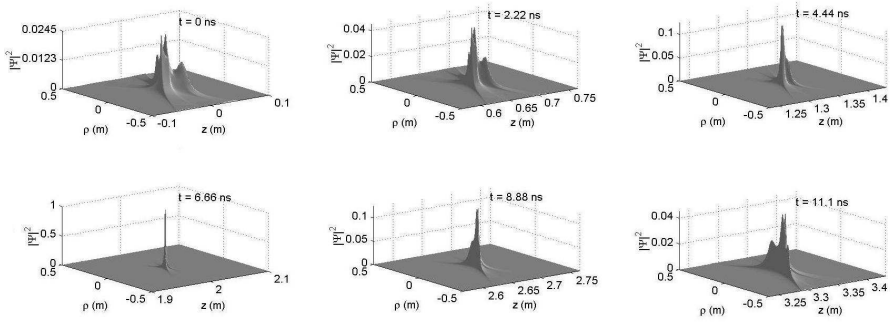}}
\end{center}
\caption{Space-time evolution of the Superluminal pulse
represented by eq.(\ref{S3solsec3}); the chosen parameter values
are $a=10^{-12}\;$s; \ $V_\minrm = 1.001 \; c$; \ $V_\maxrm =
1.005 \; c$ while the focusing point is at $z_\frm = 200\;$cm. \
One can see that this solution is associated with a rather good
spatio-temporal focusing. \ The field amplitude at $z = z_\frm$ is
40.82 times larger than the initial one. \ The field amplitude is
normalized at the space-time point $\rho = 0, \ z = z_\frm, \ t =
t_\frm$.} \label{S3MHRFig1sec3}
\end{figure}

\bb
\begin{array}{clcr}
\Psi(\rho,z,t) \!\!&=
\dis{\frac{a}{P}}\,\dis{\left(\sqrt{PV_\maxrm ^2 + QV_\maxrm + R}
- \sqrt{PV_\minrm ^2 + QV_\minrm  + R}\,\right)}\\
\\
&\;\;\; \dis{ + \frac{a\,Q}{2P^{3/2}}\,{\rm
ln}\left(\frac{2\,\sqrt{P(PV_\minrm ^2 + QV_\minrm  + R)} +
2PV_\minrm  + Q}{2\,\sqrt{P(PV_\maxrm ^2 + QV_\maxrm  + R)} +
2PV_\maxrm  + Q}\right)} \ , \label{S3solsec3}
\end{array}
\ee

where $P$, $Q$ and $R$ are given in eq.(\ref{S3P}). A
3-dimensional (3D) plot of this function is provided in
Fig.\ref{S3MHRFig1sec3}; where we have chosen $a=10^{-12}$ s, \
$V_\minrm =1.001\; c$, \ $V_\maxrm =1.005\; c$ and $z_\frm
=200\;$cm. It can be seen that this solution exhibits a rather
evident space-time focusing. An initially spread-out pulse (shown
for $t=0$) becomes highly localized at $t=t_\frm=z_\frm
/V_\minrm=6.66\;$ns, the pulse peak amplitude at $z_\frm$ being
$40.82$ times greater than the initial one. In addition, at the
focusing time $t_\frm$ the field is much more localized than at
any other times. The velocity of this pulse is approximately
$V=1.003\; c$.

\

\

\emph{Second example:}

\

In this case we choose $A(V) = 1/V \;$s/m, and, using the
identity 2.261 in ref.\cite{[19]}, eq.(\ref{S3intx}) yields

\bb \Psi(\rho,z,t) \ug \dis{\frac{a}{\sqrt{P}}\,{\rm ln}
\left(\frac{2\,\sqrt{P(PV_\maxrm ^2 + QV_\maxrm  + R)} +
2PV_\maxrm  + Q} {2\,\sqrt{P(PV_\minrm ^2 + QV_\minrm  + R)} +
2PV_\minrm  + Q} \right)} \ . \label{S3dois} \ee

\

Other exact closed-form solutions can be obtained\cite{[21]} by
considering, for instance, velocity distributions like
$A(V)=1/V^2$ and $A(V)=1/V^3$.

\h Once more, we can actually construct many others spatio-temporally
focused pulses from the above solutions, just by taking their time
derivatives (of any order). It is possible to show\cite{[21]}
that also the new solutions obtained in this way have their spectra
shifted towards higher frequencies.

\

\

\section{Chirped Optical X-Type Pulses in Material Media} 

The theory of the localized waves was initially developed for free
space (vacuum). In 1996, S\~{o}najalg et al.\cite{[5]} showed that
the localized wave theory can be extended to include (unbounded)
dispersive media. This was obtained by making the axicon angle of
the Bessel beams (BBs) vary with the frequency[117-119] in such
a way that a suitable frequency superposition of these beams does
compensate for the material dispersion. Soon after this idea was
reported, many interesting nondiffracting/nondispersive pulses
were obtained theoretically[117-119] and
experimentally\cite{[5]}. \ In spite of such an extended method to be
of remarkable importance, working well in theory, its experimental
implementation is not so simple\footnote{We refer the interested
reader to quotations[117-119] for obtaining a description,
theoretical and experimental, of that extended method}.

\h In 2004 Zamboni-Rached et al.\cite{[22]} developed a simpler
way to obtain pulses capable of recovering their spatial shape,
both transversally and longitudinally, after some propagation. It
consisted in using chirped optical X-typed pulses, while keeping
the axicon angle fixed. Let us recall that, by contrast, chirped
Gaussian pulses in unbounded material media may recover only their
longitudinal shape, since they undergo a progressing transverse
spreading while propagating.

\h The present Section is devoted to this approach. \
Let us start with an axially-symmetric Bessel beam in a material
medium with refractive index $n(\om)$:

\bb \psi(\rho,z,t)\ug J_0(k_{\rho}\rho) \, \exp(i\be z) \,
\exp(-i\om t) \; ,
\label{S4bb}\ee  

where it must be obeyed the condition \
$k_{\rho}^2=n^2(\om)\om^2/c^2 - \be^2$, \ which connects among
themselves the transverse and longitudinal wave numbers $k_{\rho}$
and $\be$, and the angular frequency $\om$.  In addition, we
impose that $k_{\rho}^2 \geq 0$ and $\om / \be \geq 0$, to avoid a
nonphysical behaviour of the Bessel function $J_0(.)$ and to
confine ourselves to forward propagation only. \
Once the conditions above are satisfied, we have the liberty of
writing the longitudinal wave number as $\be = (n(\om)\om
\cos\theta)/c$ and, therefore, $k_{\rho} = (n(\om)\om
\sin\theta)/c$; where (as in the free space case) $\theta$ is the
axicon angle of the Bessel beam.

\h Now we can obtain an X-shaped pulse by performing a frequency
superposition of these Bessel beams [BB], with $\be$ and $k_{\rho}$ given by the
previous relations:

\bb \Psi(\rho,z,t) \ug \int_{-\infi}^{\infi}\,
S(\om)\,J_0\left(\frac{n(\om)\om}{c} \sin\theta\,\rho\right)\,
\exp[i\be(\om)z]\, \exp(-i\om t)\,\drm \om \; ,
\label{S4geral}\ee        

where $S(\om)$ is the frequency spectrum, and the axicon angle is
kept constant. \ One can see that the phase velocity of each BB in our
superposition (\ref{S4geral}) is different, and given by $V_{\rm
phase} = c/(n(\om)\cos\theta)$. So, the pulse represented by
eq.(\ref{S4geral}) will suffer dispersion during its propagation.

\h As we said, the method developed by S\~onajalg et al.\cite{[5]},
and explored by others[118,119], to overcome this problem
consisted in regarding the axicon angle $\theta$ as a function of
the frequency, in order to obtain a linear relationship between
$\be$ and $\om$.

\h Here, however, we choose to work with a {\em fixed} axicon angle,
and we have to find out another way for avoiding dispersion and
diffraction all along a certain propagation distance. To do that, we
might choose a chirped gaussian spectrum  $S(\om)$ in
eq.(\ref{S4geral})

\bb S(\om) \ug
\frac{T_0}{\sqrt{2\pi(1+iC)}}\,\,\exp[-q^2(\om-\om_0)^2]
\label{S4S}\;\;\;\;{\rm with}\;\;\;\;\; q^2 \ug
\frac{T_0^2}{2(1+iC)} \; , \label{S4s}\ee           

where $\om_0$ is the central frequency of the spectrum, $T_0$ is
a constant related with the initial temporal width,  and $C$ is
the chirp parameter (we chose as temporal width the half-width of
the relevant gaussian curve when its heigth equals $1/e$ times its
full heigth).  Unfortunately, there is no analytic solution to
eq.(\ref{S4geral}) with $S(\om)$ given by eq.(\ref{S4s}), so that
some approximations are to be made. \ Then, let us assume  that
the spectrum $S(\om)$, in the
surroundings of the carrier frequency $\om_0$ , is narrow enough to guarantee
that $\Delta\om/\om_0<<1$, so to ensure that $\be(\om)$ can be
approximated by the first three terms of its Taylor expansion in
the vicinity of $\om_0$: That is, \ $\be(\om)\approx \be(\om_0) +
\be'(\om)|_{\om_0}\,(\om - \om_0) + (1/2) \be''(\om)|_{\om_ 0}\,
(\om - \om_0)^2$; \ when, after using $\be =n(\om)\om
\cos\theta/c$, it results that

\bb \frac{\pa \be}{\pa\om} \ug \frac{\cos\theta}{c}\left[n(\om) +
\om\,\frac{\pa n}{\pa\om} \right]\,; \ \ \ \frac{\pa^2
\be}{\pa\om^2} \ug \frac{\cos\theta}{c}\left[ 2\frac{\pa
n}{\pa\om} + \om \frac{\pa^2 n}{\pa\om^2} \right] \; .
\label{S4b1}\ee

\

\h As we know, $\be'(\om)$  is related to the pulse group-velocity by
the relation $Vg = 1/ \be'(\om)$.  Here we can see the difference
between the group-velocity of the X-type pulse (with a fixed
axicon angle) and that of a standard gaussian pulse.  Such a
difference is due to the factor $\cos\theta$ in eq.(\ref{S4b1}).
Because of it, the group-velocity of our X-type pulse is always
greater than the gaussian's. In other words, $(V_\grm)_\X =
(1/\cos\theta)(V_\grm)_{{\rm gauss}}$. \ We also know that the second
derivative of $\be(\om)$ is related
to the group-velocity dispersion (GVD) $\be_2$ by $\be_2 =
\be''(\om)$. \
The GVD is responsible for the temporal (longitudinal) spreading
of the pulse. Here one can see that the GVD of the X-type pulse is
always smaller than that of the standard gaussian pulses, due the
factor $\cos\theta$ in eq.(\ref{S4b1}).  Namely: $(\be_{2})_\X =
\cos\theta (\be_{2})_{\rm gauss}$.

\h On using the above results, we can write

\

\bb \begin{array}{clcr}
 \Psi(\rho,z,t) \!\!&= \dis{\frac{T_0\,\,\exp[i\be(\om_0)z]\,
\exp(-i\om_0 t)}{\sqrt{2\pi(1+iC)}}
\,\int_{-\infi}^{\infi}\drm \om\,J_0\left(\frac{n(\om)\om}{c}
\sin\theta\,\rho\right)}\\
\\
&\;\;\;\times\,\dis{\exp\left\{i\frac{(\om-\om_0)}{V_\grm}\left[z
- V_\grm t \right] \right\} \,
\exp\left\{(\om-\om_0)^2\left[\frac{i\be_2}{2}z - q^2
\right] \right\}} \; . \label{S4geral2} \end{array} \ee  

\

The integral in eq.(\ref{S4geral2}) cannot be evaluated analytically,
but for us it is sufficient to obtain the pulse behaviour. Let us
analyse the pulse for $\rho=0$. In this case we get

\bb \Psi(\rho=0,z,t) \ug \dis{\frac{T_0\,\exp[i\be(\om_0)z]\,
\exp(-i\om_0 t)}{\sqrt{T_0^2 -
i\be_2(1+iC)z}}\,\exp\left[\frac{-(z-V_\grm
t)^2(1+iC)}{2V_\grm^2[T_0^2 - i\be_2(1+iC)z]}\right]} \; .
\label{S4sol1} \ee

From eq.(\ref{S4sol1}) one can immediately see that the initial
temporal width of the pulse intensity is $T_0$ and that, after
a propagation distance $z$, the time-width $T_1$ becomes

\bb \frac{T_1}{T_0} \ug \left[\left(1+\frac{C\be_2z}{T_0^2}
\right)^2 + \left(\frac{\be_2z}{T_0^2}\right)^2\right]^{1/2} \; .
\label{S4T1}\ee

Relation (\ref{S4T1}) describes the pulse spreading-behaviour. One
can easily show that such a  behaviour depends on the sign
(positive or negative) of the product $\be_2 C$, as is well known to happen
for the standard gaussian pulses\cite{[23]}. \ In the case $\be_2C
> 0$,  the pulse will monotonically become broader and broader
with the distance $z$.  On the other hand, if $\be_2C < 0$ the
pulse will suffer, in a first stage, a narrowing, and then
(during the rest of its propagation) it will spread . So, there will be a
certain propagation distance AT after which the pulse will recover its
initial temporal width ($T_1=T_0$). From relation (\ref{S4T1}), we
can find such a distance $Z_{T1=T_0}$ (considering $\be_2 C < 0$) to
be

\bb Z_{T_1=T_0} \ug \frac{-2CT_0^2}{\be_2(C^2+1)} \; . \label{S4Z}
\ee

One may notice that the maximum distance at which our chirped
pulse, with given $T_0$ and $\be_2$,  may recover its initial
temporal width can  be easily evaluated from eq.(\ref{S4Z}), and
it results to be $L_{\rm disp} = T_0^2 / \be_2$.  We shall call
such a maximum value $L_{\rm disp}$ the ``dispersion length": It
is the maximum distance the X-type pulse may travel while
recovering its initial longitudinal shape. Obviously, if we want
the pulse to reassume its longitudinal shape at some desired
distance $z < L_{\rm disp}$, we have just to suitably choose a new
value for the chirp parameter.

\h Let us emphasize that the property of recovering its own initial
temporal (or longitudinal) width may be verified to exist also in
the case of chirped standard gaussian pulses. However, the latter
will suffer a progressing transverse spreading, which will not be
reversible. The distance at which a gaussian pulse doubles its
initial transverse width $w_0$ \ is \ $z_{\rm diff} = \sqrt{3}\pi
w_0^2/\lambda_0$, where $\lambda_0$ is the carrier wavelength.
Thus, one can see that optical gaussian pulses with great
transverse localization will get spoiled in a few centimeters or
even less.

\h Now we shall show that it is possible to recover also the
transverse shape of  the chirped X-type pulse intensity; actually,
it is possible to recover its entire spatial shape after a
distance $Z_{T_1=T_0}$. \
To see this, let us go back to our integral solution
(\ref{S4geral2}), and perform the change of coordinates \ $(z,t)
\rightarrow (\Delta z, t_c = z_c/V_\grm)$, with

\bb\left\{\begin{array}{l} z \ug z_c + \Delta z\\
\\
 \dis{t=t_c\equiv \frac{z_c}{V_\grm}} \end{array}\right. \label{S4zc}\ee

where $z_c$ is the center of the pulse ($\Delta z$ is the distance
from such a point), and $t_c$ is the time at which the pulse
center is located at $z_c$. What we are going to do is comparing
our integral solution (\ref{S4geral2}), when $z_c=0$ (initial
pulse), with that when $z_c=Z_{T_1=T_0}=-2CT_0^2/(\be_2(C_2+1))$.
\ In this way, solution (\ref{S4geral2}) can be written, when
$z_c = 0$, as

\bb \begin{array}{clcr} \Psi(\rho,z_c=0,\Delta z)
\!\!&=\dis{\frac{T_0\,\,\, \exp(i\be_0\Delta
z)}{\sqrt{2\pi(1+iC)}}}\,\int_{-\infi}^{\infi}\drm \om\,
J_0(k_{\rho}(\om)\rho)\,\exp\left[\frac{-T_0^2\,(\om-\om_0)^2}{2(1+C^2)}\right]\\
\\
\\ &\;\;\;\dis{\times\,\exp\left\{i\left[\frac{(\om-\om_0)\Delta
z}{V_\grm} + \frac{(\om-\om_0)^2\be_2\Delta z}{2} +
\frac{(\om-\om_0)^2T_0^2C}{2(1+C^2)} \right]
\right\}}\\
\ \label{S4zc0} \end{array}\ee      

where we have taken the value $q$ given by eq.(\ref{S4s}). \
To verify that the pulse intensity recovers its entire original
form at $z_c = Z_{T_1=T_0} = -2\,CT_0^2/ [\be_2(C^2+1)]$, we can
analyse our integral solution at that point, obtaining

\

\bb \begin{array}{l} \Psi(\rho,z_c=Z_{T_1=T_0},\Delta z) \ugg
\dis{\frac{T_0\,\,\dis{\exp\left\{i\be_0 \left[z_c-\Delta z'-
\frac{cz_c}{\cos\theta\,n(\om_0)V_\grm}\right]\right\}}}{\sqrt{2\pi(1+iC)}}}\\
\\
\\
\times \dis{\int_{-\infi}^{\infi}} \dis{
\drm \om\,J_0(k_{\rho}(\om)\rho)\,\exp\left[\frac{-T_0^2\,(\om-\om_0)^2}{2(1+C^2)}
\right]}\\
\\
\\
\dis{\times\,\exp\left\{-i\left[\frac{(\om-\om_0)\Delta
z'}{V_\grm} + \frac{(\om-\om_0)^2\be_2\Delta z'}{2} +
\frac{(\om-\om_0)^2T_0^2C}{2(1+C^2)} \right]
\right\}}\\
\label{S4zt1}\end{array}\ee               

where we put  $\Delta z = -\Delta z'$. \ In this way, one
immediately sees that

\

\bb |\Psi(\rho,z_c=0,\Delta z)|^2 =
|\Psi(\rho,z_c=Z_{T_1=T_0},-\Delta z)|^2 \; . \label{S4inten}
\ee                                        

\

Therefore, from eq.(\ref{S4inten}) it is clear that the intensity of a chirped
optical X-type pulse is able to recover its original three-dimensional shape,
just with a longitudinal inversion at the
pulse center. The present method results to be, therefore, a simple and
effective procedure for compensating diffraction
and dispersion in an unbounded material medium; and a method
simpler than the one of varying the axicon angle with the
frequency.

\h Let us stress again that one can determine the distance $z = Z_{T_1=T_0}\leq
L_{\rm disp}$ at which the pulse takes on again its spatial
shape by choosing a suitable value of the chirp parameter.

\h We have shown that the chirped X-type pulse recovers its
three-dimensional shape after some distance, and we have also
obtained an analytic description of the pulse {\em longitudinal}
behaviour (for $\rho=0$) during propagation, by means of
eq.(\ref{S4sol1}). However, we have not got yet the same information
about the pulse transverse behaviour: We just learned, till now, that it
will be recovered at $z=Z_{T_1=T_0}$.

\h So, to complete the picture, we should find out
also the {\em transverse} behaviour in the plane of the
pulse center $z=V_\grm t$: \ We would then obtain
quantitative information about the evolution of the pulse-shape
during its entire propagation. \ But we are not going to expound all the
relevant mathematical details here; let us only state that the transverse
behaviour of the pulse (in the plane $z=z_c=V_\grm t$), during its
whole propagation, can approximately be described by

\bb \begin{array}{l} \Psi(\rho,z=z_c,t=z_c/V_\grm) \, \approx \,
 \dis{\frac{T_0\,\,\, \exp[i\be(\om_0)z]\, \exp(-i\om_0
t)}{\sqrt{2\pi(1+iC)}}\,\,\frac{\exp\dis{\left[\frac{-\tan^2\theta\,\rho^2}
{8\,V_\grm^2(-i\be_2z_c/2 \, + q^2)}
\right]}}{\sqrt{-i\be_2 z_c/2 \, + q^2}}}\\
\\
\\
\times\,\dis{\left[\Gamma(1/2)J_0\left(\frac{n(\om_0)\,\om_0\,\sin\theta\,\rho}
{c}\right)I_0\left(
\frac{\tan^2\theta\,\rho^2}{8\,V_\grm^2(-i\be_2 z_c/2 \, +
q^2)}\right)
\right.}\\
\\
\\
\left. + \dis{
2\sum_{p=1}^{\infty}\frac{2^p\Gamma(p+1/2)\Gamma(p+1)}{\Gamma(2p+1)}\,\,
J_{2p}\left(\frac{n(\om_0)\,\om_0\,\sin\theta\,\rho}{c}\right)I_{2p}\left(
\frac{\tan^2\theta\,\rho^2}{8\,V_\grm^2(-i\be_2 z_c/2 \, +
q^2)}\right)} \right]
\\
\ \label{S4trans1}
\end{array}\ee         

where $I_p(.)$ is the modified Bessel function of the first kind
of order $p$, quantity $\Gamma(.)$ being the gamma function and
$q$ being given by (\ref{S4s}). \ The interested reader can check
ref.\cite{[22]} for details on how eq.(\ref{S4trans1}) is obtained
from eq.(\ref{S4geral2}).

\h At a first sight, this solution appears to be  very complicated,
but {\em the series in its r.h.s. gives a negligible
contribution}. This circumstance renders our solution (\ref{S4trans1}) of
important practical interest, and we will use it in the following.
\ For additional information  about the transverse pulse evolution
(to be extracted from eq.(\ref{S4trans1})), the reader can consult again
ref.\cite{[22]}. In the same paper, it is analysed how the
generation by a finite aperture affects the chirped X-type pulses.

\h The valuable methods developed in ref.\cite{[22]}, and that we are
partially revisiting in this Section, are of general interest, and
work is in progress for applying them, e.g., also to the (different)
case of the Schroedinger equation.

\

\

\

\subsection{An example: Chirped Optical X-typed pulse in bulk fused
Silica}

\

For a bulk fused Silica, the refractive index $n(\om)$ can be
approximated by the Sellmeier equation\cite{[23]}

\

\bb n^2(\om) \ug
\dis{1\,+\,\sum_{j=1}^{N}\,\frac{B_j\,\om_j^2}{\om_j^2- \om^2}} \;
, \label{S4Selm}\ee \

where $\om_j$ are the resonance frequencies, $B_j$ the strength of
the $j$-th resonance, and $N$ the total number of the material
resonances that appear in the frequency range of interest. For our
purposes it is appropriate to choose $N=3$, which yields, for bulk
fused silica\cite{[23]}, the values \ $B_1=0.6961663$; \
$B_2=0.4079426$; \ $B_3=0.8974794$; \ $\lambda_1=0.0684043 \; \mu
\;$m; \ $\lambda_2=0.1162414 \; \mu\;$m; \ and $\lambda_3=9.896161
\; \mu\;$m.

\begin{figure}[!h]
\begin{center}
\scalebox{1.4}{\includegraphics{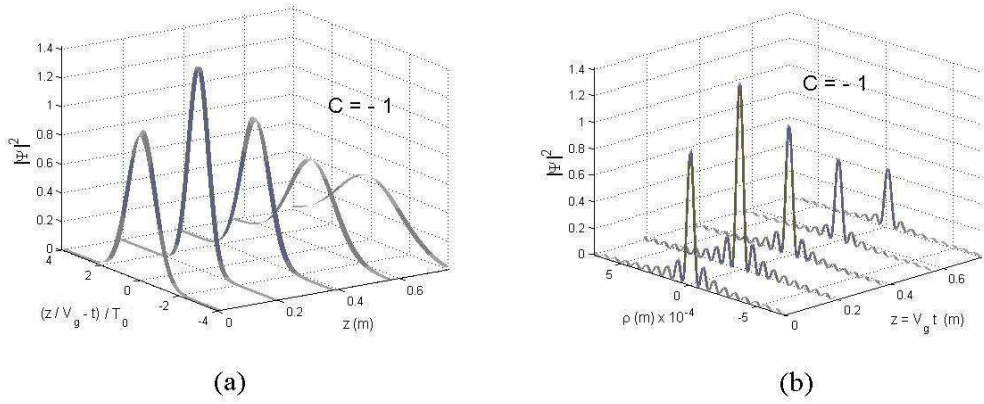}}
\end{center}
\caption{(a): Longitudinal-shape evolution of a chirped X-type
pulse, propagating in fused silica with $\lambda_0=0.2\mu \;$m, \
$T_0=0.4 \;$ps, \ $C=-1$, and axicon angle $\theta=0.00084 \;$rad,
which correspond to an initial transverse width of
$\Delta\rho_0=0.117 \;$mm. \ \ (b): Transverse-shape evolution for
the same pulse.} \label{S4MHRFig4sec4}
\end{figure}

\h Now, let us consider in this medium a chirped X-type pulse, with
$\lambda_0=0.2\,\mu$m, \ $T_0=0.4 \;$ps, \ $C=-1$, and with
axicon angle $\theta=0.00084 \;$rad: That correspond to an
initial central spot with $\Delta\rho_0=0.117 \;$mm. \
We get from eqs. (\ref{S4sol1}) and (\ref{S4trans1}) the
longitudinal and transverse pulse evolution, which are represented
in Fig.\ref{S4MHRFig4sec4}.

\h From Fig.(\ref{S4MHRFig4sec4}\,a), we can observe that
the pulse suffers initially a longitudinal narrowing with an increase of
intensity, till the position $z=T_0^2/2\be_2=0.186\;$m. After that
point, the pulse starts broadening and decreasing its intensity, while
recovering its entire longitudinal shape (width and intensity) at
the point $z=T_0^2/\be_2=0.373$m, as it was predicted. \
At the same time, from Fig.(\ref{S4MHRFig4sec4}\,b), one can notice that
the pulse maintains its transverse width
$\Delta\rho=2.4\,c/(n(\om_0)\om_0 \sin\theta)=0.117 \;$mm \
(because $T_0\om_0>>1$) during its entire propagation.
The same does not occur, however, with the pulse intensity: Initially, the
pulse suffers an increase of intensity, till position
$z_c=T_0^2/2\be_2=0.186\;$m; after that point the intensity starts
decreasing, and the pulse recovers its entire transverse shape at
point $z_c=T_0^2/\be_2=0.373$m, as expected. \ In the calculations
we could skip the series in the r.h.s. of
eq.(\ref{S4trans1}), because, as we already said, it yields a
negligible contribution.

\h Summarizing, from Fig.\ref{S4MHRFig4sec4}, we can see that the
chirped X-type pulse recovers totally its longitudinal and
transverse shape at position $z = L_{\rm disp} =
T^2_0/\be_2=0.373 \;$m, as we expected. \
Let us recall that a \emph{chirped gaussian pulse} may just
recover its longitudinal width, but with an intensity decrease, at
the position given by $z = Z_{T_1=T_0}=L_{\rm disp} =
T^2_0/\be_2$. Its transverse width, on the other hand, suffers a
progressing and irreversible spreading.

\

\h In the following Third Part we are going to ``complete" our review by
investigating also the (not less interesting) case of the {\em subluminal}
Localized Solutions to the wave equations, which, among the others, will
allow us to set forth remarkable considerations about the role of (extended)
Special Relativity. For instance, the various Superluminal and subluminal
LWs are expected to be transformed one into the other by suitable Lorentz
transformations. \ We shall start by studying, in terms of various different
approaches, the very peculiar topic of zero-speed waves: Namely, the question of
constructing localized fields with a {\em static}
envelope; consisting, for example, in ``light at rest" endowed with
zero peak-velocity. We called {\em Frozen Waves} such solutions: They
can have a lot of applications.

\newpage

\

\cent{\Huge{\bf THIRD  PART}}

\

\

\cent{\Large{\bf ``FROZEN WAVES", }}

\

\cent{\Large{\bf AND THE SUBLUMINAL WAVE-BULLETS}}

\

\

\

\section{Modeling the Shape of
Stationary Wave Fields: Frozen Waves}  

As just mentioned, we {\em start} this Third Part by studying the very peculiar
topic of zero-speed waves: Namely, the question of constructing localized fields
with a
{\em static} envelope (for example, consisting in ``light at rest" endowed
with null peak-velocity). We called {\em Frozen Waves} such solutions: They
permit a priori a lot of applications, as we are going to see.

\h In the present Section we develop a very simple {\em first} method[122,128,129],
{\em based on our Second Part,} by having
recourse to superpositions of forward propagating and
\emph{equal-frequency} Bessel beams, that allows one controlling
the {\em longitudinal} beam-intensity shape within a chosen
interval $0\leq z \leq L$, where $z$ is the propagation axis and
$L$ can be much greater than the wavelength $\lambda$ of the
monochromatic light (or sound) which is being used. Inside such a
space interval, indeed, we succeed in constructing a {\em
stationary} envelope whose longitudinal intensity pattern can
approximately assume any desired shape, including, for instance,
one or more high-intensity peaks (with distances between them much
larger than $\lambda$); and which ---in addition--- results to be
naturally endowed also with a good transverse localization. Since
the intensity envelopes remains static, i.e., with velocity $V=0$,
we called ``Frozen Waves" (FW) such new solutions[122,128,129] to the
wave equations.

\h Although we are dealing here with exact solutions of the scalar
wave equation, vectorial solutions of the same kind for the
electromagnetic field can be worked out: Indeed, solutions to
Maxwell's equations may be naturally inferred even from the scalar
wave-equation solutions[136-138].

\h We present first the method referring to lossless media[128,129]
while, in the second part of this Section, we extend the method to
absorbing media\cite{[10]}.

\subsection{Stationary wavefields
with arbitrary longitudinal shape in lossless media, obtained by
superposing equal-frequency Bessel beams}

Let us start from the well-known axis-symmetric zeroth order
Bessel beam solution to the wave equation:

\bb \psi(\rho,z,t)\ug J_0(k_{\rho}\rho)e^{i\be z}e^{-i\om t}
\label{S5bb}\ee with

\bb k_{\rho}^2=\frac{\om^2}{c^2} - \be^2 \; , \label{S5k}  \ee

where $\om$, $k_{\rho}$ and $\be$ are the angular frequency, the
transverse and the longitudinal wave numbers, respectively. We
also impose the conditions

\bb \om/\be > 0 \;\;\; {\rm and}\;\;\; k_{\rho}^2\geq 0
\label{S5c2} \ee (which imply $\om/\be \geq c$) to ensure forward
propagation only (with no evanescent waves), as well as a physical
behaviour of the Bessel function $J_0$.

\h Now, let us make a superposition of $2N + 1$ Bessel beams with the
same frequency $\om_0$, but with {\em different} (and still
unknown) longitudinal wave numbers $\be_m$:

\bb \dis{\Psi(\rho,z,t) \ug e^{-i\,\om_0\,t}\,\sum_{m=-N}^{N}
A_m\,J_0(k_{\rho\,m}\rho)\,e^{i\,\be_m\,z} } \; , \label{S5soma}
\ee

where the $m$ represent integer numbers and the $A_m$ are constant
coefficients. For each $m$, the parameters $\om_0$, $k_{\rho\,m}$
and $\beta_m$ must satisfy eq.(\ref{S5k}), and, because of conditions
(\ref{S5c2}), when considering $\om_0 > 0$ we must have

\bb 0 \leq \be_m \leq \frac{\om_0}{c} \; . \label{S5be} \ee

Let us now suppose that we wish $|\Psi(\rho,z,t)|^2$, given by
eq.(\ref{S5soma}), to assume on the axis $\rho=0$ the pattern
represented by a function $|F(z)|^2$, inside the chosen interval
$0 \leq z \leq L$. In this case, the function $F(z)$ can be
expanded, as usual, in a Fourier series:

\

$$F(z) \ug
\sum_{m=-\infty}^{\infty}\,B_m\,e^{i\,\frac{2\pi}{L}\,m\,z} \; ,$$

\

where

\

$$B_m \ug \frac{1}{L} \dis{
\int_{0}^{L}\,F(z)\,e^{-i\,\frac{2\pi}{L}\,m\,z}\,\drm \,z } \ .$$

\

More precisely, our goal now is finding out the values of the
longitudinal wave numbers $\be_m$ and the coefficients $A_m$ of
(\ref{S5soma}), in order to reproduce approximately, within the
said interval $0 \leq z \leq L$ (for $\rho=0$), the predetermined
longitudinal intensity-pattern $|F(z)|^2$. \ Namely, we wish to
have

\bb \left|\,\sum_{m=-N}^{N} A_m e^{i\,\be_m\,z}\right|^{\,2}
\approx |F(z)|^{\,2} \;\;\;\; {\rm with}\;\;\; 0\leq z \leq L  \;
. \label{S5soma1} \ee

\

\h Looking at eq.(\ref{S5soma1}), one might be tempted to take $\be_m
= 2\pi m/L$, thus obtaining a truncated Fourier series, expected
to represent approximately the desired pattern $F(z)$. \
Superpositions of Bessel beams with $\be_m = 2\pi m/L$ have been
actually used in some works to obtain a large set of {\it
transverse} amplitude profiles\cite{[26]}. However, for our purposes,
this choice is not appropriate, due to two principal reasons: \ 1)
It yields negative values for $\be_m$ (when $m<0$), which implies
backward propagating components (since $\om_0 > 0$); \ 2) In the
cases when $L>>\lambda_0$, which are of our interest here, the
main terms of the series correspond to very small values of
$\be_m$, which results in a very short field-depth of the
corresponding Bessel beams (when generated by finite apertures),
preventing the creation of the desired envelopes far form the
source.

\h Therefore, we need to make a better choice for the values of
$\be_m$, which permits forward propagation components only, and a
good depth of field. \ This problem can be solved by putting

\bb \be_m \ug Q + \frac{2\,\pi}{L}\,m \; , \label{S5be2}  \ee

where $Q>0$ is a value to be chosen (as we shall see) according to
the given experimental situation and the desired degree of {\em
transverse} field localization. \ Due to eq.(\ref{S5be}), one gets

\bb 0\leq Q \pm \frac{2\,\pi}{L}\,N \leq \frac{\om_0}{c} \; .
\label{S5N} \ee

Inequality (\ref{S5N}), can be used to determine the maximum value
of $m$, that we call $N_{\rm max}$, once $Q$, $L$ and $\om_0$ have
been chosen.

\h As a consequence, for getting a longitudinal intensity pattern
approximately equal to the desired one, $|F(z)|^2$, in the
interval $0\leq z \leq L $, eq.(\ref{S5soma}) has to be rewritten
as

\bb \dis{\Psi(\rho=0,z,t) \ug
e^{-i\,\om_0\,t}\,e^{i\,Q\,z}\,\sum_{m=-N}^{N}
A_m\,e^{i\,\frac{2\pi}{L}m\,z} } \; , \label{S5soma2} \ee

with

\bb A_m \ug \frac{1}{L} \dis{
\int_{0}^{L}\,F(z)\,e^{-i\,\frac{2\pi}{L}\,m\,z}\,\drm \,z } \; .
\label{S5An} \ee

Obviously, one obtains only an approximation to the desired
longitudinal pattern, because the trigonometric series
(\ref{S5soma2}) is necessarily truncated ($N \leq N_{\rm max}$).
Its total number of terms, let us repeat, is fixed once the
values of $Q$, $L$ and $\om_0$ have been chosen.

\h When $\rho \neq 0$, the wavefield $\Psi(\rho,z,t)$ becomes

\bb \dis{\Psi(\rho,z,t) \ug
e^{-i\,\om_0\,t}\,e^{i\,Q\,z}\,\sum_{m=-N}^{N}
A_m\,J_0(k_{\rho\,m}\,\rho)\,e^{i\,\frac{2\pi}{L}m\,z} } \; ,
\label{S5soma3} \ee with

\bb k_{\rho\,m}^2 \ug \om_0^2 - \left(Q + \frac{2\pi\,m}{L}
\right)^2 \; . \label{S5krn} \ee

The coefficients $A_m$ will yield {\em the amplitudes} and {\em
the relative phases} of each Bessel beam in the superposition.

\h Because we are adding together zero-order Bessel functions, we can
expect a {\em high} field concentration around $\rho=0$. Moreover,
due to the known non-diffractive behaviour of the Bessel beams, we
expect that the resulting wavefield will preserve its transverse
pattern in the entire interval $0\leq z \leq L $.

\h The present methodology addresses itself to the longitudinal
intensity pattern control. Obviously, we cannot get a total 3D
control, due the fact that the field must obey the wave equation.
However, we can use two ways to have some control over the
transverse behaviour too. The first is through the parameter $Q$ of
eq.(\ref{S5be2}). \ Actually, we have some freedom in the choice
of this parameter, and FWs representing the same longitudinal
intensity pattern can possess different values of $Q$. The
important point is that, in superposition (\ref{S5soma3}), using a
smaller value of $Q$ makes the Bessel beams to have a higher
transverse concentration (because, on decreasing the value of $Q$,
one increases the value of the Bessel beams transverse wave
numbers), and this will reflect in the resulting field, which will
present a narrower central transverse spot. The second way to
control the transverse intensity pattern is using higher order
Bessel beams, and we shall show this in Section 5.1.1.

\h Now, let us present a few examples of our methodology.

\

\emph{First example:}

\

Let us suppose that we want an optical wavefield with $\lambda_0 =
0.632\;\mu$m, i.e. with $\om_0 = 2.98 \times 10^{15}\;$Hz, whose
longitudinal pattern (along its $z$-axis) in the range $0 \leq z
\leq L$ is given by the function

 \bb
 F(z) \ug \left\{\begin{array}{clr}
 -4\,\,\dis{\frac{(z-l_1)(z-l_2)}{(l_2 - l_1)^2}} \;\;\; & {\rm
for}\;\;\; l_1 \leq z \leq l_2  \\
\\
 \;\;\;\;\;\;\;\;1 & {\rm for}\;\;\; l_3 \leq z \leq l_4 \\
\\
 -4\,\,\dis{\frac{(z-l_5)(z-l_6)}{(l_6 - l_5)^2}} & {\rm for}\;\;\; l_5
\leq z \leq
 l_6 \\
 \\
 \;\;\;\;\;\;\;\; 0  & \mbox{elsewhere} \ ,
\end{array} \right. \label{S5Fz1}
 \ee

where $l_1=L/5-\Delta z_{12}$ \ and \ $l_2=L/5+\Delta z_{12}$ \ with
$\Delta z_{12}=L/50$; while $l_3=L/2-\Delta z_{34}$ and
$l_4=L/2+\Delta z_{34}$ with $\Delta z_{34}=L/10$; \ and, at last,
$l_5=4L/5-\Delta z_{56}$ \ and \ $l_6=4L/5+\Delta z_{56}$ \ with $\Delta
z_{56}=L/50$. In other words, the desired longitudinal shape, in
the range $0 \leq z \leq L$, is a parabolic function for $l_1 \leq
z \leq l_2$, a unitary step function for $l_3 \leq z \leq l_4$,
and again a parabola in the interval $l_5 \leq z \leq l_6$, being
zero elsewhere (within the interval $0 \leq z \leq L$, as we
said). In this example, let us put $L=0.2\;$m.

\h We can then easily calculate the coefficients $A_m$, which
appear in superposition (\ref{S5soma3}), by inserting
eq.(\ref{S5Fz1}) into eq.(\ref{S5An}). Let us choose, for instance,
$Q=0.999\,\om_0/c$. This choice yields for $m$ a maximum value $N_{\rm
max}=316$, as one can infer from eq.(\ref{S5N}). Let us
underline that one is not compelled to use just $N=316$, but can
adopt for $N$ any values {\em smaller} than it; more generally,
any value smaller than that calculated via inequality (\ref{S5N}).
Of course, when using the maximum value allowed for $N$, one gets a
better result.

\

\begin{figure}[!h]
\begin{center}
 \scalebox{1}{\includegraphics{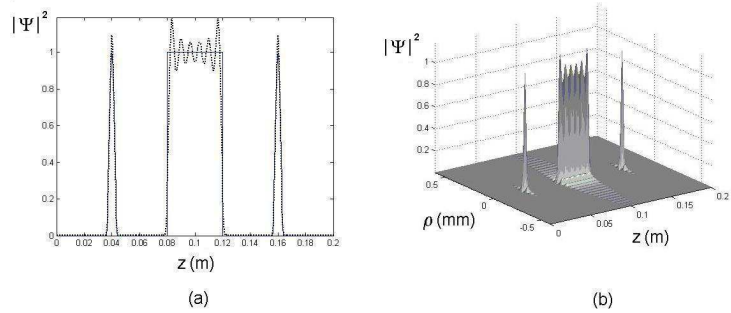}}
\end{center}
\caption{\textbf{(a)} Comparison between the intensity of the
desired longitudinal function $F(z)$ and that of our Frozen Wave
(FW), \ $\Psi(\rho=0,z,t)$, \ obtained from eq.(\ref{S5soma2}).
The solid line represents the function $F(z)$, and the dotted one
our FW. \ \textbf{(b)} 3D-plot of the field-intensity of the FW
chosen in this case by us.} \label{S5MHRFig1sec5}
\end{figure}

\h In the present case, let us adopt the value $N=30$. In
Fig.(\ref{S5MHRFig1sec5}\,a) we compare the intensity of the desired
longitudinal function $F(z)$ with that of the Frozen Wave, \
$\Psi(\rho=0,z,t)$, \ obtained from eq.(\ref{S5soma2}) by adopting
the mentioned value $N=30$.

\h One can verify that a good agreement between the desired
longitudinal behaviour and our approximate FW is already got for
$N=30$. The use of higher values for $N$ can only improve the
approximation. \ Figure (\ref{S5MHRFig1sec5}\,b) shows the
3D-intensity of our FW, given by eq.(\ref{S5soma3}). One can
observe that this field possesses the desired longitudinal
pattern, while being endowed with a good transverse localization.

\

\emph{Second example} (controlling the transverse shape too):

\

We wish to take advantage of this example for addressing an
important question: \ We can expect that, for a desired
longitudinal pattern of the field intensity, by choosing smaller
values of the parameter $Q$ one will get FWs with narrower {\em
transverse} width [for the same number of terms in the series
entering eq.(\ref{S5soma3})], because of the fact that the Bessel
beams in eq.(\ref{S5soma3}) will possess larger transverse wave
numbers, and, consequently, higher transverse concentrations. \ We
can verify this expectation by considering, for instance, inside
the usual range $0 \leq z \leq L$, the longitudinal pattern
represented by the function

 \bb
 F(z) \ug \left\{\begin{array}{clr}
 -4\,\,\dis{\frac{(z-l_1)(z-l_2)}{(l_2 - l_1)^2}} \;\;\; &
 {\rm for}\;\;\; l_1 \leq z \leq l_2  \\

 \\
 \;\;\;\;\;\;\;\; 0  & \mbox{elsewhere}
\end{array} \right. \; , \label{S5Fz2}
 \ee

with $l_1=L/2-\Delta z$ and $l_2=L/2+\Delta z$.  Such a function
has a parabolic shape, with its peak centered at $L/2$ and with
longitudinal width $2 \Delta z/\sqrt{2}$.  By adopting $\lambda_0
= 0.632\;\mu$m (that is, $\om_0 = 2.98 \times 10^{15}\;$Hz), let
us use superposition (\ref{S5soma3}) with {\em two} different
values of $Q$: \ We shall obtain two different FWs that, in spite
of having the same longitudinal intensity pattern, possess
different transverse localizations. Namely, let us consider
$L=0.06\,$m and $\Delta z = L/100$, and the two values
$Q=0.999\,\om_0/c$ and $Q=0.995\,\om_0/c$. In both cases the
coefficients $A_m$ will be the same, calculated from
eq.(\ref{S5An}) using this time the value $N=45$ in superposition
(\ref{S5soma3}). The results are shown in Figs.(32a) and (32b). Both
FWs have the same longitudinal intensity pattern, but the one with
the smaller $Q$ is endowed with a narrower transverse width.

\begin{figure}[!h]
\begin{center}
\scalebox{1}{\includegraphics{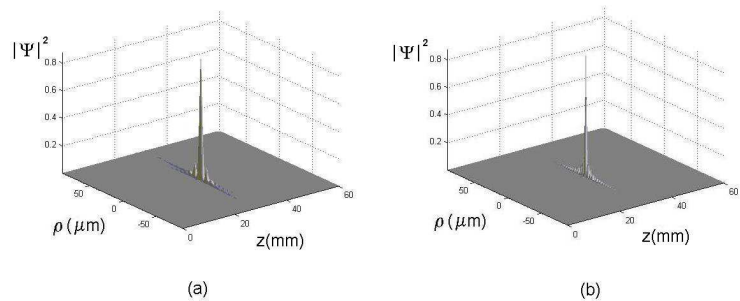}}
\end{center}
\caption{\textbf{(a)} The Frozen Wave with $Q=0.999 \, \om_0/c$ and
$N=45$, approximately reproducing the chosen longitudinal pattern
represented by eq.(\ref{S5Fz2}). \ \textbf{(b)} A different Frozen
wave, now with $Q=0.995 \om_0/c$ (but still with $N=45$)
forwarding the same longitudinal pattern. We can observe that in
this case (with a lower value for $Q$) a higher transverse
localization is obtained.} \label{S5MHRFig3sec5}
\end{figure}

\h In this way, we can get some control on the transverse spot size
through the parameter $Q$. \ Actually, eq.(\ref{S5soma3}), which
defines our FW, is a superposition of zero-order Bessel beams,
and, due to this fact, the resulting field is expected to possess
a transverse localization around $\rho=0$. Each Bessel beam in
superposition (\ref{S5soma3}) is associated with a central spot
with transverse size, or width, $\Delta\rho_m \approx
2.4/k_{\rho\,m}$. On the basis of the expected convergence of
the series (\ref{S5soma3}), we can estimate the width of the
transverse spot of the resulting beam as being

\bb \Delta \rho \approx \frac{2.4}{k_{\rho, \, m=0}} \ug
\frac{2.4}{\sqrt{\om_0^2/c^2 - Q^2}} \; , \label{S5tspot} \ee

which is the same value as that for the transverse spot of the
Bessel beam with $m=0$ in superposition (\ref{S5soma3}). Relation
(\ref{S5tspot}) can be useful: Once we have chosen the desired
longitudinal intensity pattern, we can even choose the size
of the transverse spot, and use relation (\ref{S5tspot}) for
evaluating the corresponding needed value of parameter $Q$. \
For a more detailed analysis concerning the spatial resolution and
residual intensity of the Frozen Waves, we refer the reader to
ref.\cite{[17]}.

\h The Frozen Waves, corresponding to zero group-velocity, are a
particular case of the {\em subluminal} Localized Waves. \
Actually, like in the Superluminal case, the (more orthodox, in a
sense) subluminal LWs can be obtained by suitable superpositions
of Bessel beams. They have been till now almost neglected,
however, for the mathematical difficulties met in getting analytic
expressions for them, difficulties associated with the fact that
the superposition integral runs now over a finite interval. \ In
Ref.\cite{subArt1} we have shown, by contrast,  that one can
indeed arrive at exact (analytic) solutions also in the case of
general subluminal LWs, and both in the case of integration over
the Bessel beams' angular frequency $\om$, and in the case of
integration over their longitudinal wavenumber $k_z$.  \ We shall
come back to this point in the following.

\

\subsubsection{Increasing the control on the transverse shape by
using higher-order Bessel beams}

\

Here, we are going to argue that it is possible to increase even
more our control on the transverse shape by using higher-order
Bessel beams in our fundamental superposition (\ref{S5soma3}). \
This new approach can be understood and accepted on the basis of
simple and intuitive arguments, which are not presented here, but
can be found in ref.\cite{[17]}. A brief description of that approach follows
below.

\h The basic idea is obtaining the desired longitudinal intensity
pattern not along the axis $\rho=0$, but on a cylindrical surface
corresponding to $\rho=\rho'>0$. \
To do that, we first proceed as before: Once we have chosen the
desired longitudinal intensity pattern $F(z)$, within the interval
$0 \leq z \leq L$, we calculate the coefficients $A_m$ as before,
i.e., \ $A_m = (1/L) \int_{0}^{L}\,F(z)\,{\rm exp}(-i 2 \pi m z
/L)\,\drm z$, \ and \ $k_{\rho\,m} = \sqrt{\om_0^2 - \left(Q + 2 \pi
m/L \right)^2}$. \
Afterwards, we just replace the zero-order Bessel beams
$J_0(k_{\rho\,m}\rho)$, in superposition (\ref{S5soma3}), with
higher-order Bessel beams, $J_{\mu}(k_{\rho\,m}\rho)$, to get

\bb \dis{\Psi(\rho,z,t) \ug
e^{-i\,\om_0\,t}\,e^{i\,Q\,z}\,\sum_{m=-N}^{N}
A_m\,J_{\mu}(k_{\rho\,m}\,\rho)\,e^{i\,\frac{2\pi}{L}m\,z} } \; ,
\label{S5soma4} \ee

From this result, and on the basis of intuitive arguments\cite{[17]}, we can expect
that the desired longitudinal intensity pattern, initially
constructed for $\rho=0$, will approximately shift to $\rho =
\rho'$, where $\rho'$ represents the position of the first maximum
of the Bessel function, i.e., the first positive root of the
equation \
$\drm\,J_{\mu}(k_{\rho, \, m=0}\,\rho)/\drm\rho)|_{\rho'}=0$.

\h By such a procedure, one can obtain very interesting stationary
configurations of field intensity, as ``donuts", cylindrical
surfaces, and much more.

\h In the following example, we show how to obtain, e.g., a
cylindrical surface of stationary {\em light}. \ To get it, within
the interval $0 \leq z \leq L$, let us first select the
longitudinal intensity pattern given by eq.(\ref{S5Fz2}), with
$l_1=L/2-\Delta z$ \ and \ $l_2=L/2+\Delta z$, and with $\Delta z =
L/300$. Moreover, let us choose $L=0.05\,$m, $Q=0.998\,\om_0/c$,
and use $N=150$.

\h Then, after calculating the coefficients $A_m$ by
eq.(\ref{S5An}), we use to superposition
(\ref{S5soma4}), choosing, in this case, $\mu =4$. \ According to
the previous discussion, one can expect the desired longitudinal
intensity pattern to appear shifted to $\rho ' \approx
5.318/k_{\rho, \, m=0}=8.47\,\mu$m, where 5.318 is the value of
$k_{\rho, \, m=0}\,\rho$ for which the Bessel function
$J_{4}(k_{\rho, \, m=0}\,\rho)$ assumes its maximum value, with
$k_{\rho, \, m=0} = \sqrt{\om_0^2 - Q^2}$. \ The figure
\ref{S5MHRFig3sec5} below shows the resulting intensity field.

\h In Fig.(\ref{S5MHRFig3sec5}\,a) the transverse section of the
resulting beam for $z=L/2$ is shown. The transverse peak intensity
is located at $\rho=7.75\,\mu$m, with a $8.5\%$ difference w.r.t.
the predicted value of $8.47\,\mu$m. Figure (\ref{S5MHRFig3sec5}\,b)
shows the orthogonal projection of the resulting field, which
corresponds to nothing but a cylindrical surface of stationary
light (or other fields).

\begin{figure}[!h]
\begin{center}
\scalebox{.6}{\includegraphics{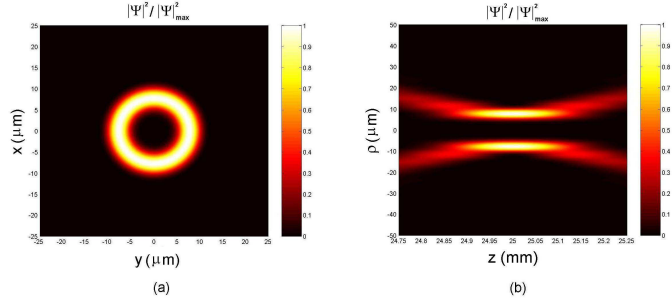}}
\end{center}
\caption{\textbf{(a)} Transverse section at $z=L/2$ of the
considered higher-order FW. \ \textbf{(b)} Orthogonal projection
of the three-dimensional intensity pattern of the same
higher-order FW.} \label{S5MHRFig3sec5}
\end{figure}

\h We can see that the desired longitudinal intensity pattern has
been approximately obtained, shifted, as desired, from $\rho=0$
to $\rho = 7.75\,\mu$m; and the resulting field resembles a
cylindrical surface of stationary light with radius $7.75\,\mu$m
and length $238\,\mu$m. Donut-like configurations of light (or
sound) are also possible.

\

\subsection{Stationary wavefields
with arbitrary longitudinal shape in absorbing media: An extension of
the method.}

\

When propagating in a non-absorbing medium, the so-called
nondiffracting waves maintain their spatial shape for long
distances. However, the situation is not the same when dealing
with absorbing media. In such cases, both the ordinary and the
nondiffracting beams (and pulses) will suffer the same effect: an
exponential attenuation along the propagation axis. \
We shall present an extension\cite{[10]} of the method given
above, with the aim of showing that, through suitable superpositions of
equal-frequency Bessel beams, it is possible to obtain even
in {\it absorbing media} nondiffracting beams , whose longitudinal
intensity pattern can assume any desired shape within a chosen
interval $ 0 \leq z \leq L$ of the propagation axis $z$.

\h As a particular example, we are going to obtain new nondiffracting beams
capable to resist the loss effects, maintaining amplitude and spot
size of their central core for long distances.

\h It is important to stress that the energy absorption by the medium
continues to occur normally, but the new beams have an initial
transverse field distribution, such to reconstruct
(notwithstanding the presence of absorption) their central cores for
distances considerably longer than the penetration depths of
ordinary (nondiffracting or diffracting) beams. In this sense, the
new method can be regarded as extending, for absorbing media,
the self-reconstruction properties\cite{[28]} that usual Localized
Waves are known to possess in loss-less media.

\h In the same way as for lossless media, we construct a Bessel beam
with angular frequency $\om$ and axicon angle $\theta$ in the
absorbing materials by superposing plane waves, with the same
angular frequency $\om$, and whose wave vectors lie on the surface
of a cone with vertex angle $\theta$. The refractive index of the
medium can be written as $n(\om) = n_{\Rrm}(\om) + in_{\Irm}(\om)$, quantity
$n_{\Rrm}$ being the real part of the complex refraction index and
$n_{\Irm}$ the imaginary one, responsible for the absorbtion effects.
For a plane wave, the penetration depth $\delta$ for the
frequency $\om$ is given by $\delta = 1/\alpha= c/2\om n_{\Irm}$, where
$\alpha$ is the absorption coefficient. \
Therefore, a zero-order Bessel beam in dissipative media can be
written as \ $\psi = J_0(k_{\rho}\rho){\rm exp}(i\beta z){\rm
exp}(-i\om t)$ \ with \ $\beta = n(\om)\om \cos\theta/c = n_{\Rrm}\om
\cos\theta/c + in_{\Irm}\om \cos\theta/c \equiv \beta_{\Rrm} + i\beta_{\Irm}$;
$k_{\rho}=n_{\Rrm}\om \sin\theta/c + in_{\Irm}\om \sin\theta/c \equiv
k_{\rho \Rrm} + ik_{\rho \Irm}$, \ and so \ $k_{\rho}^2 = n^2\om^2/c^2 -
\beta^2$. \ Thus, it results \  $\psi=J_0((k_{\rho \Rrm} + ik_{\rho
\Irm})\rho){\rm exp}(i\beta_{\Rrm} z){\rm exp}(-i\om t){\rm exp}(-\beta_{\Irm}
z)$, \  where $\beta_{\Rrm}$, $k_{\rho \Rrm}$ are the real parts of the
longitudinal and transverse wave numbers, and $\beta_{\Irm}$, $k_{\rho
\Irm}$ are the imaginary ones, while the absorption coefficient of a
Bessel beam with axicon angle $\theta$ is given by
$\alpha_{\theta}=2\beta_{\Irm}=2n_{\Irm}\om \cos\theta/c$, its penetration
depth being $\delta_{\theta}=1/\alpha_{\theta}=c/2\om
n_{\Irm}\cos\theta$.

\h Due to the fact that $k_{\rho}$ is complex, the amplitude of the
Bessel function $J_0(k_{\rho}\rho)$ starts decreasing from
$\rho=0$ till the transverse distance $\rho=1/2k_{\rho \Irm}$, and
afterwards it starts growing exponentially.  This behaviour is not
physically acceptable, but one must remember that it occurs only
because of the fact that an ideal Bessel beam needs an infinite
aperture to be generated. However, in any real situation, when a
Bessel beam is generated by finite apertures, that exponential
growth in the transverse direction, starting after
$\rho=1/2k_{\rho \Irm}$, will {\it not} occur indefinitely, stopping
at a given value of $\rho$. Let us moreover emphasize that, when
generated by a finite aperture of radius $R$, the truncated Bessel
beam\cite{[17]} possesses a depth of field $Z = R/\tan\theta$, and can
be approximately described by the solution given in the previous
paragraph, for $\rho<R$ and $z < Z$.

\h Experimentally, to guarantee that the mentioned exponential
growth in the transverse direction {\it does not} even start, so
as to meet only a decreasing transverse intensity, the radius $R$
of the aperture used for generating the Bessel beam should be $R
\leq 1/2k_{\rho \Irm}$. However, as noted by Durnin et al., the same
aperture has to satisfy also the relation $R \geq 2\pi/k_{\rho
\Rrm}$. From these two conditions, one can infer that, in an
absorbing medium, a Bessel beam with just a decreasing transverse
intensity can be generated only when the absorption coefficient is
$ \alpha < 2/\lambda$, i.e., if the penetration depth is $\delta >
\lambda /2$. The present method does refer to these cases, i.e.,
it is always possible to choose a suitable finite aperture size in
such a way that the truncated versions of all solutions, including
the general one given by eq.(111), will not develop any unphysical
behaviour. Let us now outline the method\cite{subArt1}.

\h Let us consider an absorbing medium with the complex refraction index
$n(\om) = n_{\Rrm}(\om) + in_{\Irm}(\om)$, and the following superposition
of $2N + 1$ Bessel beams with the same frequency $\om$:

\bb
\begin{array}{clr}
\Psi(\rho,z,t) = \dis{\sum_{m=-N}^{N}} A_m\,J_0\left((k_{\rho \Rrm_m}
+ ik_{\rho
\Irm_m})\rho\right)\,e^{i\,\be_{\Rrm_m}z}\,e^{-i\,\om\,t}\,e^{-\be_{\Irm_m}z}
\; ,
\end{array} \label{S52soma1} \ee

where the $m$ are integer numbers, the $A_m$ are constant coefficients
(yet unknown), quantities $\be_{\Rrm_m}$ and $k_{\rho \Rrm_m}$ ($\be_{\Irm_m}$ and
$k_{\rho \Irm_m}$) are the real (the imaginary) parts of the
complex longitudinal and transverse wave numbers of the $m$-th
Bessel beam in superposition (\ref{S52soma1}); the following
relations being satisfied

\bb
 k_{\rho_m}^2 \ug n^2\frac{\om^2}{c^2} - \be_{m}^2 \label{S52kr}
\ee

\bb \frac{\be_{\Rrm_m}}{\be_{\Irm_m}} \ug \frac{n_{\Rrm}}{n_{\Irm}} \label{S52bei}
\ee

where  $\be_m = \be_{\Rrm_m} + i\be_{\Irm_m}$, $k_{\rho_m} =
k_{\rho \Rrm_m} + ik_{\rho \Irm_m}$, \ with \ $k_{\rho \Rrm_m}/k_{\rho
\Irm_m}=n_{\Rrm}/n_{\Irm}$.

\h Our goal is finding out the values of the longitudinal wave
numbers $\be_m$ and the coefficients $A_m$ in order to reproduce
approximately, inside the interval $0 \leq z \leq L$ (on the axis
$\rho=0$), a {\it freely chosen} longitudinal intensity pattern
that we call $|F(z)|^2$. \
The problem for the particular case of lossless media[128,129],
i.e., when $n_{\Irm}=0 \rightarrow \be_{\Irm_m}=0$, was solved in the
previous subsection. For those cases, it was shown that the choice
$\beta=Q+2\pi m/L$, \ with $A_m = \int_{0}^{L} F(z){\rm exp}(-i2\pi
mz/L)/L\,\,\drm z$ can be used to provide approximately the desired
longitudinal intensity pattern $|F(z)|^2$ in the interval $0\leq z \leq L$,
and, at the same time, to
regulate the spot size of the resulting beam by means of the
parameter $Q$. \ Such parameter, incidentally, can be also used to obtain large field-depths
and moreover to inforce the linear polarization approximation to the
electric field for the TE electromagnetic wave (see details in
refs.[128,129]).

\h However, when dealing with absorbing media, the procedure
described in the last paragraph does not work, due to the presence
of the functions $\exp (-\be_{\Irm_m}z)$ in the superposition
(\ref{S52soma1}), because in this case that series does not became a
Fourier series when $\rho=0$. \
On attempting to overcome this limitation, let us write the real
part of the longitudinal wave number, in superposition
(\ref{S52soma1}), as

\bb \be_{\Rrm_m} \ug Q + \frac{2\pi m}{L} \label{S52br} \ee

with

\bb 0 \leq Q + \frac{2\pi m}{L} \leq n_{\Rrm} \frac{\om}{c} \; .
\label{S52cond} \ee

where inequality (\ref{S52cond}) guarantees forward
propagation only, with no evanescent waves. \
In this way, the superposition (\ref{S52soma1}) can be written

\bb
\begin{array}{clr}
\Psi(\rho,z,t) = e^{-i\,\om\,t}\,e^{i\,Qz}\, \dis{\sum_{m=-N}^{N}}
A_m\,J_0\left((k_{\rho \Rrm_m} + ik_{\rho \Irm_m})\rho\right)
\,e^{i\,\frac{2\pi m}{L}z}\,e^{-\be_{\Irm_m}z}  \; ,
\end{array} \label{S52soma2} \ee

where, by using eq.(\ref{S52bei}), we have \ $\be_{\Irm_m}=(Q + 2\pi
m/L)n_{\Irm}/n_{\Rrm}$, \ and \ $k_{\rho_m}=k_{\rho \Rrm_m} + ik_{\rho \Irm_m}$ is
given by eq.(\ref{S52kr}). Obviously, the discrete superposition
(\ref{S52soma2}) could be written as a continuous one (i.e., as an
integral over $\be_{\Rrm_m}$) by taking $L \rightarrow \infty$, but
we prefer the discrete sum due to the difficulty of obtaining
closed-form solutions to the integral form.

\h Now, let us examine the imaginary part of the longitudinal wave
numbers. The minimum and maximum values among the $\be_{\Irm_m}$ are
$(\be_{\Irm})_{\rm min}=(Q-2\pi N/L)n_{\Irm}/n_{\Rrm}$ and $(\be_{\Irm})_{\rm
max}=(Q+2\pi N/L)n_{\Irm}/n_{\Rrm}$, the central one being given by
$\overline{\be}_{\Irm} \equiv (\be_{\Irm})_{m=0} = Q n_{\Irm}/n_{\Rrm} $. With this in
mind, let us evaluate the ratio $\Delta = [(\be_{\Irm})_{\rm max} -
(\be_{\Irm})_{\rm min}]/{\overline{\be}_{\Irm}} = 4 \pi N/LQ$. \
Thus, when $\Delta <<1$, there are no considerable differences
among the various $\be_{\Irm_m}$, because $\be_{\Irm_m} \approx
\overline{\be}_{\Irm}$ holds for all $m$. In the same way, there are no
considerable differences among the exponential attenuation
factors, since $\exp (-\be_{\Irm_m}z) \approx \exp (-\overline{\be}_{\Irm}
z)$. So, when $\rho=0$ the series in the r.h.s. of
eq.(\ref{S52soma2}) can be approximately considered a truncated
Fourier series {\it multiplied by} the function $\exp
(-\overline{\be}_{\Irm} z)$, and, therefore, superposition
(\ref{S52soma2}) can be used to reproduce approximately the
desired longitudinal intensity pattern $|F(z)|^2$ (on $\rho=0$),
within $0\leq z \leq L$, when the coefficients $A_m$ are given by

\bb A_m \ug \frac{1}{L}\,\int_{0}^{L} F(z)\,e^{\overline{\be}_{\Irm}
z}e^{-i\,\frac{2\pi m}{L}z}\,\drm z \; ,\label{S52am} \ee

the presence of the factor $\exp (\overline{\be}_{\Irm} z)$ in
the integrand being necessary to compensate for the factors
$\exp (-\be_{\Irm_m}z)$ in superposition (\ref{S52soma2}). \
Since we are adding together zero-order Bessel functions, we can
expect a good field concentration around $\rho=0$.

\h In short, we have shown in this Section how one can get, in an
{\it absorbing medium}, a {\it stationary} wave-field with a good
transverse concentration, and whose longitudinal intensity pattern
(on $\rho=0$) can approximately assume {\it any desired shape}
$|F(z)|^2$ within the predetermined interval $0 \leq z \leq L$.
The method followed above ---let us resume--- is a generalization of a
previous one[128,129], and
consists in the superposition in eq.(\ref{S52soma2}) of Bessel beams
whose longitudinal wave numbers are individuated by the real and imaginary parts
given in eqs.(\ref{S52br}) and (\ref{S52bei}), respectively, while their
complex transverse wave numbers are given by eq.(\ref{S52kr}). Finally,
the coefficients of the superposition are given by
eq.(\ref{S52am}). The method is justified when $4\pi N/LQ << 1$:
happily enough, this condition is satisfied in a great number of
situations.

\h Regarding the generation of these new beams, once we have an apparatus
capable of generating a single Bessel beam, we can just use an array of
such apparatuses to generate a sum of Bessel beams, with the appropriate
longitudinal wave numbers and amplitudes/phases [as specified by
our method], thus producing the desired final beam. For instance, we can
use[128,129] a laser illuminating an array of concentric annular
apertures (located at the focus of a convergent lens), with the
appropriate radii and transfer functions in order to be able to yield
both the required longitudinal wave numbers (once a value for $Q$ has been
chosen) and the coefficients $A_n$ of the fundamental
superposition (\ref{S52soma2}).

\subsubsection{Some Examples}

For generality's sake, let us consider a hypothetical medium in
which a typical XeCl excimer laser ($\lambda = 308 {\rm nm}
\rightarrow \om = 6.12\times 10^{15}$Hz) has a penetration depth
of 5 cm; i.e. an absorption coefficient $\alpha = 20 {\rm
m}^{-1}$, and therefore $n_{\Irm} = 0.49\times 10^{-6} $. Besides this,
let us suppose that the real part of the refraction index for this
wavelength is $n_{\Rrm} = 1.5$ and therefore $n = n_{\Rrm} + in_{\Irm} = 1.5 +
i\,0.49\times 10^{-6}$. Note that the value of the real part of
the refractive index is not so important for us, since we are
dealing with monochromatic wave fields.

\h A Bessel beam with $\om  = 6.12\times 10^{15}$Hz and with an
axicon angle $\theta = 0.0141 \;$rad (thus, with a transverse spot of
radius $8.4\,\mu$m), when generated by an aperture, say, of radius
$R=3.5\;$mm, can travel in vacuum a distance
equal to $Z=R/\tan\theta=25\;$cm while resisting the diffraction
effects. However, in the material medium considered here, the
penetration depth of this Bessel beam would be only $z_p = 5\;$cm.
Now, let us set forth two interesting applications of the present method.

\

\textbf{First Example}: Almost Undistorted Beams in Absorbing
Media.

\

We can use the extended method to obtain, in the same medium
and for the same wavelength, an almost undistorted beam capable of
preserving its spot size and the intensity of its {\it central
core} for a distance many times larger than the typical
penetration depth of an ordinary beam (nondiffracting or not). \
To this purpose, let us suppose that, in the considered material medium,
we want a beam (with $\om  = 6.12\times 10^{15}$Hz) that maintains
amplitude and spot size of its central core for a distance of
$25\;$cm, i.e., a distance 5 times greater than the penetration
depth of an ordinary beam with the same frequency. We can model
this beam by choosing the desired longitudinal intensity pattern
$|F(z)|^2$ (on $\rho=0$), within $0 \leq z \leq L$, to be given
by the function

\bb
 F(z) \ug \left\{\begin{array}{clr}
&1 \;\;\; {\rm for}\;\;\; 0 \leq z \leq Z  \\

&0 \;\;\; \mbox{elsewhere} ,
\end{array} \right.  \label{S52Fz}
 \ee

and by putting $Z=25\;$cm, with, for example, $L=33\;$cm.

\h Now, one can use the Bessel beam superposition (\ref{S52soma2}) to
reproduce approximately the selected intensity pattern. \ Let us
choose $Q=0.9999\om / c$ for the $\be_{\Rrm_m}$ in eq.(\ref{S52br}),
and $N=20$ (notice that, according to inequality
(\ref{S52cond}), $N$ could assume a maximum value of $158$.) \
After having chosen the values of $Q$, $L$ and $N$, the values of
the complex longitudinal and transverse wave numbers of the Bessel
beams happen to be defined by relations (\ref{S52br}), (\ref{S52bei})
and (\ref{S52kr}). Eventually, we can have recourse to eq.(\ref{S52am}), and
find out the coefficients $A_m$ of the fundamental superposition
(\ref{S52soma2}), that defines the resulting stationary
wave-field. \
Let us just note that the condition $4\pi N/LQ << 1$ is perfectly
satisfied in this case.

\begin{figure}[!h]
\begin{center}
\scalebox{.8}{\includegraphics{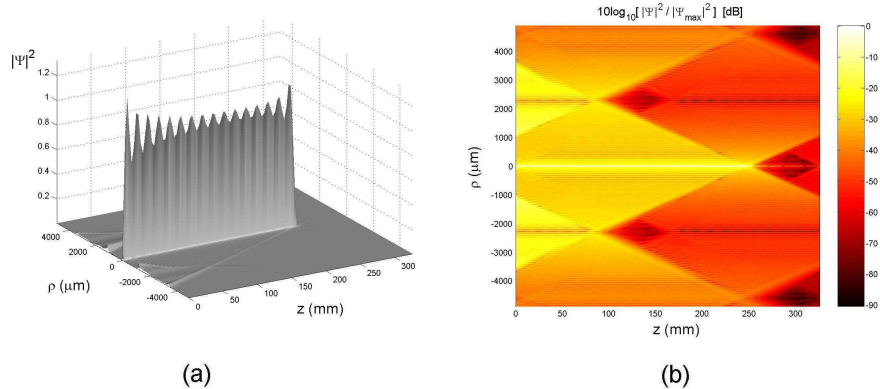}}
\end{center}
\caption{\textbf{(a)} Three-dimensional field-intensity of the
resulting beam. \textbf{(b)} The resulting beam, in an orthogonal
projection and in {\bf logaritmic} scale.} \label{S52MHRFig5sec5}
\end{figure}

\h In Fig.(\ref{S52MHRFig5sec5}\, a) we can see the 3D
field-intensity of the resulting beam. One can observe that the field
possesses a good transverse localization (with a spot size smaller
than $10 \; \mu$m), and is capable of maintaining spot size and
{\it intensity} of its central core till the desired distance (a
better result could be reached by using a higher value of $N$).

\h It is interesting to note that at that distance (25 cm), an
ordinary beam would have got its initial field-intensity
attenuated $148$ times.

\h As we said above, the energy absorption by the
medium continues to occur normally; the difference is that these
new beams have an initial transverse field distribution
sophisticated enough to be able to reconstruct (even in the
presence of absorption) their central cores, till a certain
distance. For a better visualization of this field-intensity
distribution and of the energy flux, we show in Fig.(\ref{S52MHRFig5sec5}\,b)
the resulting beam, in an
orthogonal projection and in {\em logarithmic} scale. It appears clear
that the energy comes from the lateral regions, in order to
reconstruct the central core of the beam. On the plane $z=0$,
within the region $\rho \leq R=3.5\,$mm, there is an uncommon field
intensity distribution, it being very dispersed instead of
concentrated. This uncommon initial field intensity distribution
is responsible for the construction of the central core of the resulting
beam, and for its reconstruction all along the distance $z=25\,$cm. Due
to absorption, the beam (total) energy, flowing through
different $z$ planes, is not constant; but the energy flowing
in the beam spot area, and the beam spot size itself, are
conserved till the distance (in this case) \ $z=25\,$ cm.

\

\textbf{Second Example}: Beams in absorbing media with a growing
longitudinal field intensity.

\

Let us consider again the previous hypothetical medium, in which an
ordinary Bessel beam with $\theta = 0.0141 \; $rad and $\om  =
6.12\times 10^{15}$Hz would have a penetration depth of $5\;$cm. \ We aim
at constructing now a beam that, instead of possessing a {\it
constant} core-intensity till the position $z=25\;$cm, presents
on the contrary a (moderate) exponential {\em growth} of its
intensity, till that distance ($z=25\;$cm).

\h Let us assume we wish to get the following longitudinal intensity pattern
$|F(z)|^2$, in the interval $0<z<L$:

\bb
 F(z) \ug \left\{\begin{array}{clr}
& {\rm exp}(z/Z) \;\;\; {\rm for}\;\;\; 0 \leq z \leq Z  \\

&0 \;\;\; \mbox{elsewhere} \; ,
\end{array} \right.  \label{S52Fz2}
 \ee
\setcounter{equation}{114}
%

with $Z=25\,$cm and $L=33\;$cm. \
Using again $Q=0.9999\,\om / c$, and $N=20$, we can proceed as in the first
example, calculating the complex longitudinal and transverse
wave numbers of the Bessel beams, and finally the coefficients $A_m$ of
the fundamental superposition (\ref{S52soma2}).

\begin{figure}[!h]
\begin{center}
\scalebox{.8}{\includegraphics{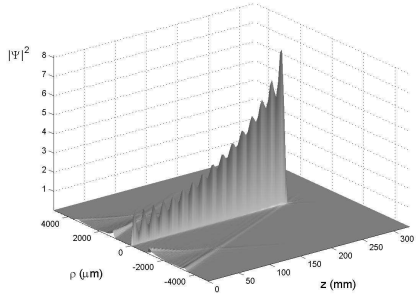}}
\end{center}
\caption{Three-dimensional field-intensity of the resulting beam,
in AN absorbing medium, with a growing longitudinal field
intensity.} \label{S52MHRFig6sec5}
\end{figure}

\h In Fig.(\ref{S52MHRFig6sec5}) we can see the 3D field-intensity
of the resulting beam. One can observe that the field presents the
desired longitudinal intensity pattern, with a good transverse
localization (a spot size smaller than $10 \; \mu$m).

\

\

\h Obviously, the amount of energy necessary to construct these new
beams is greater than that necessary to generate an ordinary beam
in a non-absorbing medium. And it is also clear that there is a
{\it limitation} on the depth of field of these new beams. In the
first example, for distances {\em larger than} 10 times the penetration
depth of an ordinary beam, the field-intensity in the lateral regions
would result to be higher than that at the core, and the field would
loose the usual characteristics of a beam (transverse field
concentration); not to speak of the greater energy demand.

\

\

\

\section{Subluminal Localized Waves (or Bullets)} 

In this Section, abandoning for a while the subject of the
so-called ``Frozen Waves", we want to face the more general
problem of obtaining, in a simple way, localized (non-diffractive) {\em
subluminal} pulses as exact analytic solutions to the wave
equations.\cite{subArt1} \  These new ideal subluminal solutions,
which propagate without distortion in any homogeneous linear
media, will be here obtained for arbitrarily chosen frequencies and
bandwidths, avoiding in particular any recourse to the non-causal
(backward moving) components that so frequently plague the
previously known localized waves. \ The new solutions are suitable
superpositions of ---zeroth-order, in general--- Bessel beams,
which can be performed either by integrating w.r.t. the angular
frequency $\om$, or by integrating w.r.t. the
longitudinal wavenumber $k_z\,$: \ Both methods are expounded in this
review. \ The first one will appear to be powerful enough; \ we shall
present the second method as well, however, since it allows dealing once
more ---from a different starting point--- also with the limiting
case of zero-speed solutions (and furnishes a {\em new} way, in
terms of {\em continuous spectra,} for obtaining our Frozen Waves,
so promising also from the point of view of applications). \ Some
attention is moreover paid to the known role of Special
Relativity, and to the fact that the localized waves are expected
to be transformed one into the other by suitable Lorentz
Transformations. \ At last, we briefly treat the case of non
axially-symmetric solutions, in terms of higher order Bessel
beams. \ The analogous pulses with intrinsic finite energy, or
merely truncated, will be considered elsewhere. \ We keep
fixing our attention especially on electromagnetism and optics:
But {\em let us repeat} that results of the same kind are valid
whenever an essential role is played by a wave-equation [like in
acoustics, seismology, geophysics, elementary particle physics (as
we verified even in the slightly different case of the
Schroedinger equation), and also gravitation (for which we have
recently got stimulating new results), and so on].

\

\

\subsection{A foreword about the Subluminal Localized Waves}

For self-consistency, let us repeat here the following considerations. 
For more than ten years, the so-called (non-diffracting)
``Localized Waves" (LW), which are new solutions to the
wave equations (scalar, vectorial, spinorial,...), are in fashion,
both in theory and in experiment.  In particular, rather well-known are
the ones with luminal or Superluminal peak-velocity\cite{Livro*}: Like the
so-called
X-shaped waves (see\cite{Lu1*,MRH*} and refs. therein; for a review,
see, e.g., ref.\cite{IEEE*}), which are supersonic in Acoustics\cite{Lu2*},
and Superluminal in Electromagnetism (see\cite{PhysicaA*} and refs. therein).

\h As we know, since Bateman\cite{Bateman*} and later on Courant \&
Hilbert\cite{Courant*}, it was recognized, e.g., that {\em luminal} LWs
exist, which are solutions to the wave equations. More
recently, some attention[9-13] started to be paid to the
(more ``orthodox", as we said) {\em
subluminal} LWs too. Let us recall that all the LWs
propagate without distortion ---and in a self-reconstructive
way[14-16]--- in a homogeneous linear medium (apart from local
variations): In the sense that their square magnitude keeps its
shape during propagation, while local variations are shown only by
its real, or imaginary, part.

\h As in the Superluminal case, the subluminal LWs can be obtained
by suitable superpositions of
Bessel beams.\cite{subArt1} \ They have been till now almost
neglected, as we know, for the mathematical difficulties met in
getting analytic expressions for them, difficulties associated
with the fact that the superposition integral runs over a finite
interval. \ We shall here re-address the question of such
subluminal LWs, showing, by contrast, that one can indeed arrive
at exact (analytic) solutions, both in the case of integration
over the Bessel beams' angular frequency $\om$, and in the case of
integration over their longitudinal wavenumber $k_z$.

\h As already claimed, the present work is devoted to the exact, analytic
solutions: i.e., to ideal solutions. \ The corresponding pulses with finite
energy, or truncated, will be presented elsewhere.

\h  Let us recall that, in the past, too much attention was not even paid
to Brittingham's 1983 paper\cite{Brittingham*}, wherein he had shown
the possibility of obtaining pulse-type solutions to the Maxwell equations,
which propagate in free space as a new kind of speed-c ``solitons".  That
lack of attention was partially due to the fact that Brittingham had been
able neither to get correct finite-energy expressions for such ``wavelets",
nor to make suggestions about their practical production.  Two years later,
however, Sezginer\cite{Sezginer*} was able to obtain
quasi-nondiffracting luminal pulses endowed with a finite energy. \
Finite-energy pulses do no longer travel undistorted, as we do know, for an infinite distance,
but can nevertheless propagate without deformation for a long field-depth,
much larger than the one achieved by ordinary pulses like the gaussian ones:
Cf., e.g., refs.[19-24] and refs. therein.

\h Only after 1985 the general theory of LWs started to be extensively
developed[25-31,2,6,3],
both in the case of beams and in the case of pulses. For reviews, see for
instance the refs.\cite{IEEE*,Birds*,Introd*,Tesi*,PIER98*} and citations therein.
 \ For the propagation of LWs in bounded regions (like {\em wave-guides\/}),
see refs.[33-36] and refs therein. \ For the focusing of
LWs, see the Second Part of this review [as well as refs.\cite{MSR*,focusing*} and
quotations therein]. \
As to the construction of general LWs propagating in {\em dispersive} media, see
refs.[39-47]; and, for {\em lossy} media, cf. ref.\cite{OpEx2*} and
refs. therein. \ Al last, for finite-energy, or truncated, solutions
see refs.[48-50,24,3,34], and work in progress.

\h By now, the LWs have been experimentally
produced\cite{Lu2*,PSaari*,Ranfagni*}, and are being applied in
fields ranging from ultrasound scanning\cite{Lu3*,Lu4*,Lu5*} to
optics (for the production, e.g., of new type of
tweezers\cite{brevetto*}). All those works have demonstrated by now that
nondiffracting pulses can travel with an arbitrary peak-speed $v$,
that is, with $0<v<\infty$; while Brittingham and Sezginer had
confined themselves to the luminal case ($v=c$) only.

\h As we were remarking, the Superluminal and luminal LWs have been, and are
being, intensively studied; whilst the subluminal ones have been
neglected: Almost all the few papers dealing with them had
recourse till now to the paraxial\cite{Molone*}
approximation\cite{paraxial*}, or to numerical
simulations\cite{SaloSalomaa*}, due to the above mentioned
mathematical difficulty in obtaining exact analytic expressions
for subluminal pulses.  Indeed, only {\em one} analytic solution
was known[9-11,28,57,58], biased by the physically unconvenient
facts that its frequency spectrum is very large, that it doesn't even
possess a well-defined central frequency, and, even more, that
backward-travelling\cite{Ziolk2*,PIER98*} components (ordinarily
called ``non-causal", since they should be {\em entering} the antenna or
generator) were needed for constructing it. \ Aim of the
next Sections is showing, on the contrary, that subluminal localized exact solutions
can be constructed with any spectra, in any frequency bands and
for any bandwidths; and {\em without}
employing\cite{MRH*,Introd*} any backward-travelling components.

\section{A first method for constructing physically \\
acceptable, subluminal Localized Pulses}  

Axially symmetric solutions to the scalar wave equation are known
to be superpositions of zero-order Bessel beams over the angular
frequency $\om$ and the longitudinal wavenumber $k_z$: \ That is, in
cylindrical co-ordinates,

\bb \Psi(\rho,z,t) \ug
\int_{0}^{\infty}\,\drm\om\,\int_{-\om/c}^{\om/c}\,\drm k_z\,
\overline{S}(\om,k_z) J_0\left(\rho\sqrt{\frac{\om^2}{c^2} -
k_z^2}\right)e^{ik_z z}e^{-i\om t} \; , \label{eq.(1)} \ee    

where \ $k_\rho^2 \equiv \om^2/c^2 - k_z^2$ \ is the transverse
wavenumber. \ Quantity $k_\rho^2$ has to be positive since
evanescent waves cannot come into the play.

\h The condition characterizing a nondiffracting wave is the
existence\cite{PIER98*,MHoptfibers*} of a linear
relation between longitudinal wavenumber $k_z$ and frequency $\om$ for
all the Bessel beams entering superposition (113); that is to say, the chosen
spectrum has to entail\cite{MRH*,Tesi*} for each Bessel beam a linear relationship
of the type:\footnote{More generally, as shown in ref.\cite{MRH*}, the chosen
spectrum has to call into the play, in the plane $\om, k_z$, if
not exactly the line (114), at least a region in the proximity of a
straight-line of that a type. It is interesting that in the latter case
one obtains solutions endowed with finite energy, but possessing a finite
``depth of field", that is to say, nondiffracting only till a certain finite
distance.}

\bb \om \ug v\,k_z + b \; \, \label{eq.(2)} \ee      

with $b \geq 0$. \ Requirement (116) can be regarded also as a
specific space-time coupling, implied by the chosen spectrum
$\overline{S}$. Equation (116) has to be obeyed by the spectra of
any one of the three possible types (subluminal, luminal or
Superluminal) of nondiffracting pulses. \ Let us mention
that with the choice in eq.(116) the pulse re-gains
its initial shape after the space-interval ${\Delta z}_1 = 2\pi
v/b$. \ But the more general case can be also
considered\cite{MRH*,Capitulo*} when $b$ assumes any values
$b_m=m\,b$ (with $m$ an integer), and the periodicity
space-interval becomes ${\Delta z}_m = {\Delta z}_1 / m\,$. \
We are referring ourselves, of course, to the real (or imaginary)
part of the pulse, since its modulus is known to be endowed with
rigid motion.

\h In the subluminal case, of interest here, the only exact solution
known till recent time, represented by eq.(124) below, was the one found by
Mackinnon\cite{Mackinnon*}. Indeed, by taking into explicit account that the
transverse wavenumber $k_\rho$ of each Bessel beam entering
eq.(115) has to be real, it can be easily shown (as first noticed by
Salo et al. for the analogous acoustic solutions\cite{SaloSalomaa*})
that in the subluminal case $b$ cannot vanish, but must be larger than zero:
 \ $b>0$. \ Then, on using conditions (116) and $b>0$, the subluminal
localized pulses can be expressed as integrals over the frequency only:

\bb \Psi(\rho,z,t) \ug  \exp{[-ib {z \over v}]} \,
\int_{\om_-}^{\om_+}\,\drm\om \; S(\om) \, J_0(\rho k_\rho)\,
\exp{[i\om {\zeta \over v}]} \; , \label{eq.(3)} \ee     

where now

\bb k_\rho \ug {1 \over v} \, \sqrt{2b\om - b^2
- (1-v^2/c^2) \om^2} \; \, \label{eq.(4)} \ee    

with

\bb \zeta \, \equiv \,  z - v t \; \, \label{eq.(5)} \ee    

and with

\hfill{$
\left\{ \begin{array}{clr}
\om_{-} \ug \dis{{b \over {1+v/c}}} \\
\\
\om_{+} \ug \dis{{b \over {1-v/c}}}
\end{array}   \right.
$\hfill} (120)        

\setcounter{equation}{120}

\

As anticipated, the Bessel beam superposition in the subluminal
case results to be an integration over a finite interval of $\om$,
which does clearly shows that the backward-travelling (non-causal)
components correspond to the interval $\om_- < \om < b$. \ It
could be noticed that eq.(117) does not represent the most general
exact solution, which on the contrary is {\em a
sum\/}\cite{Capitulo*} of such solutions for the various possible
values of $b$ mentioned above: That is, for the values $b_m=m\,b$
with spatial periodicity ${\Delta z}_m = {\Delta z}_1 / m\,$. \
But we can confine ourselves to solution (117) without any real
loss of generality, since the actual problem is evaluating in
analytic form the integral entering eq.(117). For any mathematical
and physical details, see ref.\cite{Capitulo*}.

\h Now, if one adopts the change of variable

\bb \om \, \equiv \, {b \over {1-v^2/c^2}}\;(1 + {v \over c} s) \; \,
\label{eq.(7)} \ee    

equation (117) becomes\cite{SaloSalomaa*}

\begin{eqnarray}
\lefteqn{\Psi(\rho,z,t) \ug {b \over c} \,{v \over {1-v^2/c^2}} \, \exp{[-i
{b \over v} z]} \;
\exp{\left[ i{b \over v} \, {1 \over {1-v^2/c^2}} \, \zeta \right]}} \nonumber
\\
& & {} \times \int_{-1}^{1}\,\drm s \; S(s) \, J_0\left( {b \over
c} \, {\rho \over {\sqrt{1-v^2/c^2}}} {\sqrt{1-s^2}} \right) \;
\exp{\left[ i {b \over c} {1 \over {1-v^2/c^2}} \zeta s \right]}
\; . \label{eq.(8)}
\end{eqnarray}   

In the following we shall adhere ---as it is an old habit of ours--- to some symbols
standard in  Special Relativity (since the whole topic of subluminal, luminal
and Superluminal LWs is strictly connected\cite{IEEE*,PhysicaA*,RMDartora*}
with the principles and structure of
SR [cf.\cite{Barut*,Review*} and refs. therein],
as we shall mention also in the concluding remarks which fllow below); namely:

\bb \beta \equiv {v \over c} \; ; \ \ \ \ \ \ \ \ \ \gamma \,
\equiv \, {1 \over {\sqrt{1-\beta^2}}} \; . \label{eq.(9)}
\ee    

\h As already said, eq.(122) has till now yielded {\em one}
analytic solution for $S(s) \, = \;${\em constant}, only (for
instance, $S(s)=1$); which means nothing but $S(\om) \, =
\;$constant: in this case one gets indeed {\em the Mackinnon
solution\/}\cite{Mackinnon*,Donnelly*,Lu5*,JosaMlast*}

\begin{eqnarray}
\lefteqn{ \Psi(\rho,\zeta,\eta) \ug 2{b \over c} v \, \gamma^2\,
{\exp{\left[ i{b \over c} \, \beta \gamma^2 \; \eta \right]}}}
\nonumber
 \\
& & {} \times \sinc \ {\sqrt{ {b^2 \over c^2}\, \gamma^2 \left(
\rho^2 + \gamma^2 \; \zeta^2 \right)
}}  \; , \label{eq.(10)}     
\end{eqnarray}

which however, for its above-mentioned drawbacks, is endowed with
little physical and practical interest. \ In eq.(124) the $\sinc$
function has the ordinary definition

\

$$\sinc \ x \, \equiv \, (\sin \; x)/x \; ,$$

\

and

\bb \eta \, \equiv \, z -Vt, \ \ \ \ \ \ \ {\rm with} \ \ V \equiv
{c^2 \over v}
 \; , \label{eq.(11)} \ee  \  

where $V$ and $v$ are related by the {\em de Broglie relation.} \ [Notice that
$\Psi$ in eq.(124), and in the following ones, is eventually a function
(besides of $\rho$) of $z,t$ via $\zeta$ and $\eta$, both functions of
$z$ and $t\;$].

\h {\em However,} we can construct by a very simple method new
subluminal pulses corresponding to whatever spectrum, and devoid
of backward-moving (i.e., ``entering") components, just by taking
advantage of the fact that in our equation (122) the integration
interval is finite: that it, by transforming it in a good, instead
of a harm. Let us first observe that eq.(122) doesn't admit only
the already known analytic solution corresponding to
$S(s)=\;$constant, and more in general to $S(\om) \, =
\;$constant, but it will also yield an exact, analytic solution
for {\em any} exponential spectra of the type

\bb
S(\om) \ug \exp{[{i2n\pi \om \over \Omega}]}
 \; , \label{eq.(12)}
\ee  

with $n$ any integer number, which means for any spectra of the type
$S(s)= \exp{[in\pi / \beta]} \,
\exp{[in\pi s]}$, \ as can be easily seen by checking the product of the various
exponentials entering the integrand. \ In eq.(126) we have set

\

$$ \Omega \, \equiv \, {\om_+} - {\om_-} \; . $$

\

The solution writes in this more general case:

\begin{eqnarray}
\lefteqn{ \Psi(\rho,\zeta,\eta) \ug 2b\beta\, \gamma^2 \,
\exp{\left[ i{b \over c} \, \beta \, \gamma^2 \, \eta \right]}}
\nonumber
 \\
& & {} \times \exp{[in {\pi \over \beta}]}
 \ \sinc \ {\sqrt{{b^2 \over c^2}\, \gamma^2 \, \rho^2
+ \left( {b \over c} \, \gamma^2 \, \zeta + n \pi
\right)^2}}
 \; . \label{eq.(13)}
\end{eqnarray}     

Let us explicitly notice that also in eq.(127) quantity $\eta$ is
defined as in Eqs.(125) above, where $V$ and $v$ obey the de
Broglie relation $vV=c^2$, the subluminal quantity $v$ being the
velocity of the pulse envelope, and $V$ playing the role (in the
envelope's interior) of a Superluminal phase velocity.

\h The next step, as anticipated, consists just in taking {\em advantage}
of the finiteness of the integration limits for
expanding any arbitrary spectra $S(\om)$ in a Fourier series in the interval
$\om_{-} \leq \om \leq \om_{+}\;$:

\bb S(\om) \ug \sum_{n=-\infty}^{\infty} \, A_n \,
\exp{[+in {2\pi \over \Omega} \om]} \; , \label{eq.(14)}
\ee    

where (we went back, now, from the $s$ to the $\om$ variable):

\bb A_n \ug {1 \over \Omega} \, \int_{\om_-}^{\om_+} \drm \om \, S(\om) \,
\exp{[-in {2 \pi \over \Omega} \om]}  \; \,
\label{eq.(15)} \ee    

quantity $\Omega$ being defined as above.

\h Then, on remembering the special solution (127), we can infer from expansion
(126) that, for any arbitrary spectral function $S(\om)$, one can work out
a rather general axially-symmetric analytic solution for the subluminal case:

\begin{eqnarray}
\lefteqn{ \Psi(\rho,\zeta,\eta) \ug 2b\beta\, \gamma^2 \,
{\exp{\left[ i{b \over c} \, \beta \, \gamma^2 \; \eta \right]}}}
\nonumber
 \\
& & {} \times \sum_{n=-\infty}^{\infty} \, A_n \, \exp{[in{\pi
\over \beta}]} \ \sinc \ {\sqrt{ {b^2 \over c^2}\, \gamma^2 \rho^2
+ \left( {b \over c} \gamma^2 \; \zeta + n \pi \right)^2}}
\; , \label{eq.(16)}     
\end{eqnarray}

in which the coefficients $A_n$ are still given by eq.(129). Let us
repeat that our solution is expressed in terms of
the particular equation (127), which is a ``Mackinnon-type" solution.

\h The present approach presents many advantages. We can easily choose
spectra localized within the prefixed frequency interval (optical waves, microwaves,
etc.) and endowed with the desired bandwidth.   Moreover, as we have seen,
spectra can now be chosen
such that they have zero value in the region $\om_{-} \leq \om \leq b$,
which is responsible for the backward-travelling components of the
subluminal pulse.

\h Let us stress that, even when the adopted spectrum $S(\om)$
does not possess a known Fourier series (so that the coefficients
$A_n$ cannot be exactly evaluated via eq.(129)), one can calculate
approximately such coefficients without meeting any problem, since
our general solutions (130) will still be exact solutions.

\h Let us set forth some examples.

\

\subsection{Some Examples}

\h In general, optical pulses generated in the lab possess a
spectrum centered at some frequency value, $\om_0$, called the
carrier frequency. \ The pulses can be, for instance, ultra-short,
when $\Delta\om/\om_0 \geq 1$; or quasi-monochromatic, when
$\Delta\om/\om_0 << 1$, where $\Delta\om$ is the spectrum
bandwidth.

\h These kinds of spectra can be mathematically represented by a
gaussian function, or functions with similar behaviour.

\

\h {\bf First Two Examples:}

\

Let us first consider a gaussian spectrum

\bb S(\om) \ug \frac{a}{\sqrt{\pi}}\exp{[-a^2(\om-\om_0)^2]} \label{eq.(17)} \ee   

whose values are negligible outside the frequency interval $\om_-
< \om < \om_+$ over which the Bessel beams superposition in eq.(117)
is made, it being  $\om_- = b/(1+\beta)$ and $\om_+ =
b/(1-\beta)$. \ Of course, relation (116) has still to be satisfied,
with $b>0$, for getting an ideal subluminal localized
solution. \ Notice that, with spectrum (131), the bandwidth
(actually, the FWHM) results to be $\Delta \om = 2 / a$. \ Let us
emphasize that, once $v$ and $b$ have been fixed, the values of
$a$ and $\om_0$ can afterwards be selected in order to kill the
backward-travelling components, that correspond, as we know, to $\om < b\;$.

\h The Fourier expansion in eq.(128), which yields, with the above
spectral function (131), the coefficients

\bb A_n \, \simeq \, \dis{{{1} \over {W}}} \, \exp{[-in {2\pi
\over \Omega} \om_0]} \;
\exp{[{{-n^2 \pi^2} \over {a^2 W^2}}]} \; , \label{eq.(18)} \ee   

\

constitutes an excellent representation of the gaussian spectrum (131) in
the interval $\om_- < \om < \om_+$ (provided that, as we requested, our
gaussian spectrum does get negligible values outside the frequency interval
$\om_- < \om < \om_+$).

\h In other words, on choosing a pulse velocity $v<c$ and a value
for the parameter $b$, a subluminal pulse with the above frequency
spectrum (131) can be written as eq.(129), with the coefficients
$A_n$ given by eq.(132): the evaluation of such coefficients $A_n$
being rather simple. \ Let us repeat that, even if the values of
the $A_n$ are obtained via a (rather good, by the way)
approximation, we based ourselves on the {\em exact} solution
eq.(130).

\h One can, for instance, obtain exact solutions representing
subluminal pulses for optical frequencies. Let us get the
subluminal pulse with velocity $v=0.99\;c$, carrier angular
frequency $\om_0= 2.4\times 10^{15}\;$Hz (that is,
$\la_0=0.785\;\mu$m) and bandwidth (FWHM) $\Delta\om=\om_0
/20=1.2\times 10^{14}\;$Hz, which is an optical pulse of $24$ fs
(which is the FWHM of the pulse intensity). \ For a complete pulse
characterization, one has to choose the value of the frequency
$b$: \ let it be  $b=3\times 10^{13}\;$Hz; \ as a consequence one has
$\om_{-}=1.507\times 10^{13}\;$Hz and $\om_{+}=3\times
10^{15}\;$Hz. \ [This is exactly a case in which the considered
pulse is not plagued by the presence of backward-travelling components,
since the chosen spectrum possesses totally negligible values for
$\om < b$]. \ The construction of the pulse does already result
satisfactory when considering about 51 terms ($-25 \leq n \leq
25$) in the series entering eq.(130).

\begin{figure}[!h]
\begin{center}
 \scalebox{.5}{\includegraphics{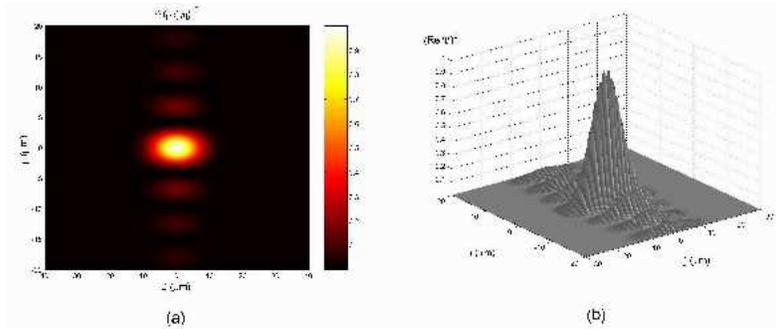}}
\end{center}
\caption{\textbf{(a)} The intensity orthogonal projection for the pulse
corresponding to eqs.(131,132) in the case of an optical frequency (see the
text); \ \textbf{(b)} The three-dimensional intensity
pattern of the \emph{real part} of the same pulse, which reveals the
carrier wave oscillations.} \label{fig36}
\end{figure}

\h Figures \ref{fig36} show our pulse, evaluated  by summing the mentioned
fifty-one terms. Namely: Fig.(36a) depicts the orthogonal projection
of the pulse intensity; \ Fig.(36b) shows the three-dimensional
intensity pattern of the \emph{real part} of the pulse, which
reveals the carrier wave oscillations.

\

\h Let us stress that the ball-like shape\footnote{It can be noted
that each term of the series in eq.(130) corresponds to an
ellipsoid or, more specifically, to a spheroid, for each velocity
$v$.} for the field intensity should be typically associated with
all the subluminal LWs, while the typical Superluminal ones are
known to be X-shaped\cite{Lu1*,PhysicaA*,RMDartora*}, as predicted
since long by special relativity in its ``non-restricted"
version: See refs.\cite{Barut*,Review*,PhysicaA*,IEEE*} and refs
therein.

\

\h A second spectrum $S(\om)$ would be, for instance, the
``inverted parabola" one, centered at the frequency $\om_0$: \ that
is,

\bb
S(\om) \ug \left\{ \begin{array}{clr} {\frac{-4 \, [\om -
(\om_0 - \Delta \om/2)] [\om - (\om_0 + \Delta \om/2)]} {\Delta
\om^2}} \;\; & {\rm for}\;\;\; \om_0 - \Delta \om/2 \leq \om \leq
\om_0 + \Delta \om/2 \\
\\
\;\;\;\;\;\; 0  & {\rm otherwise} \ , \label{eq.(19)}
\end{array} \right.    
\ee

where $\Delta \om$, the distance between the two zeros of the
parabola, can be regarded as the spectrum bandwidth. \ One can expand
$S(\om)$, given in eq.(133), in the
Fourier series (128), for $\om_{-} \leq \om \leq \om_{+}\;$, with coefficients $A_n$
that ---even if straightforwardly calculable--- results to be complicated,
so that we skip reporting them here explicitly. \ Let us here only mention
that spectrum (133) may be easily used to get, for instance, an ultrashort
(femtasecond) optical non-diffracting pulse, with satisfactory results even
when considering very few terms in expansion (128).

\

\h {\bf Third Example:}

\

As a third interesting example, let us consider the very simple
case when ---within the integration limits $\om_{-}$, $\om_{+}$---
the complex exponential spectrum (124) is replaced by the real
function (still linear in $\om$)

\bb S(\om) \ug \frac{a}{1-\exp{[-a(\om_+ - \om_-)]}}\,\exp{[a(\om
- \om_{+}]}
 \; , \label{eq.(20)}
\ee  

with $a$ a positive number [for $a=0$ one goes back to the
Mackinnon case]. Spectrum (134) is exponentially concentrated in
the proximity of $\om_{+}$, where it reaches its maximum value;
and becomes more and more concentrated (on the left of $\om_{+}$,
of course) as the arbitrarily chosen value of $a$ increases, its
frequency bandwidth being $\Delta\om=1/a$. Let us recall
that, on their turn, quantities $\om_{+}$ and $\om_{-}$
depend on the pulse velocity $v$ and on the arbitrary parameter
$b$.

\h By performing the integration as in the case of spectrum (126), instead
of solution (127) in the present case one eventually gets the solution

\begin{eqnarray}
\lefteqn{ \Psi(\rho,\zeta,\eta) \ug \frac{2ab\beta\gamma^2 \,
\exp{[ab\gamma^2]} \, \exp{[-a\om_{+}]}}{1-\exp{[-a(\om_+ -
\om_-)]}}}
 \nonumber
 \\
 \nonumber
 \\
 & & {} \times
 \exp{\left[ i{b \over c} \, \beta \, \gamma^2 \, \eta \right]}
 \ \sinc{\left[ {b \over c} \, \gamma^2 \; {\sqrt{\gamma^{-2} \, \rho^2
- (av+i\zeta)^2}} \right]} \; . \label{eq.(21)}
\end{eqnarray}     

\h After Mckinnon's, this eq.(135) appears to be the simplest
closed-form solution, since both of them do not need any recourse
to series expansions. In a sense, our solution (135) may be
regarded as the subluminal analogue of the (Superluminal) X-wave
solution; a difference being that the standard X-shaped solution
has a spectrum starting with 0, where it assumes its maximum
value, while in the present case the spectrum starts at $\om_{-}$
and gets increasing afterwards, till $\om_+$. \ More important is
to observe that the gaussian spectrum has a priori two advantages
w.r.t. eq.(134): It may be more easily centered around any value
$\om_{0}$ of $\om$, and, when increasing its concentration in the
surroundings of $\om_{0}$, the spot transverse width does not
increase indefinitely, but tends to the spot-width of a Bessel
beam with $\om=\om_0$ and $k_z=(\om_0 - b)/V$, at variance with
what happens for spectrum (134). \ Anyway, solution (135) is
noticeable, since it is really {\em the simplest} one.

\begin{figure}[!h]
\begin{center}
 \scalebox{.45}{\includegraphics{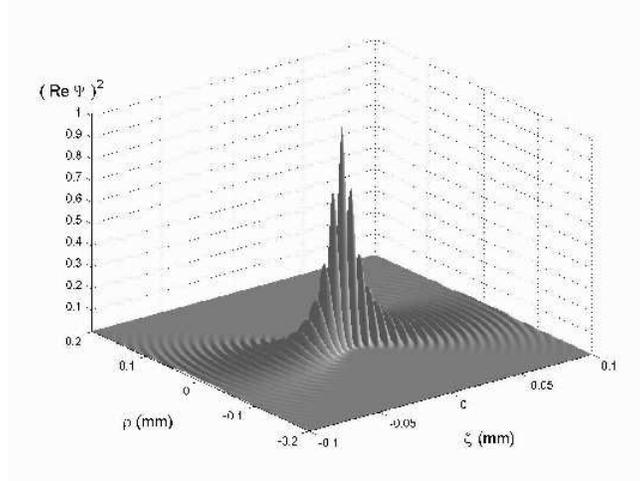}}
\end{center}
\caption{The intensity of the real part of the subluminal pulse
corresponding to spectrum (134), with $v=0.99\;c$, with
$b=3\times10^{13}\,$Hz (which result in $\om_-=1.5\times 10^{13}\;$Hz
and $\om_-=3\times 10^{15}\;$Hz), and with $\Delta\om/\om_+ = 1/100$ (i.e.,
$a=100$).} \label{fig37}
\end{figure}

\h Figure \ref{fig37} shows the intensity of the real part of the subluminal
pulse corresponding to this spectrum, with $v=0.99\;c$, with
$b=3\times10^{13}\;$Hz (which result in $\om_-=1.5\times 10^{13}\;$Hz
and $\om_+=3\times 10^{15}\;$Hz), and with $\Delta\om/\om_+ = 1/100$ (i.e.,
$a=100$). This is an optical pulse of $0.2$ ps.

\

\section{A second method for constructing\\
subluminal Localized Pulses} 

The previous method appears to be very efficient for finding out analytic
subluminal LWs, but it looses its validity in the limiting case $v
\rightarrow 0$, since for $v=0$ it is $\om_{-} \equiv \om_{+}$ and the
integral in eq.(117) degenerates, furnishing a null value. \ By
contrast, we are interested also in the $v=0$ case, since it corresponds,
as we already know, to some of the most interesting, and potentially useful, LWs: \
Namely, to the stationary solutions to the wave equations endowed with
a static envelope, and that we have called ``Frozen Waves".

\h Before going on, let us recall that the theory of Frozen Waves was
initially developed in refs.\cite{OpEx1*,FWart2*}, by having recourse to discrete
superpositions in order to bypass the need of numerical simulations. \
In the case of continuous superpositions, some numerical
simulations were performed in refs.\cite{Dartora12*}. \ However,
the method presented in this Section does allow finding out analytic
exact solutions (without any further need of numerical simulations)
even for Frozen Waves consisting in {\em continuous superpositions}. Actually,
we are going to see that the present method works whatever is the
chosen field-intensity shape, also in regions with size of the order
of the wavelength.

\h It is possible to get such results by starting again from eq.(115),
with constraint (114), but going on ---this time--- to integrals over
$k_z$, instead of over $\om$. \ It is enough to write relation (116)
in the form

\hfill{$
k_z \ug \dis{{1 \over v} \, (\om - b)}
$\hfill} (116')        

\

for expressing the exact solutions (115) as

\bb \Psi(\rho,z,t) \ug  \exp{[-ibt]} \;
\int_{k_z{\rm min}}^{k_z{\rm max}}\,\drm k_z \; S(k_z) \, J_0(\rho k_\rho)\,
\exp{[i \zeta k_z]} \; , \label{eq.(22)} \ee     

with

\hfill{$
\left\{ \begin{array}{clr}
k_z{\rm min} \ug \dis{ {-b \over c} \, {1 \over {1+\beta}}} \\
\\
k_z{\rm max} \ug \dis{ {b \over c} \, {1 \over {1-\beta}} }
\end{array}   \right.
$\hfill} (135)        

\setcounter{equation}{137}

and with

\bb {k_\rho}^2 \ug -{k_z^2 \over \gamma^2} + 2 {b \over c} \beta k_z +
{b^2 \over c^2} \; , \label{eq.(24)} \ee    

where quantity $\zeta$ is still defined according to eq.(119), always with
$v<c$.

\h One can show that the unique exact solution previously
known\cite{Mackinnon*} may be rewritten in form (136) {\em with}
$S(k_z) = \;$constant. \ Then, on following the same procedure exploited
in our first method (previous Section), one can find out new exact solutions
corresponding to

\bb
S(k_z) \ug \exp{[{i2n\pi k_z \over K}]}
 \; , \label{eq.(25)}
\ee  

where

\

$$ K \, \equiv \, k_z{\rm max} - {k_z{\rm min}} \; , $$

\

by performing the change of variable [analogous, in its finality, to the
one in eq.(121)]

\bb k_z \, \equiv \, {b \over c} \, \gamma^2 \, (s + \beta) \; .
\label{eq.(26)} \ee    

\h At the end, the exact subluminal solution corresponding to
spectrum (139) results to be

\begin{eqnarray}
\lefteqn{ \Psi(\rho,\zeta,\eta) \ug 2\,{b \over c} \, \gamma^2 \,
\exp{[i{b \over c} \, \beta \, \gamma^2 \, \eta]}}  \nonumber
 \\
& & {} \times \exp{[in \pi \beta]}
 \ \sinc \ {\sqrt{{b^2 \over c^2}\, \gamma^2 \, \rho^2 +
\left( {b \over c} \, \gamma^2 \, \zeta + n \pi \right)^2}}
 \; , \label{eq.(27)}
\end{eqnarray}          

\h We can again observe that any spectra $S(k_z)$ can be expanded,
in the interval $k_z{\rm min} < k_z < k_z{\rm max}$, in the Fourier
series:

\bb S(k_z) \ug \sum_{n=-\infty}^{\infty} \, A_n \,
\exp{[+in {2\pi \over K} k_z]} \; , \label{eq.(28)}
\ee    

with coefficients given now by

\bb A_n \ug {1 \over K} \, \int_{k_z{\rm min}}^{k_z{\rm max}} \drm k_z \,
S(k_z) \, \exp{[-in {2 \pi \over K} k_z]} \; \,
\label{eq.(29)} \ee    

quantity $K$ having been defined before.

\h At the end of the whole procedure, the general exact solution
representing a subluminal LW, for any spectra $S(k_z)$, can be
eventually written

\begin{eqnarray}
\lefteqn{ \Psi(\rho,\zeta,\eta) \ug 2\,{b \over c} \, \gamma^2 \,
\exp{[i{b \over c} \, \beta \, \gamma^2 \, \eta]}}  \nonumber
 \\
& & {} \times \sum_{n=-\infty}^{\infty} \, A_n \, \exp{[in \pi \beta]}
 \ \sinc \ {\sqrt{{b^2 \over c^2}\, \gamma^2 \, \rho^2 +
\left( {b \over c} \, \gamma^2 \, \zeta + n \pi \right)^2}}
 \; , \label{eq.(30)}
\end{eqnarray}     

whose coefficients are expressed in eq.(143), and where quantity $\eta$
is defined as above, in eq.(125).

\h Interesting examples could be easily worked out, as we did at the end of
the previous Section.

\section{\bf Stationary solutions with zero-speed envelopes}  

Here, we shall refer ourselves to the (second) method, expounded
in the previous Section. \ Our solution (144), for the case of
envelopes {\em at rest}, that is, in the case $v=0$ [which implies
$\zeta = z$], becomes

\

\hfill{$
\Psi(\rho,z,t) \ug \dis{ 2\,{b \over c} \; \exp{[-i b t]} \; \sum_{n=-\infty}^{\infty} \, A_n \
\sinc \ {\sqrt{{b^2 \over c^2}\, \rho^2 +
\left( {b \over c} \, z + n \pi \right)^2}}} \; ,
$\hfill} (145)        

\

with coefficients $A_n$ given by eq.(143) with $v=0$, so that its integration
limits simplify into $-b/c$ and $b/c$, respectively. Thus, one gets

\

\hfill{$ A_n \ug \dis{{c \over {2b}} \, \int_{-b/c}^{b/c} \drm k_z \,
S(k_z) \, \exp{[-in {c \pi \over b} k_z]}} \; .
$\hfill} (143')        

\

\h Equation (145) is a new exact solution, corresponding to
stationary beams with a {\em static} intensity envelope. \ Let us
observe, however, that even in this case one has energy
propagation, as it can be easily verified from the power flux
$\Sbf_{\rm s} = -\nabf\Psi_{\cal R} \ \pa\Psi_{\cal R}/\pa t$
(scalar case) or from the Poynting vector $\Sbf_{\rm v} = (\Ebf
\wedge \Hbf)$ (vectorial case: the condition being that
$\Psi_{\cal R}$ be a single component, $A_z$, of the vector
potential $\Abf$).\cite{PhysicaA*} \ We have indicated by
$\Psi_{\cal R}$ the real part of $\Psi$. \ For $v=0$, eq.(116)
becomes

$$ \om \ug b \, \equiv \, \om_0 \; ,$$

so that the particular subluminal waves endowed with null velocity are
actually monochromatic beams.

\h It may be stressed that the present (second) method does yield
{\em exact} solutions, without any need of the paraxial approximation,
which, on the contrary, is so often used when looking for
expressions representing beams, like the gaussian ones. \ Let us
recall that, when having recourse to the paraxial approximation,
the obtained beam expressions are valid only when the envelope
sizes (e.g., the beam spot) vary in space much more slowly than
the beam wavelength.  For instance, the usual expression for a
gaussian beam\cite{Molone*} holds only when the beam spot $\Delta
\rho$ is much larger than $\lambda_0 \equiv \om_0 / (2\pi c) =
b/(2\pi c)$: so that those beams {\em cannot} be very much
localized. \ By contrast, our method overcomes such problems,
since we have seen that it yields exact
expressions for (well localized) beams with sizes {\em of the
order} of their wavelength. \ Notice, moreover, that the
already-known exact solutions ---for instance, the Bessel beams---
are nothing but particular cases of our solution (145).

\

{\em An example:} \ On choosing (with $0 \leq q_- < q_+ \leq 1$)
the spectral double-step function

\

\hfill{$
S(k_z) \ug \left\{ \begin{array}{clr}
\dis{\frac{c}{\om_0(q_+ - q_-)}} \ \ \ \ \ \ \ \ \ \ & {\rm for} \ q_-\om_0/c \leq k_z \leq q_+\om_0/c \\
\\
0 \ \ \ \ \ \ \ \ \ \ & {\rm elsewhere} \; ,
\end{array}   \right.
$\hfill} (146)        

\

the coefficients of eq.(145) become

\

\hfill{$ A_n \ug \dis{\frac{ic}{2\pi n\om_0(q_+ - q_-)} \; \left[
e^{-iq_+\pi n} - e^{-iq_-\pi n} \right]}
 \; .
$\hfill} (147)        
%

\

\h The double-step spectrum (146) corresponds, with regard to the longitudinal wave number,
to the mean value $\overline{k}_z = \om_0(q_++q_-)/2c$ \ and to
the width $\Delta k_z=\om_0(q_+-q_-)/c$. From such relations, it follows
that $\Delta k_z/\overline{k}_z = 2(q_+-q_-)/(q_++q_-)$.

\h For values of $q_-$ and $q_+$ that do not satisfy the inequality $\Delta
k_z / \overline{k}_z << 1$, the resulting solution will be a {\em non-paraxial}
beam.

\h Figures 38 show the exact solution corresponding to $\om_0 =
1.88\times 10^{15}\;$Hz (i.e., $\lambda_0=1\;\mu$m), and to $q_- =
0.3$, and to $q_+ = 0.9$: \ It results to be a beam with a spot-size
diameter of $0.6\;\mu$m, and, moreover, with a rather good
longitudinal localization. \ In the case of Eqs.(144, 145), about 21
terms ($-10 \leq n \leq 10$) in the sum entering eq.(143) are quite
enough for a good evaluation of the series. The beam considered in
this example is highly non-paraxial (with $\Delta
k_z/\overline{k}_z = 1$ ), and therefore couldn't have been
obtained by ordinary gaussian beam solutions (which are valid in
the paraxial regime only)\footnote{We are considering here only
scalar wave fields. In the case of non-paraxial optical beams, the
vector character of the field has to be taken into account}.

\

\begin{figure}[!h]
\begin{center}
 \scalebox{.45}{\includegraphics{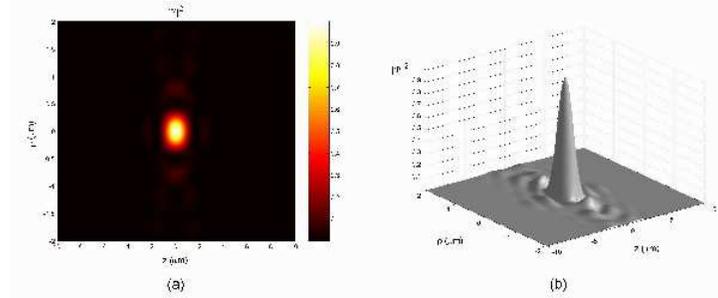}}
\end{center}
\caption{\textbf{(a)} Orthogonal projection of the
three-dimensional intensity pattern of the beam (a null-speed
subluminal wave) corresponding to spectrum (146); \ \textbf{(b)}
3D plot of the field intensity. The beam considered in this
example is highly non-paraxial.} \label{fig38}
\end{figure}

\subsection{A new approach to the ``Frozen Waves"}

\h Let us now emphasize that a noticeable property of the present
method is that it allows a spatial modeling even of monochromatic
fields (that correspond to envelopes {\em at rest\/}; so that, in the
electromagnetic cases, one can speak, e.g., of the modeling of
``light-fields at rest").

\h Let us recall that such a modelling
---rather interesting, especially for applications\cite{brevetto*}---
was already performed in
refs.\cite{OpEx1*,FWart2*,OpEx2*}, and has been exploited at the beginning
of our Third Part [cf. eqs.(99, 100)], in terms of discrete superpositions
of Bessel beams. And the stationary fields with
static envelopes have been called ``Frozen Waves" (FW) by us.

\h But the method presented in the last Sections allows us to make use of
{\em continuous} superpositions, in order to get a predetermined longitudinal
(on-axis) intensity pattern, inside a desired space interval
$0<z<L$. \ In fact, the continuous superposition, analogous to eq.(100),
now writes

\

\hfill{$ \Psi(\rho,z,t) \ug \dis{ e^{-i\om_0 t} \;
\int_{-\om_0/c}^{\om_0/c} \drm k_z \ S(k_z) \ J_0(\rho k_\rho) \
e^{iz k_z} } \; ,
$\hfill} (148)    

\

which is nothing but the previous eq.(136) with $v=0$
(and therefore $\zeta = z$). \ In other words, eq.(148) does just
represent a {\em null-speed} subluminal wave. \ To be clearer, let
us recall once more that the FWs were expressed in the past as discrete
superposition because it was not known at that time how to treat
analytically a continuous superposition like (148). \ We are now able, however,
to extend the previous approach to FWs to the case of
integrals: without numerical simulations, but in terms once more
of analytic solutions.

\h Indeed, the exact solution of eq.(148) is given by eq.(145), with
coefficients (143');  and one can choose the spectral function
$S(k_z)$ in such a way that $\Psi$ assumes the on-axis pre-chosen
static intensity pattern $|F(z)|^2$. \ Namely, the equation to be
satisfied by $S(k_z)$, to such an aim, comes out by associating
eq.(148) with the requirement $|\Psi(\rho=0,z,t)|^2 = |F(z)|^2$,
which entails the integral relation

\

\hfill{$  \dis{\int_{-\om_0/c}^{\om_0/c} \drm k_z \; S(k_z) \;
e^{i z k_z}} \ug F(z) \; .
$\hfill} (149)    

\

Equation (149) would be trivially solvable in the case of an
integration between $-\infty$ and $+\infty$, since it would merely
be a Fourier transformation; but obviously this is not the case,
because its integration limits are finite. Actually, there are
functions $F(z)$ for which eq.(149) is not solvable at all, in the sense
that no spectra $S(k_z)$ exist obeying the last equation. \
For instance, if we consider the {\em Fourier} expansion

\

\hfill{$ F(z) \ug \dis{\int_{-\infty}^{\infty}} \drm k_z \;
\widetilde{S}(k_z) \; e^{iz k_z} \; ,
$\hfill}

\

when $\widetilde{S}(k_z)$ does assume non-negligible values outside
the interval $-\om_0/c < k_z < \om_0/c$, then in eq.(149) {\em no} $S(k_z)$
can forward that particular $F(z)$ as a result.

\h {\em However}, way-outs can be devised, such that one can
nevertheless find out a function $S(k_z)$ that approximately
(but satisfactorily) complies with eq.(149).

\

\h {\em The first way-out} consists in writing $S(k_z)$ in the form

\

\hfill{$ S(k_z) \ug \dis{ {1 \over K} \; \sum_{n=-\infty}^{\infty}
\, F \left( {{2n\pi} \over K} \right) \; e^{-i2n\pi k_z / K} } \; ,
$\hfill} (150)    

\

where, as before, $K=2\om_0/c$. Then, one can easily verify
eq.(150) to guarantee that the integral in eq.(149) yields the
values of the desired $F(z)$ {\em at the discrete points} \ $z =
2n\pi / K$. \ Indeed, the Fourier expansion (150) is already of the
same type as eq.(142), so that in this case the coefficients $A_n$
of our solution (145), appearing in eq.(143'), do simply become

\

\hfill{$
A_n \ug \dis{{1 \over K} \; F(-{{2n\pi} \over K})} \; .
$\hfill} (151)    

\

\h This is a powerful way for obtaining a desired longitudinal (on-axis)
intensity pattern, especially for tiny spatial
regions, because it is not necessary to solve any integral to find
out the coefficients $A_n$, which by contrast are given directly by eq.(151).

\begin{figure}[!h]
\begin{center}
 \scalebox{.7}{\includegraphics{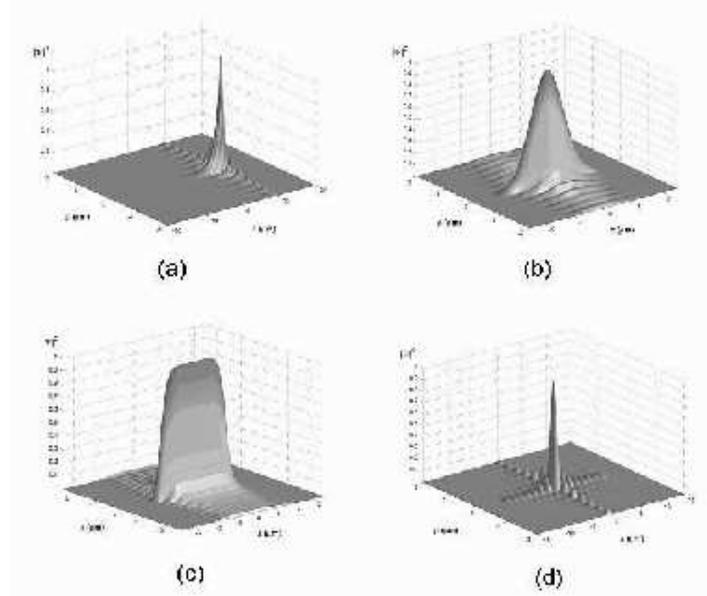}}
\end{center}
\caption{Frozen Waves with the on-axis longitudinal field pattern
chosen as: \textbf{(a)} Exponential; \ \textbf{(b)} Gaussian;  \
\textbf{(c)} Super-gaussian;  \ \textbf{(d)} Zero order \ Bessel
function} \label{fig39}
\end{figure}

\h Figures \ref{fig39} depict some interesting applications of this method.
A few  desired longitudinal intensity patterns $|F(z)|^2$ have been chosen,
and the corresponding Frozen Waves calculated by using eq.(145) with the
coefficients $A_n$ given in eq.(151). The desired patterns are enforced
to exist within very small spatial intervals only, in order to show the
capability of the method to model\cite{subArt1} the field intensity shape
also under such strict requirements.

\h In the four examples below, we considered a wavelength
$\lambda=0.6\;\mu$m, which corresponds to $\om_0=b=3.14\times
10^{15}\;$Hz.

\h The first longitudinal (on-axis) pattern considered by us is that given
by

\

$$
 F(z) \ug \left\{\begin{array}{clr}
 \dis{e^{a(z-Z)}} \;\;\; &
 {\rm for}\;\;\; 0 \leq z \leq Z  \\

 \\
 \;\;\;\;\;\;\;\; 0  & \mbox{elsewhere} \; ,
\end{array} \right. \label{Fz1}
 $$

i.e., a pattern with an exponential increase, starting from $z=0$
untill $z=Z$. The chosen values of $a$ and $Z$ are $Z=10\;\mu$m
and $a=3/Z$. The intensity of the corresponding Frozen Wave is
shown in Fig.(39a).

\h The second longitudinal pattern (on-axis) taken into consideration
is the gaussian one, given by

\

$$
 F(z) \ug \left\{\begin{array}{clr}
\dis{e^{-q(\frac{z}{Z})^2}} \;\;\; &
 {\rm for}\;\;\; -Z \leq z \leq Z  \\

 \\
 \;\;\;\;\;\;\;\; 0  & \mbox{elsewhere} \; ,
\end{array} \right.  \label{Fz1}
 $$

 \

with $q=2$ and $Z=1.6\;\mu$m. The intensity of the corresponding
Frozen Wave is shown in Fig.(39b).

\h In the third example, the desired longitudinal pattern is supposed to
be a super-gaussian:

\

$$
 F(z) \ug \left\{\begin{array}{clr}
 \dis{e^{-q(\frac{z}{Z})^{2m}}} \;\;\; &
 {\rm for}\;\;\; -Z \leq z \leq Z  \\

 \\
 \;\;\;\;\;\;\;\; 0  & \mbox{elsewhere} \; ,
\end{array} \right.  \label{Fz1}
 $$

 \

where $m$ controls the edge sharpness. Here we have chosen
$q=2$, $m=4$ and $Z=2 \ \mu$m. \ The intensity of the Frozen Wave
obtained in this case is shown in Fig.(39c).

\h Finally, in the fourth example, let us choose the longitudinal
pattern as being the zero-order Bessel function

\

$$
 F(z) \ug \left\{\begin{array}{clr}
 J_0(q\,z) \;\;\; &
 {\rm for}\;\;\; -Z \leq z \leq Z  \\

 \\
 \;\;\;\;\;\;\;\; 0  & \mbox{elsewhere} \; ,
\end{array} \right.  \label{Fz1}
 $$

with $q=1.6\times 10^{6}\;\rm{m}^{-1}$ and $Z=15\;\mu$m. The intensity
of the corresponding Frozen Wave is shown in Fig.(39d).

\h Let us observe that any static envelopes of this type can
be easily transformed into propagating pulses by the mere application
of Lorentz transformations.

\

\h {\em Another way-out} exists for evaluating $S(k_z)$, based on the assumption
that

\

\hfill{$
S(k_z) \, \simeq \, \widetilde{S}(k_z)  \; ,
$\hfill} (152)    
\setcounter{equation}{152}

\

which consitutes a good approximation whenever $\widetilde{S}(k_z)$ assumes
{\em negligible} values outside the interval $[-\om_0/c, \ \om_0/c]$. \ In
such a case, one can have recourse to the method associated with eq.(142)
and expand $\widetilde{S}(k_z)$ itself in a Fourier series, getting
eventually the relevant coefficients $A_n$ by eq.(143). \ Let us recall
that it is still $K \equiv k_z{\rm max} - k_z{\rm min} = 2\om_0/c$.

\h It is worthwhile to call attention to the circumstance
that, when constructing FWs in terms of a sum of discrete
superpositions of Bessel beams (as it has been done by us in the
Second part of this work, and in
refs.\cite{OpEx1*,FWart2*,OpEx2*,brevetto*}), it was easy to obtain
extended envelopes like, e.g., ``cigars": Where ``easy" means by
using only a few few terms of the sum. By contrast, when we construct
FWs ---following this Section--- as continuous superpositions,
then it is easy to get highly localized (concentrated) envelopes.
\ Let us explicitly mention, moreover, that the method presented
in this Section furnishes FWs that are no longer periodic along
the $z$-axis (a situation that, with our old
method\cite{OpEx1*,FWart2*,OpEx2*}, was obtainable only when the
periodicity interval tended to infinity).

\

\

\section{The role of Special Relativity, and of Lorentz Transformations}

Strict connections exist between, on one hand, the principles and
structure of Special Relativity and, on the other hand, the whole
subject of subluminal, luminal, Superluminal Localized Waves, in
the sense that it is expected since long time that a priori they
are transformable one into the other via suitable Lorentz
transformations (cf. refs.\cite{Barut*,Review*,NCim*}, besides
work of ours in progress).$^{\heartsuit}$

\footnotetext{$^{\heartsuit}$ Let us call attention to a paper by
Saari et al.\cite{Saari2004*}, noticed by us only recently,
wherein the relativistic connections among the LWs were also
investigated in terms of suitable LTs. We are actually glad in
quoting here such a reference because it appears inspired by a
philosophy, which
---going back in part to papers like refs.\cite{Barut*,Review*}---
has been constantly shared by our old, and recent, papers.  Let us
mention also a further interesting article noticed by us recently,
by Besieris et al.\cite{BASC}, wherein ordinary LTs were already
used, correctly, in the context of subluminal LWs [whilst the
Superluminal LTs used in that paper seem, however, to be partially
defective].}

\h Let us first confine ourselves to the cases faced in this Third
Part. Our subluminal localized pulses, that may be called ``wave
bullets", behave as {\em particles\/}: Indeed, our subluminal
pulses [as well as the luminal and Superluminal (X-shaped) ones,
that have been amply investigated in the past literature] do exist
as solutions of any wave equations, ranging from electromagnetism
and acoustics or geophysics, to elementary particle physics (and
even, as we discovered recently, to gravitation physics). \ From
the kinematical point of view, the velocity composition
relativistic law holds also for them. \ The same is true, more in
general, for any localized waves (pulses or beams).

\h Let us start for simplicity by considering, in an initial reference-frame
O, just a ($\nu$-order) Bessel beam:

\bb
\Psi(\rho,\phi,z,t) \ug J_\nu(\rho k_\rho) \; e^{i\nu\phi} \; e^{iz k_z} \;
e^{-i\om t}
\; . \label{eq.(41)} \ee    

In a second reference-frame O', moving with respect to (w.r.t.) O
with speed $u$ ---along the positive z-axis and in the positive
direction, for simplicity's sake---, it will be observed\cite{Saari2004*} the new Bessel beam

\bb \Psi(\rho ',\phi ',z ',t ') \ug J_\nu(\rho ' {k'}_{\rho'}) \;
e^{i\nu\phi '} \; e^{iz' {k'}_{z'}} \; e^{-i\om ' t'} \; ,
\label{eq.(42)} \ee    

obtained by applying the appropriate Lorentz transformation (a Lorentz
``boost") with \ $\gamma = [\sqrt{1-u^2/c^2}]^{-1}$:

\bb {k'}_{\rho'} = k_\rho; \ \ \ {k'}_{z'} = \ga (k_z-u\om/c^2); \
\ \ \om ' = \ga (\om - uk_z)  \; ;
\label{eq.(43)} \ee    

this can be easily seen, e.g., by putting

\bb
\rho = \rho '; \ \ \  z = \ga (z'+ut'); \ \ \  t = \ga (t'+ uz'/c^2)
\label{eq.(44)} \ee    

directly into eq.(154).

\h Let us now pass to subluminal {\em pulses}. We can investigate
the action of a Lorentz transformation (LT), by expressing them
either via the first method (Section 12) or via the second one
(Section 13). \ Let us consider for instance, in the frame O, a
$v$-speed (subluminal) pulse, given by eq.(155) of our Section 15.
\ When we go on to a second observer O' moving {\em with the same
speed} $v$ \ w.r.t. frame O, and, still for the sake of
simplicity, passing through the origin $O$ of the initial frame at
time $t=0$, the new observer O' will see the
pulse\cite{Saari2004*}

\bb \Psi(\rho ',z ',t ') \ug e^{-i t' {\om'}_0} \,
\int_{\om_{-}}^{\om_{+}} \drm \om \; S(\om) \; J_0(\rho '
{k'}_{\rho'}) \; e^{i z' {k'}_{z'}}
\; , \label{eq.(45)} \ee     

with

\bb
{k'}_{z'} = {\ga}^{-1} \om/v - \ga b/v; \ \ \ \om ' = \ga b; \ \ \
{k'}_{\rho'} = {\om'}_0/c^2 - {{k'}_{z'}}^2 \; ,
\label{eq.(46)} \ee    

as one gets from the Lorentz transformation in eq.(155), or in eq.(156),
with $u = v$ \ [and $\gamma$ given by Eqs.(123)]. \ Notice that ${k'}_{z'}$
is a function of $\om$, as expressed by the first one of the three
relations in the previous Eqs.(158); and that here $\om'$ is a constant.

\h If we explicitly insert into eq.(157) the relation \ $ \om =
\ga (v{{k'}_{z'}} + \ga b)$, \ which is nothing but a re-writing of the first
one of Eqs.(158), then eq.(157) becomes\cite{Saari2004*}

\bb \Psi(\rho ',z ',t ') \ug \ga v \; e^{-i t' \om_0} \;
\int_{-\om_0/c}^{\om_0/c} \drm {k'}_{z'} \;
\overline{S}({k'}_{z'}) \; J_0(\rho ' {k'}_{\rho'}) \; e^{i z'
{k'}_{z'}} \; ,
\label{eq.(47)} \ee     

where $\overline{S}$ is expressed in terms of the previous function $S(\om)$,
entering eq.(157), as follows:

\bb
\overline{S}({k'}_{z'}) \ug S(\ga v {k'}_{z'} + \ga^2 b)
\; .
\label{eq.(48)}
\ee     

Equation (159) describes monochromatic beams with axial symmetry
(and does coincide also with what derived within our second
method, in Section 13, when posing $v=0$).

\h The remarkable conclusion is that a subluminal pulse, given by
our eq.(117), which appears as a $v$-speed {\em pulse} in a frame
O, will appear\cite{Saari2004*} in another frame O' (travelling
w.r.t. observer O with the same speed $v$ in the same direction
$z$) just as the {\em monochromatic beam} in eq.(159) endowed with
angular frequency ${\om'}_0 = \ga b$, whatever be the pulse
spectral function in the initial frame O: \ even if the kind of
monochromatic beam, one arrives to, does of course depend$^{+}$
\footnotetext{$^{+}$ One gets in particular a Bessel-type beam when $S$
is a Dirac's delta-function: $S(\om)= \delta(\om-\om_0)$.
Moreover, let us notice that, on applying a LT to a Bessel beam,
one obtains another Bessel beam, with a different axicon-angle.}
on the chosen $S(\om)$. The vice-versa is also true, in general.

\h Let us set forth explicitly an observation that up to now has been
noticed only in ref.\cite{subArt1}.  Namely, let us mention
that, when starting not from eq.(117) but from the most general
solutions which
---as we have already seen--- are {\em sums} of solutions (117) over
the various values $b_m$ of $b$, then a Lorentz transformation
will lead us to {\em a sum} of monochromatic beams: actually, of
harmonics (rather than to a single monochromatic beam). \ In
particular, if one wants to obtain a sum of harmonic beams, one
has to apply a LT to more general subluminal pulses.

\begin{figure}[!h]
\begin{center}
 \scalebox{.576}{\includegraphics{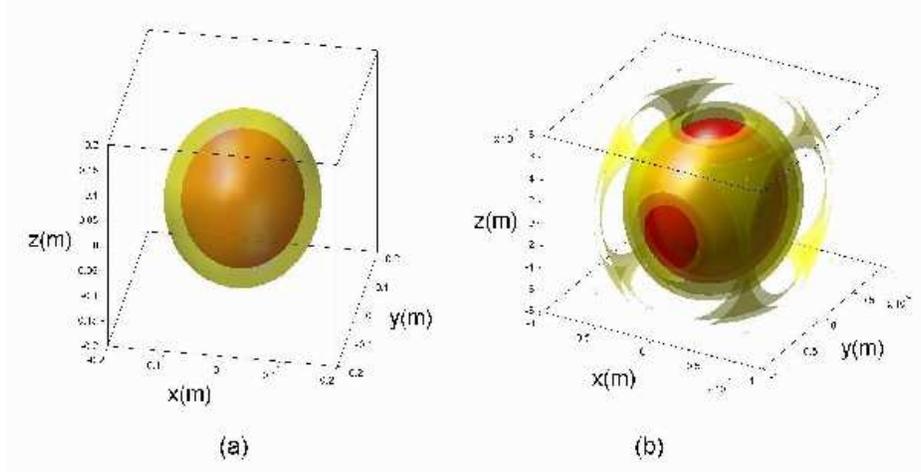}}
\end{center}
\caption{In the previous Figure we have seen how SR, in its
non-restricted version (ER), predicted\cite{Barut*,Review*} that,
while the {\em simplest} subluminal object is obviously a sphere (or, in
the limit, a space point), the simplest Superluminal object is on
the contrary an X-shaped pulse (or, in the limit, a double cone).
\ The circumstance that the Localized Solutions to the
{\em wave equations } do follow the same pattern is rather interesting,
and is expected to be useful ---in the case, e.g., of elementary
particles and quantum physics--- for a deeper comprehension of de
Broglie's and Schroedinger's wave mechanics. With regard to the
fact that the {\em simplest} subluminal LWs, solutions to the wave
equations, are ``ball-like", let us depict by these figures, in
the ordinary 3D space, the general shape of the Mackinnon's
solutions as expressed by eq.(124), numerically evaluated for
$v<<c$.  In figures (a) and (b) we graphically represent the field
iso-intensity surfaces, which in the considered case result to be
(as expected) just spherical.} \label{fig40}
\end{figure}

\h Let us add that {\em also} the various Superluminal localized
pulses get transformed\cite{Saari2004*} one into the other by the
mere application of ordinary LTs; while it may be expected that
the subluminal and the Superluminal LWs are to be linked (apart
from some known technical difficulties, that require a particular
caution) by the Superluminal Lorentz ``transformations" expounded
long ago, e.g., in refs.\cite{Review*,JWeber*,NCim*,Barut*} and refs.
therein.$^{++}$
\footnotetext{$^{++}$  One should pay attention that, as we were
saying, the topic of Superluminal LTs is a
delicate\cite{Review*,JWeber*,NCim*,Barut*} one, at the extent that
the majority of the {\em recent} attempts to re-address this question
and its applications seem to be defective (sometimes they do not
even keep the necessary covariance of the wave equation itself).}
\ Let us recall once more that, in the years 1980-82, special
relativity, in its non-restricted version, predicted that, while
the {\em simplest} subluminal {\em object} is obviously a sphere (or, in the
limit, a space point), the simplest Superluminal object is on the
contrary an X-shaped pulse (or, in the limit, a double cone): cf.
Fig.\ref{fig11}, taken$^{\diamond}$
\footnotetext{$^{\diamond}$  Let us recall, more
specifically, that Fig.11 depicts the following. Let us start from an
object that be intrinsically spherical, i.e., that is a sphere in
its rest-frame (Panel (a)). Then, after a generic subluminal LT
along $x$, i.e., under a subluminal $x$-boost, it is predicted by
Special Relativity (SR) to appear as ellipsoidal due to Lorentz
contraction (Panel (b)). After a Superluminal
$x$-boost\cite{Review*,JWeber*,NCim*} (namely, when this object
moves\cite{RMDartora*} with Superluminal speed $V$), it is
predicted by SR ---in its non-restricted version (ER)--- to
appear\cite{Barut*} as in Panel (d), i.e., as occupying the
cylindrically symmetric region bounded by a two-sheeted rotation
hyperboloid and an indefinite double cone. The whole structure,
according to ER, is expected to move {\em rigidly} and, of
course, with the speed $V$, the cotangent square of the cone
semi-angle being $(V/c)^2 - 1$. \ Panel (c) refers to the limiting
case when the boost-speed tends to $c$, either from the left or
from the right (for simplicity, a space axis is skipped). It is
remarkable that the shape of the localized (subluminal and
Superluminal) pulses, solutions to the {\em wave equations}, appears to
follow the same behaviour; this can have a role for a better
comprehension even of de Broglie and Schroedinger wave-mechanics.
\ See also Fig.40.}
from refs.\cite{Barut*,Review*}. \ The
circumstance that the localized solutions to the
{\em wave equations} follow the same behaviour is rather interesting,
and is expected to be useful ---in the case, e.g., of elementary
particles and quantum physics--- for a deeper comprehension of de
Broglie's and Schroedinger's wave mechanics. With regard to the
fact that the {\em simplest} subluminal LWs, solutions to the wave
equation, are ``ball-like", let us present in Figs.\ref{fig40}, in
ordinary 3D space, the general shape of the Mackinnon's
solutions, as expressed by eq.(124) for $v<<c$: In such figures we
graphically depict the field iso-intensity surfaces, which  result
to be (as expected) just spherical in the considered case.

\

\h We have also seen, among the others, that, even if our first method
(Section 12) cannot {\em directly} yield zero-speed envelopes, such envelopes
``at rest", eq.(145), can be however obtained by applying a $v$-speed LT
to eq.(130). In this way, one starts from many frequencies [eq.(130)] and
ends up with one frequency only [eq.(145)], since $b$ gets transformed into
{\em the} frequency of the monochromatic beam.

\

\

\

\

\section{Non-axially symmetric solutions: The case of higher-order
Bessel beams} 

\h Let us stress that till now we paid attention to exact
solutions representing axially-symmetric (subluminal) pulses only:
that is to say, to pulses obtained by suitable superpositions of
zero-order Bessel beams.

\h It is however interesting to look also for analytic solutions
representing {\em non}-axially symmetric subluminal pulses, which
can be constructed in terms of superpositions of $\nu$-order
Bessel beams, with $\nu$ a positive integer ($\nu>0$). \ This can be
attempted both in the case of Sect.12 (first method), and in
the case of Sect.13 (second method).

\h For brevity's sake, let us take only the first method (Sect.12)
into consideration. \ One is immediately confronted with the difficulty
that {\em no} exact solutions are known for the integral in eq.(122)
when $J_0(.)$ is replaced with $J_\nu (.)$.

\h One can overcome this difficulty by following a simple method,
which allows obtaining ``higher-order" subluminal waves in
terms of the axially-symmetric ones. \ Indeed, it is well-known
that, if $\Psi(x,y,z,t)$ is an exact solution to the ordinary wave
equation, then \ $\partial \Psi / \partial x$ \ and \
$\partial \Psi / \partial y$ \ are also exact
solutions.$^{\bullet}$
\footnotetext{$^{\bullet}$  Let us mention
that even \ $\partial^n \Psi /
\partial z^n$ \ and \ $\partial^n \Psi / \partial t^n$ will be
exact solutions.} \ By contrast, when
working in cylindrical co-ordinates, if $\Psi(\rho,\phi,z,t)$ is a
solution to the wave equation, quantities $\partial \Psi / \partial \rho$ \
and \ $\partial \Psi /
\partial \phi$ are {\em not} solutions, in general. \ Nevertheless, it
is not difficult at all to reach the noticeable conclusion that,
once \ $\Psi(\rho,\phi,z,t)$ \ is a solution, then also

\bb \overline{\Psi}(\rho,\phi,z,t) \ug e^{i\phi}\left(
\frac{\partial \Psi}{\partial \rho} +
\frac{i}{\rho}\frac{\partial\Psi}{\partial \phi}\right) \label{ho}
\ee  

\

is an exact solution! For instance, for an axially-symmetric solution
of the type $\Psi=J_0(k_{\rho}\rho)\,\exp[ik_z]\,\exp[-i\om t]$, equation
(\ref{ho}) yields $\overline{\Psi}=-k_{\rho}
\,J_1(k_{\rho}\rho)\,\exp[i\phi]\,\exp[ik_z]\,\exp[-i\om t]$,
which is actually one more analytic solution.

\h In other words, it is enough to start for simplicity from a
zero-order Bessel beam, and to apply eq.(\ref{ho}), successively,
$\nu$ times, in order to get as a new solution \ $\overline{\Psi}
= (-k_{\rho})^\nu \, J_\nu(k_{\rho}\rho)\,\exp[i \nu
\phi]\,\exp[ik_z]\,\exp[-i\om t]$, \ which is a $\nu$-order Bessel
beam. \

\h In such a way, when applying $\nu$ times eq.(\ref{ho}) to the
(axially-symmetric)
subluminal solution $\Psi(\rho,z,t)$ in Eqs.(130,129,128) \
[obtained from eq.(117) with spectral function $S(\om)$], we
get the subluminal non-axially symmetric pulses \
$\Psi_{\nu}(\rho,\phi,z,t)$ \ as new analytic solutions,
consisting as expected in superpositions of $\nu$-order Bessel
beams:

\bb  \Psi_{n}(\rho,\phi,z,t) \ug \int_{\om_-}^{\om_+} \drm \om \;
S'(\om)\;J_\nu(k_{\rho}\rho) \; e^{i \nu \phi} \; e^{ik_z z} \; e^{-i \om t} \; ,
\label{supho} \ee   

where $k_{\rho} (\om)$ is given by eq.(118), and quantities $S'(\om) =
(-k_{\rho}(\om))^\nu S(\om)$ are the spectra of the new pulses.
\ If $S(\om)$ is centered at a certain carrier frequency (it is a
gaussian spectrum, for instance), then $S'(\om)$ too will
approximately result to be of the same type.

\h Now, if we wish the new solution $\Psi_{\nu}(\rho,\phi,z,t)$ to
possess a pre-defined spectrum $S'(\om) = F(\om)$, we can first
take eq.(117) and put \ $S(\om) = F(\om) / (-k_{\rho}(\om))^\nu$ \
in its solution (130), and afterwards apply to it, $\nu$ times,
the operator \ $U \equiv \exp[i\phi] \; [\partial /
\partial\rho + (i / \rho) \partial /
\partial\phi)] \, $: \ As a result, we will obtain the desired
pulse, $\Psi_{\nu}(\rho,\phi,z,t)$, endowed with $S'(\om) =
F(\om)$.

\

{\bf An example:}

\

On starting from the subluminal axially-symmetric pulse
$\Psi(\rho,z,t)$, given by eq.(130) with the {\em gaussian}
spectrum (131), we can get the subluminal, non-axially symmetric,
exact solution $\Psi_{1}(\rho,\phi,z,t)$ by simply calculating

\bb \Psi_{1}(\rho,\phi,z,t) \ug {{\partial \Psi} \over
{\partial \rho}} \ \; e^{i\phi} \; , \label{ho1} \ee  

which actually yields the ``first-order" pulse $\Psi_{1}(\rho,\phi,z,t)$,
which can be more compactly written in the form:

\bb \Psi_{1}(\rho,\phi,\eta,\zeta) \ug 2 \, {b \over c} \, v \,
\gamma^2 \; {\exp{\left[ i{b \over c} \, \beta \, \gamma^2 \; \eta
\right]}} \sum_{n=-\infty}^{\infty} \; A_n \; \exp{[in{\pi \over
\beta}]} \ \; \psi_{1n}
\label{eq.(52)} \ee   

with

\bb \psi_{1n}(\rho,\phi,\eta,\zeta) \equiv \dis{ {b^2 \over c^2}
\, \gamma^2 \rho  \ \; Z^{-3} \ [ Z \, \cos Z - \sin Z ] \ \;
e^{i\phi} }
 \; , \label{eq.(53)} \ee  

where

\bb Z \equiv \dis{ \sqrt{{b^2 \over c^2} \, \gamma^2 \rho^2 +
\left({b \over c} \, \gamma^2 \zeta + n \pi  \right)^2 }} \; .
\label{eq.(54)}
\ee  

\begin{figure}[!h]
\begin{center}
 \scalebox{.35}{\includegraphics{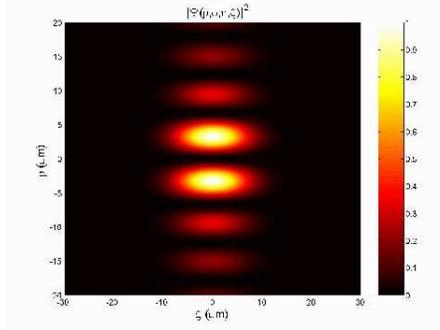}}
\end{center}
\caption{Orthogonal projection of the field intensity
corresponding to the higher order subluminal {\em pulse} represented by
the exact solution eq.(163), quantity $\Psi$ being given by eq.(128) with
the gaussian spectrum (131). The pulse intensity happens to have this
time a ``donut"-like shape.} \label{fig41}
\end{figure}

This exact solution, let us repeat, corresponds to
superposition (\ref{supho}), \ with \ $S'(\om) = k_{\rho}(\om)
S(\om)$, \ quantity $S(\om)$ being given by eq.(\ref{eq.(17)}). \
It is represented in Figure \ref{fig41}. The {\em pulse} intensity has
a ``donut-like" shape.

\

\

\subsection{A few concluding remarks about the Third Part}

In this Third Part we {\em started} developing, by suitable
superpositions of equal-frequency Bessel beams, a first
theoretical and experimental methodology to obtain localized {\em
stationary} wave fields, with high transverse localization, {\em
whose longitudinal intensity pattern can approximately assume any
desired shape} within a chosen interval $0\leq z \leq L$ of the
propagation axis $z$. Their intensity envelope remains static,
i.e., with velocity $v=0$; \ so that we named ``Frozen Waves" (FW)
these new solutions to the wave equations (and, in particular, to
the Maxwell equations). Inside the envelope of a FW only the
carrier wave does propagate: And the longitudinal shape, within
the interval $0\leq z \leq L$, can be chosen in such a way that no
nonnegligible field exists outside the pre-determined region
(consisting, e.g., in one or more high intensity peaks). Such
solutions are noticeable also for the different and interesting
applications they can have, especially in electromagnetism and
acoustics, such as optical tweezers, atom guides, optical or
acoustic bistouries, various important medical apparatus
(mainly for destroying cancerous cells), etc.

\h {\em Afterwards,} we have addressed the more general subject of the
{\em subluminal} Localized Waves, and shown that
---like in the well-known Superluminal case\cite{Livro*}---
the subluminal solutions can be obtained by superposing Bessel
beams\cite{subArt1}. Such solutions have been scarcely considered in the past,
for the reason that the superposition integral has to run in this
case over a finite interval (which makes mathematically difficult
to work out analytic expressions for them). \ We have shown,
however, how it is possible to obtain, in a simple way, non-diffracting
subluminal pulses as exact
analytic solutions to the wave equations:  For arbitrarily chosen
frequencies and bandwidths, and avoiding any recourse to the
backward-travelling components.

\h Indeed, till recent times only {\em one} closed-form subluminal LW solution,
$\psi_{\rm cf}$, to the wave equations was known\cite{Mackinnon*}: obtained
by choosing in the relevant integration a constant
weight-function $S(\om)$; whilst all other solutions had been previously
got only by numerical simulations.  On the contrary, a subluminal LW
can be obtained in closed form by adopting, for instance, any spectra
$S(\om)$ that be expansions
in terms of $\psi_{\rm cf}$. \ In fact, the initial disadvantage,
of having to deal with a limited bandwidth, may be turned into an advantage,
since in the case of ``truncated" integrals the spectrum $S(\om)$ can be
expanded in a Fourier series. \ More in general, it has been shown how
can one arrive at exact solutions both by integration over the Bessel
beams' angular frequency $\om$, and by integration over their
longitudinal wavenumber $k_z$. Both methods have been expounded above.
The first one appears to be comprehensive enough; we have
studied the second method as well, however, since it furnishes a new way,
in terms of continuous spectra, for tackling also the limiting case of zero-speed
solutions (i.e., for obtaining the Frozen Waves).

\h We have briefly treated the case, moreover, of non axially-symmetric
solutions, that is, of higher order Bessel beams.

\h At last, some attention has been paid to the role of Special
Relativity, and to the fact that the localized waves are expected to be
transformable one into the other by suitable Lorentz Transformations. \
Moreover, our results seem to show that in the subluminal case the {\em
simplest} LW solutions are (for $v<<c$) ``ball"-like, as expected
since long\cite{Barut*} on the mere basis of special
relativity\cite{Review*}. \ [Indeed, let us recall once more that already in the years
 1980-82 it had been predicted that, if the {\em simplest} subluminal
 object is a sphere (or, in the limit, a space point), then the simplest
 Superluminal object is an X-shaped pulse (or, in the limit, a double-cone);
 and viceversa: Cf. Figs.11.  It is
rather interesting that the same pattern appears to be followed by the
localized solutions of the {\em wave equations}].  For the subluminal case,
see, e.g., Figs.40.

 \h The subluminal localized pulses, endowed with a finite energy, or merely
 truncated, will be presented elsewhere.

\

\

\

{\bf Acknowledgements}\\

The authors are very grateful to Peter Hawkes for his kind invitation to
write this paper and encouragement, and to H.E.Hern\'andez-Figueroa
for his continuous help and collaboration over the years. \ For
useful discussions they wish to thank, among the others, also
I.M.Besieris, R.Bonifacio, R.Chiao, C.Cocca, C.Conti,
A.Friberg, F.Fontana, M.Ibison, G.Kurizki, I.Licata, J-y.Lu,
A.Loredo, M.Mattiuzzi, P.Milonni, S.Narayana, P.Nelli, P.Saari, A.M.Shaarawi,
E.C.G.Sudarshan, M.Tygel, and R.Ziolkowski. \ Finally, thanks are due to
M. Ten\'orio de Vasconselos and J. Marchi Madureira for their willingness
and patient cooperation.

\

\end{document}